\documentclass[useAMS,usenatbib]{mn2e}

\usepackage[pdftex,pdfpagemode={UseOutlines},bookmarks,bookmarksopen,colorlinks,linkcolor={blue},citecolor={blue},urlcolor={red}]{hyperref}
\usepackage{float}
\usepackage{graphicx}
\usepackage{caption}
\usepackage{subcaption}
\usepackage{multicol}
\usepackage{longtable}
\usepackage{natbib}
\usepackage{rotating,longtable,lscape}

\newcommand{\teff}{\mbox{$T_{\rm eff}$}}
\newcommand{\logg}{\mbox{$\log g_\star$}}

\newcommand{\mictrb}{\mbox{$\xi_{\rm t}$}}

\newcommand{\kms}{\mbox{km\,s$^{-1}$}}
\newcommand{\halpha}{\mbox{$H_\alpha$}}

\title[The WASP-53 and 81 systems]{Peculiar architectures for the WASP-53 and WASP-81 planet-hosting systems\thanks{using data collected at ESO's La Silla Observatory, Chile: HARPS on the ESO 3.6m (Prog IDs 087.C-0649, 089.C-0151, 090.C-0540, 091.C-0184 \& 093.C-0474), the ESO NTT (Prog ID 088.C-0204), the Swiss {\it Euler} telescope, and TRAPPIST. The data is publicly available at the \textit{CDS} Strasbourg and on demand to the main author.}}
\author[Amaury H.~M.~J. Triaud et al.]
{Amaury H.~M.~J. Triaud$^{1,2,3,4}$\thanks{E-mail: \href{mailto:aht34@cam.ac.uk}{aht34@cam.ac.uk}}, 
Marion Neveu-VanMalle$^{1,5}$,
Monika Lendl$^{6,1}$,
\newauthor
David R. Anderson$^{7}$,
Andrew Collier Cameron$^{8}$,
Laetitia Delrez$^{9}$,
Amanda Doyle$^{10}$,
\newauthor
Micha\"el Gillon$^{9}$,
Coel Hellier$^{7}$,
Emman\"uel Jehin$^{9}$,
Pierre F.~L. Maxted$^{7}$,
\newauthor
Damien S\'egransan$^{1}$,
Barry Smalley$^{7}$,
Didier Queloz$^{5,1}$,
Don Pollacco$^{10}$,
\newauthor
John Southworth$^{7}$,
Jeremy Tregloan-Reed$^{11,7}$,
St\'ephane Udry$^{1}$,
Richard West$^{10}$\\
$^{1}$Observatoire Astronomique de l'Universit\'e de Gen\`eve, Chemin des Maillettes 51, CH-1290 Sauverny, Switzerland\\
$^{2}$Institute of Astronomy, University of Cambridge, Madingley Road, CB3 0HA, Cambridge, United Kingdom\\
$^{3}$Centre for Planetary Sciences, University of Toronto at Scarborough, 1265 Military Trail, Toronto, ON, M1C 1A4, Canada\\
$^{4}$Department of Astronomy \& Astrophysics, University of Toronto, Toronto, ON, M5S 3H4, Canada\\
$^{5}$Cavendish Laboratory, J J Thomson Avenue, Cambridge CB3 0HE, UK\\
$^{6}$Space Research Institute, Austrian Academy of Sciences, Schmiedlstr. 6, 8042 Graz, Austria\\
$^{7}$Astrophysics Group, Keele University, Staffordshire, ST5 5BG, UK\\
$^{8}$SUPA, School of Physics \& Astronomy, University of St. Andrews, North Haugh, KY16 9SS, St. Andrews, Fife, Scotland, UK\\
$^{9}$Institut d'Astrophysique et de G\'eophysique, Universit\'e de Li\`ege, All\'ee du 6 Ao\^ut 17, Sart Tilman, 4000 Li\`ege 1, Belgium\\
$^{10}$Department of Physics, University of Warwick, Coventry CV4 7AL, UK\\
$^{11}$SETI Institute, Mountain View, CA, 94043, USA\\
}
\begin{document}

\date{Accepted ?. Received ?; in original form ?}

\pagerange{\pageref{firstpage}--\pageref{lastpage}} \pubyear{2014}

\maketitle

\label{firstpage}

\begin{abstract}
We report the detection of two new systems containing transiting planets. Both were identified by WASP as worthy transiting planet candidates. Radial-velocity observations quickly verified that the photometric signals were indeed produced by two transiting hot Jupiters. Our observations also show the presence of additional Doppler signals. In addition to short-period hot Jupiters, we find that the WASP-53 and WASP-81 systems also host brown dwarfs, on fairly eccentric orbits with semi-major axes of a few astronomical units. WASP-53c is over 16 $M_{\rm Jup} \sin i_{\rm c}$ and WASP-81c is 57 $M_{\rm Jup} \sin i_{\rm c}$. The presence of these tight, massive companions restricts theories of how the inner planets were assembled. We propose two alternative interpretations: a formation of the hot Jupiters within the snow line, or the late dynamical arrival of the brown dwarfs after disc-dispersal.

We also attempted to measure the Rossiter--McLaughlin effect for both hot Jupiters. In the case of WASP-81b we fail to detect a signal. For WASP-53b we find that the planet is aligned with respect to the stellar spin axis. In addition we explore the prospect of transit timing variations, and of using {\it Gaia}'s astrometry to measure the true masses of both brown dwarfs and also their relative inclination with respect to the inner transiting hot Jupiters.

\end{abstract}

\begin{keywords}
planetary systems -- planets and satellites: individual: WASP-81, WASP-53 -- binaries: eclipsing -- brown dwarfs
\end{keywords}

\section{Forewords}

\begin{figure*}
\center
\begin{subfigure}[b]{0.49\textwidth}
	\caption{WASP-53}\label{fig:wasp53lc}
	\includegraphics[width= \textwidth]{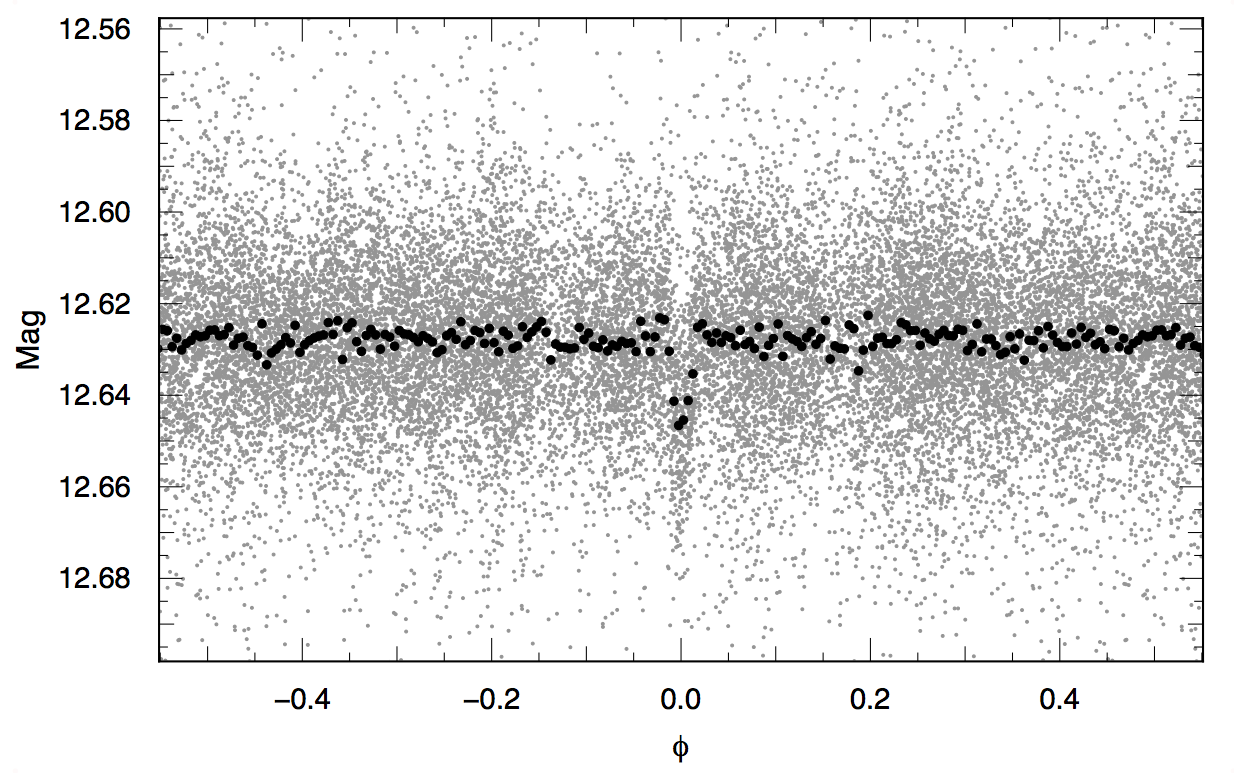}
\end{subfigure}
\begin{subfigure}[b]{0.49\textwidth}
	\caption{WASP-81}\label{fig:wasp81lc}
	\includegraphics[width= \textwidth]{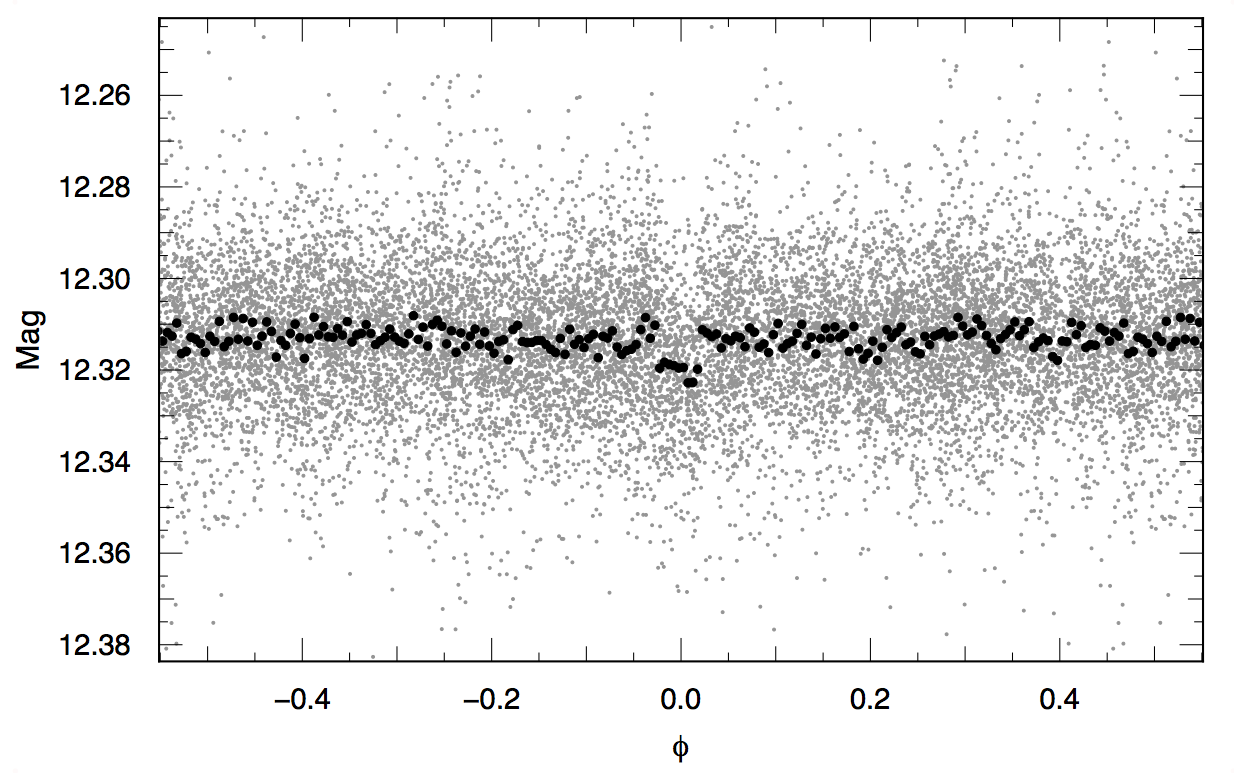}
\end{subfigure}
\caption{Variations in the magnitude of WASP-53 and WASP-81 leading to their  identification as transiting planet candidates. The grey dots show individual WASP measurements, whereas the black dots show the median magnitude within each of 200 phase bins.}\label{fig:wasp}  
\end{figure*} 

The discovery of 51 Peg b \citep{Mayor:1995uq} initiated a debate about the origin (formation and evolution) of hot Jupiters that continues to rage to this day. The question is whether they formed in situ, within the so-called {\it snow line} \citep{Bodenheimer:2000lr,Batygin:2015lr,Lee:2015kq}, or beyond, followed by inward migration \citep{Pollack:1996uq,Alibert:2005fk,Helled:2014kq}. The second hypothesis needs an explanation of how hot Jupiters have migrated, either via angular momentum transfer with their protoplanetary disc \citep{Lin:1996yq,Ward:1997kx,Baruteau:2014rp} or thanks to dynamical interactions followed by tidal circularisation (high-eccentricity migration) \citep{Rasio:1996ly,Wu:2007ve,Naoz:2011lr,Petrovich:2015qf}. Any framework has to explain why gas giants are found both close and far from their host star (not least Jupiter and Saturn), and also that they frequently orbit on planes that are inclined, sometimes retrograde, with respect to the equatorial plane of their host star \citep{Hebrard:2008mz,Winn:2009lr,Schlaufman:2010fk,Anderson:2010fj,Triaud:2010fr,Albrecht:2012lr,Lendl:2014yu,Winn:2015lr}.

Observations of the spin--orbit angle, thanks to the Rossiter--McLaughlin effect \citep{Queloz:2000rt}, were initially thought to provide a clean test between disc-driven migration and dynamical and tidal migration \citep{Gaudi:2007vn}. While some studies indicate that observations are compatible with planets undergoing orbital realignment \citep{Triaud:2011fk,Albrecht:2012lp,Dawson:2014kx}, additional theoretical arguments imply that misalignments can arise via a variety of processes. They can be primordial, with the planets forming on inclined planes \citep{Thies:2011yq,Lai:2011ul,Batygin:2012hl,Lai:2014nx,Spalding:2015ve}, or they can arise later \citep{Cebron:2011lr,Rogers:2013la}. Thus inclined hot Jupiters are compatible with both dynamical interactions and with disc-driven migration. 


Obviously, disc-driven migration and high-eccentricity migration could both be correct and each produce a fraction of the hot Jupiters, as recent observation appear to imply \citep{Anderson:2015lr}. Finally, results from \citet{Guillochon:2011fk} suggest that high-eccentricity migration requires in most cases some amount of disc-driven migration.

In this paper, we spend the first sections to describe the discovery of two new planetary systems. Both present a tight hierarchical architecture, composed of an inner, transiting, hot Jupiter, and an outer brown-dwarf companion. WASP-53 \& WASP-81 are reminiscent of a challenge proposed by \citet{Hatzes:2005ys} to test disc-driven migration. If both brown dwarfs were on their current orbits during the protoplanetary phase, they would have truncated the disc within the {\it snow line} thus preventing disc-driven migration and only allowing {\it in-situ} formation.  

In the final sections we describe a number of formation and evolution scenarios, some involving disc-driven migration and others not. We conclude that a number of scenarios will become testable soon, with the arrival of precise astrometric measurements produced by ESA's {\it Gaia} satellite. We also compute the equilibrium eccentricities of the inner gas giants \citep{Mardling:2007lr} and find it is unlikely that their $k_2$ Love number \citep{Batygin:2009qy} will be measured soon.

\section{WASP identification}\label{sec:wasp}

The WASP survey\footnote{ \href{https://wasp-planets.net}{https://wasp-planets.net}} \citep{Pollacco:2006fj} consists of two sets of eight 11-cm refractive telescope mounted together. One set is located at the Observatorio del Roque de Los Muchachos, La Palma (Spain), while the other is installed in Sutherland, hosted by the South African Astronomical Observatory. WASP has observed in excess of 30 million stars since 2004, thousands of times each. The photometric data reduction and the candidate selection are described in \citet{Collier-Cameron:2007pb}. 


WASP-53 (2MASS\,J02073820--2039426; K3, $J=10.959$) and WASP-81 (2MASS\,J20164989+0317385; F9, $J=11.263$) have been observed 21\,120 and 13\,292 times by WASP. They were two unremarkable and anonymous stars before a short, box-like photometric signals were identified at $P = 3.309866$ d and $P = 2.716554$ d, respectively. WASP-53b was sent for radial-velocity verification in 2010-08-10, and WASP-81b in 2011-05-09, with the first spectra acquired on 2010-12-05 and 2011-09-29. The WASP data are shown in Fig.~\ref{fig:wasp}.

\section{Photometric observations}\label{sec:phot}

All follow-up photometric timeseries, their dates, filters, number of data points and the detrending functions that were used in the analysis, are detailed in Table~\ref{tab:phot}. Figure~\ref{fig:phot53} shows the corrected photometry for WASP-53, and Figure~\ref{fig:phot81} shows WASP-81. Raw fluxes and residuals are shown in Figure~\ref{fig:phot53raw} \& \ref{fig:phot81raw}, respectively. Below, we provide some details on the observations and reduction, although we encourage readers to refer to cited papers for fuller information. 

\subsection{EulerCam}

EulerCam is mounted at the 1.2m {\it Euler} Swiss telescope located at ESO La Silla Observatory (Chile) It has a pixel scale of 0.215'' for a field of view of 14.7$\times$14.5'. The telescope is an alt-azimuthal design. EulerCam is mounted behind a field de-rotator. To ensure the best photometric precision, each star is kept on the same pixels, using a digital feedback scheme that compares the newly acquired frame with a composite of earlier frames and their offset from a recorded position.

We obtained four transits of WASP-53b, all using an r'-Gunn filter, and with a slightly defocused telescope to improve the observing efficiency and PSF sampling. Two of the transit, on 2011-09-22 and 2012-12-02 were scheduled to coincide with the radial-velocity time-series obtained using HARPS (see Sect~\ref{subsec:rv}).
WASP-81 was observed with EulerCam throughout three transits, one of which, on 2013-08-05, was obtained while HARPS collected a radial-velocity timeseries. Our 2012-06-07 observations were obtained using an $I$-Cousins filter, and a focused telescope. The telescope was defocused slightly for latter two observations, for which a z'-Gunn (2012-09-24) and an r'-Gunn (2013-08-05) were used.

All EulerCam images were reduced using standard image correction methods and lightcurves were obtained using differential aperture photometry, with a careful selection of aperture and reference stars. For further details on the EulerCam instrument and the associated data reduction procedures, please refer to \citet{Lendl:2012qy}. The times of observations are provided in JD(UTC), changed later to BJD(TDB) during the global analysis.

\subsubsection{A nearby stellar source}

\begin{figure}  
\center
\includegraphics[width= 0.45\textwidth]{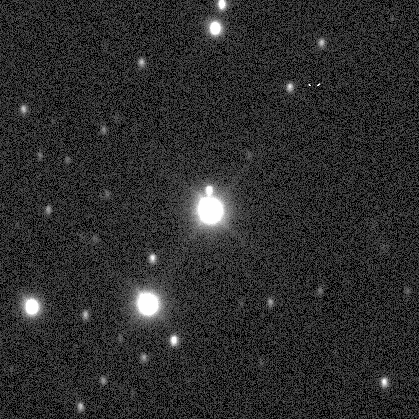}
\caption{A 1.5'x1.5' field of view centred on WASP-81, obtained with EulerCam, clearly showing a visual companion to the North. The pair is most likely unrelated.
}\label{fig:w81b_comp}  
\end{figure}

During spectroscopic observations an additional light source was noticed near WASP-81A on CORALIE's guiding camera images. On 2014-04-29 we obtained focused images of the WASP-81 system with EulerCam through Geneva $B$ (6 images), Geneva $V$ (3 images) and $r'$-Gunn (3 images) filters (Fig.~\ref{fig:w81b_comp}). 
We used \href{http://astrometry.net}{astronomy.net} \citep{Lang:2010yq} to calculate a precise astrometric solution and performed PSF fitting on the images using {\sc daophot} \citep{Stetson:1987kl}. The results are in Table~\ref{tab:W81B}.

We extracted magnitudes for the visual companion relative to the primary target. Combining with apparent magnitudes for WASP-81A (Table~\ref{tab:stellarparams}), and using a E($B-V$)=0.05 (Sect.~\ref{sec:spec}), we obtained apparent magnitudes for the visual companion. We find that the companion has colours consistent with a K3--K4 spectral type. If it is a dwarf, this implies a distance modulus of order 10.2 (1.1 kpc), compared to 8.0 (400 pc) for WASP-81A; { if} it were a giant it would be further away still.  If we place the companion on a main-sequence isochrone \citep{Marigo:2008lr}, and if WASP-81A were at the same distance, WASP-81A would have to be a 2-Gyr old, 1.4--1.6 $M_\odot$ star, contradicting our spectroscopic analysis as well as the mean stellar density obtained from the tranait. If, instead,  WASP-81A is on the main sequence, then the companion has to be below the main sequence to be at the same distance. 
The two objects are most therefore likely unrelated. We placed an HR diagram of the pair into the appendices.

For future reference we provide here the position of the visual companion. {\it Gaia} will soon produce parallaxes and proper motions that should confirm our analysis. If instead they are found at the same distance, implying they are gravitationally bound, then the companion must be an M dwarf with a much redder $B-V$ than the one we measured.

\begin{table}
\caption{Observation parameters for the companion source to WASP-81A}
\begin{tabular}{lrrrr} \hline
Filter  & $\Delta$ Mag & Mag  & separation & PA\\ 
	  & 			 & 		  & (arcsec) & (degrees)\\ 
\hline
$BG$	&	$5.64 \pm 0.03$	&	$18.78 \pm 0.07$	&	$4.34\pm0.02$&	$3.5\pm0.2$ 	\\
$VG$	&	$5.16 \pm 0.01$	&	$17.61 \pm 0.02$	&	$4.33\pm0.01$&	$3.5\pm0.3$	\\
$r'$		&	$4.87 \pm 0.01$	&	$17.14 \pm 0.04$	&	$4.32\pm0.01$&\vspace{0.5em} $3.5\pm0.3$	\\
\multicolumn{3}{l}{\it weighted mean:} 					&	$4.33\pm0.01$&	$3.5\pm0.3$	\\
\hline
\end{tabular}
\label{tab:W81B}
\end{table}

\begin{figure*}  
\center
\begin{subfigure}[b]{0.49\textwidth}
	\includegraphics[width= \textwidth]{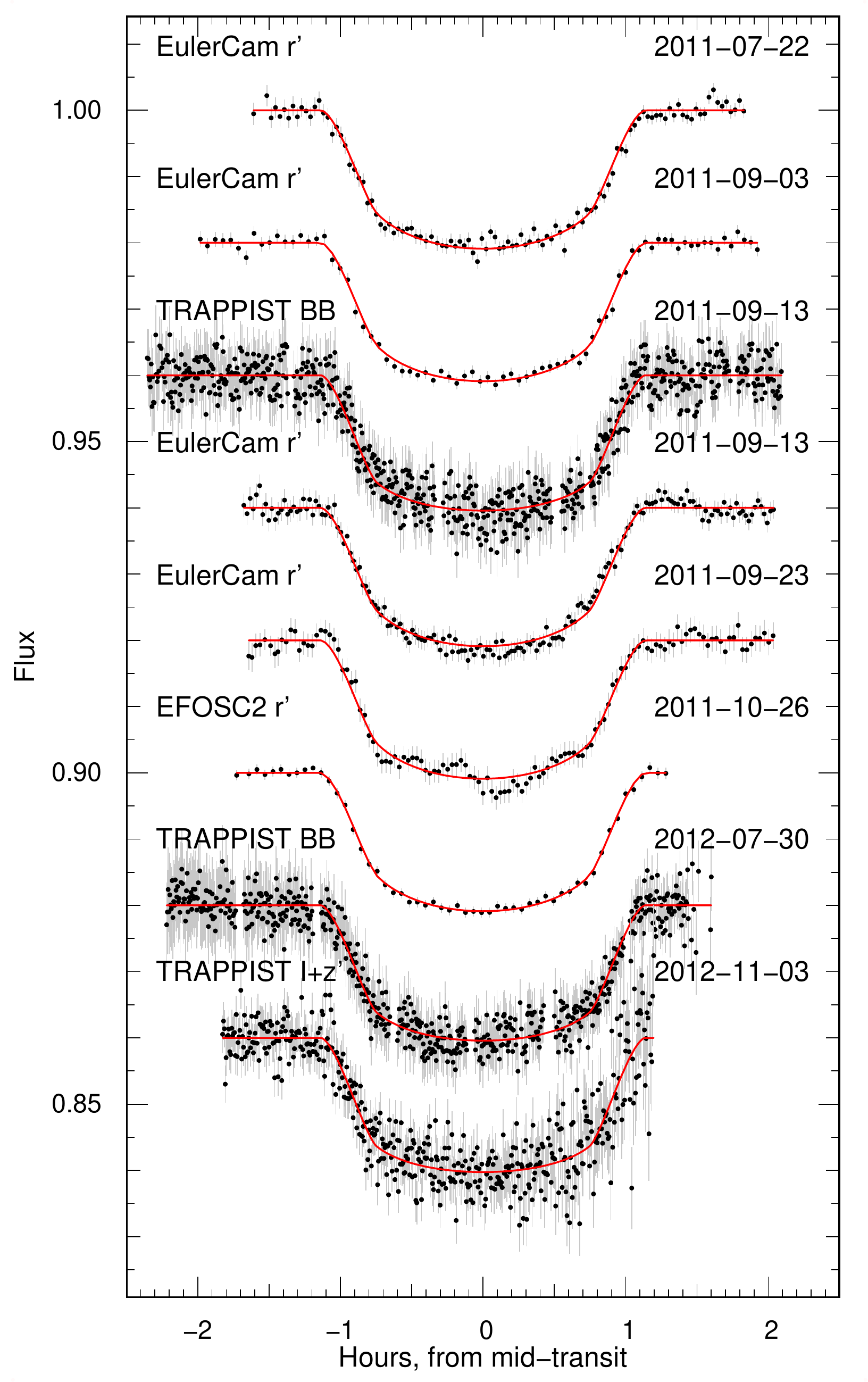}
\end{subfigure}
\begin{subfigure}[b]{0.49\textwidth}
	\includegraphics[width= \textwidth]{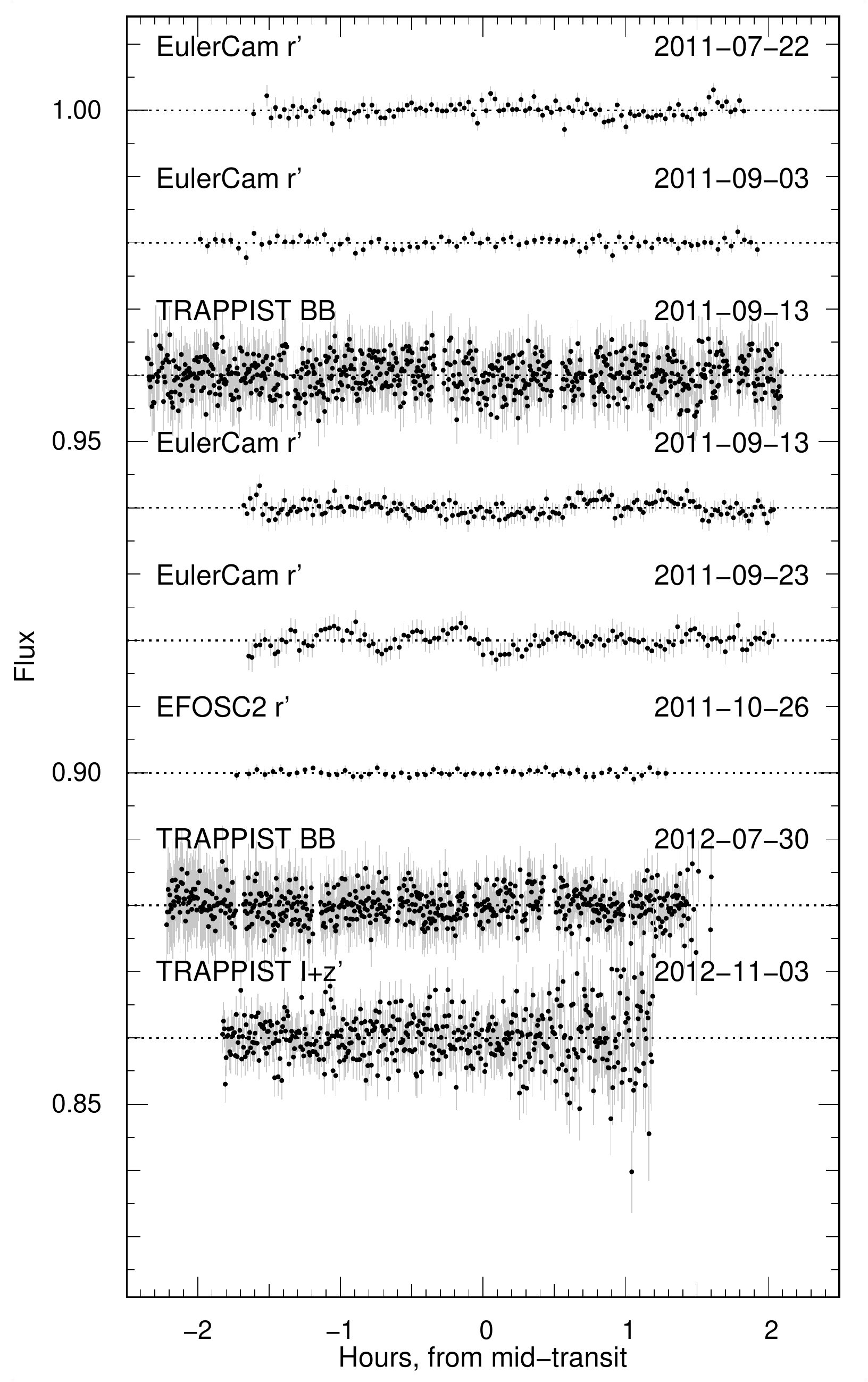}
\end{subfigure}
\caption{Photometry on WASP-53 at the time of transit, using EulerCam, TRAPPIST and the NTT. {\it Left}, we have the detrended data with, in red, the most likely model. {\it Right}, we show the residuals. Raw photometry and full models are available in Fig.~\ref{fig:phot53raw}. }\label{fig:phot53}  
\end{figure*} 

\begin{figure*}  
\center
\begin{subfigure}[b]{0.49\textwidth}
	\includegraphics[width= \textwidth]{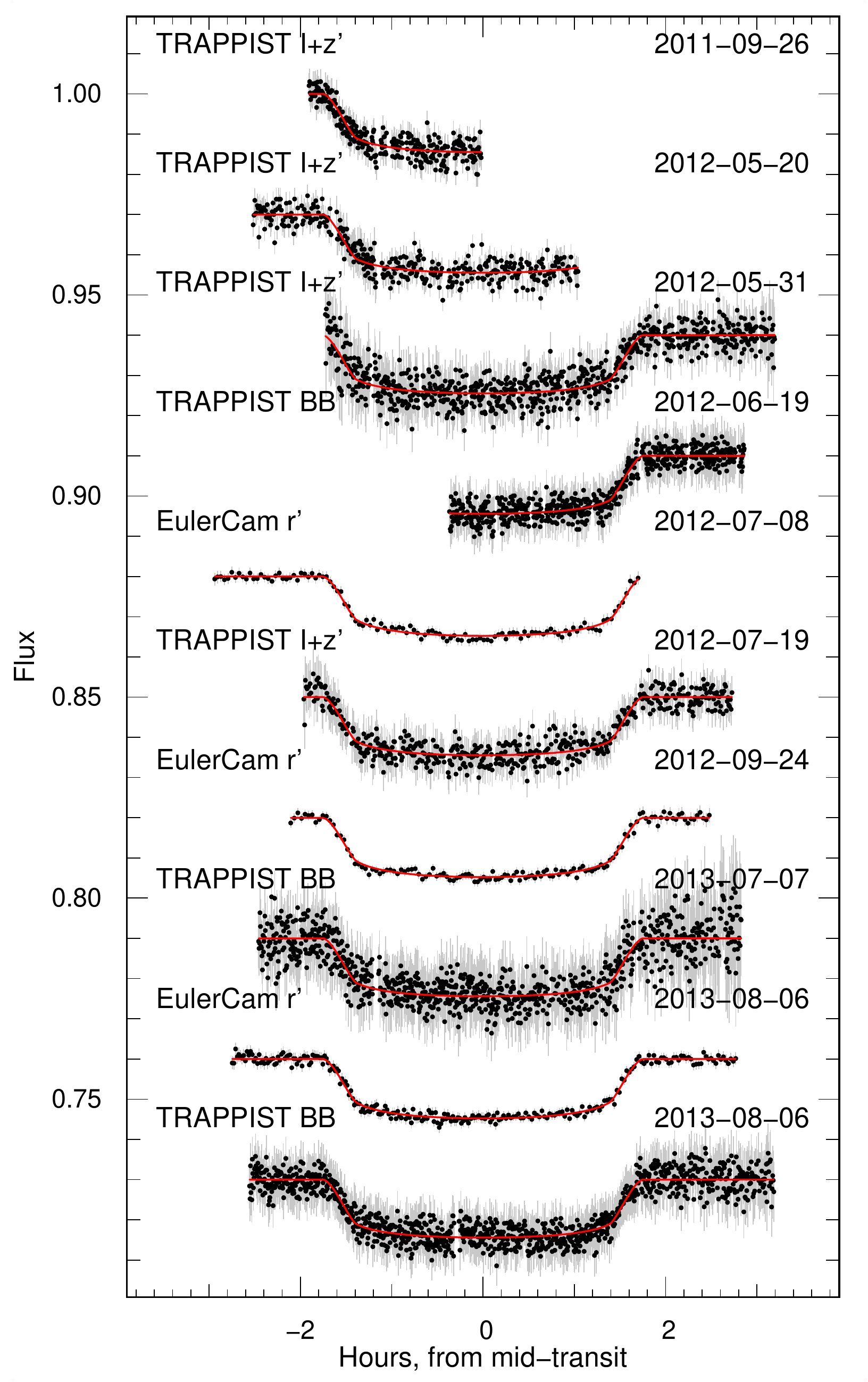}
\end{subfigure}
\begin{subfigure}[b]{0.49\textwidth}
	\includegraphics[width= \textwidth]{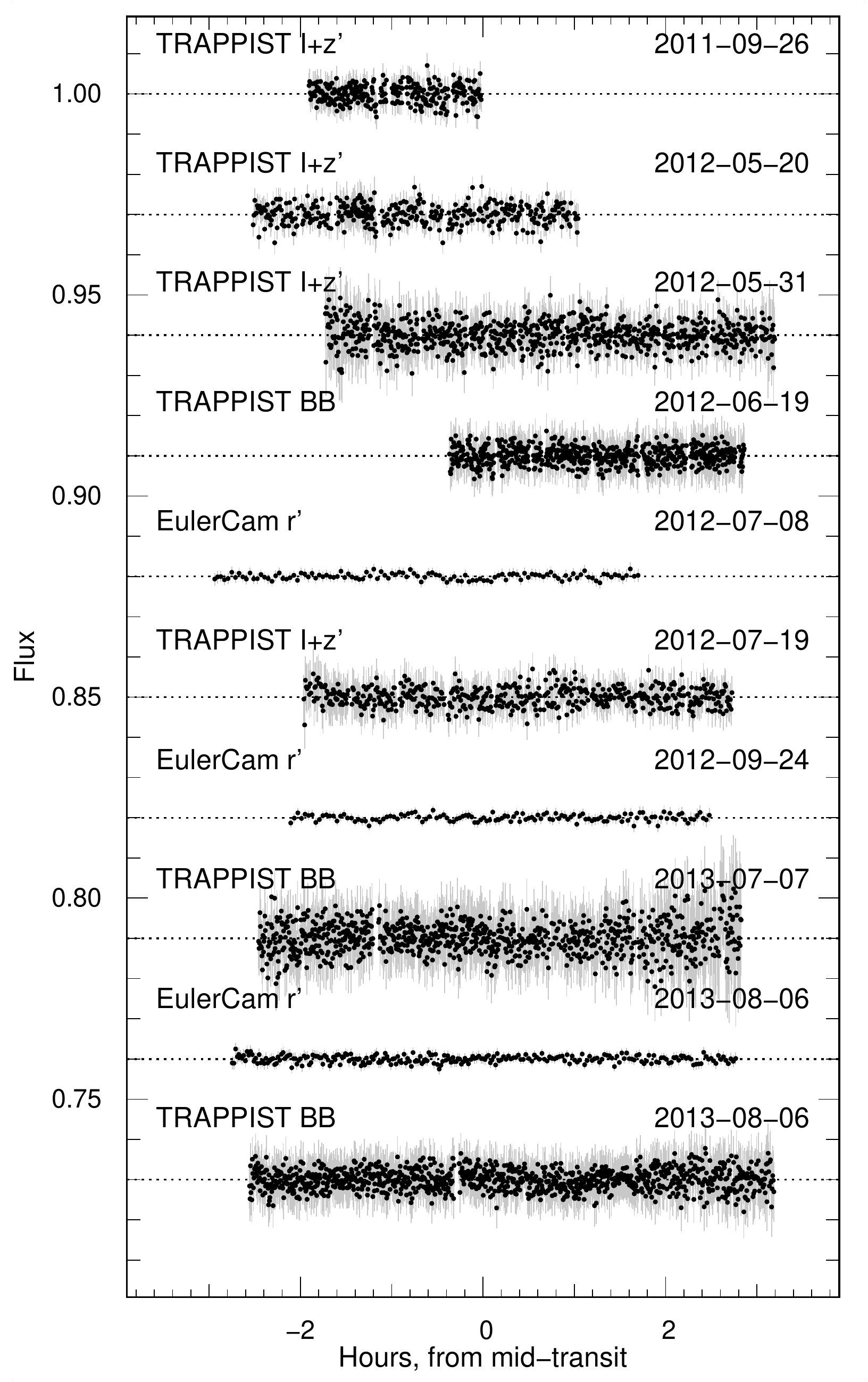}
\end{subfigure}
\caption{Photometry on WASP-81 at the time of transit, using EulerCam and TRAPPIST. {\it Left}, we show the detrended data with, in red, the most likely model. {\it Right}, we show the residuals. Raw photometry and full models are available in Fig.~\ref{fig:phot81raw}. }\label{fig:phot81}  
\end{figure*}

\subsection{TRAPPIST}
Both WASP systems were observed with the 0.6-m TRAPPIST robotic telescope (TRAnsiting Planets and PlanetesImals Small Telescope), also located at La Silla. Three transits were obtained on WASP-53, and seven on WASP-81. TRAPPIST is equipped with a thermoelectrically-cooled 2K$\times$2K CCD, which has a pixel scale of 0.6'' that translates into a 22'$\times$22' field of view. For details of TRAPPIST, see \citet{Gillon:2011qf} and \citet{Jehin:2011dk}. Two filters were used: a blue-blocking filter that has a transmittance of $> 90$\% from 500 nm to beyond 1000 nm, and an``$I + z'$" filter that has a transmittance of $> 90$\% from 750 nm to beyond 1100 nm. During the runs the positions of the stars on the chip were maintained to within a few pixels thanks to a``software guiding" system that regularly derives an astrometric solution for the most recently acquired image and sends pointing corrections to the mount if needed. After a standard pre-reduction (bias, dark, and flatfield correction), the stellar fluxes were extracted from the images using the {\sc iraf/daophot}\footnote{IRAF is distributed by the National Optical Astronomy Observatory, which is operated by the Association of Universities for Research in Astronomy, Inc., under cooperative agreement with the National Science Foundation.} aperture photometry software \citep{Stetson:1987kl}. For each light curve we tested several sets of reduction parameters and kept the one giving the most precise photometry for the stars of similar brightness as the target. After a careful selection of reference stars  the transit light curves were finally obtained using differential photometry. Some light curves were affected by a meridian flip; that is, the $180^\circ$ rotation that the German equatorial mount telescope has to undergo when the meridian is reached. This movement results in different positions of the stellar images on the detector before and after the flip, and thus in a possible jump of the differential photometry at the time of the flip. We have accounted for this in our light curve analysis by including a normalization offset in our model at the time of the flip (see Fig.~\ref{fig:phot53raw}~\&~\ref{fig:phot81raw} as well as Table~\ref{tab:phot}). More details on data acquisition and data reduction can be found in \citet{Gillon:2013yg} and \citet{Delrez:2014qf}, whose procedures were followed here as well.
The times of observations are provided in JD(UTC), changed later to BJD(TDB) during the global analysis.

\subsection{New Technology Telescope}

One transit of WASP-53 was observed using the EFOSC2 instrument on the NTT at ESO's observation of La Silla (ProgID 088.C-0204, PI Tregloan-Reed; see \citet{Tregloan-Reed:2013yq} for further details of this observing run). The instrument has a 2K$\times$2K CCD covering a 4.1'$\times$4.1' field of view, and a pixel scale of 0.12''. Observations were curtailed shortly after the end of the transit due to the onset of daytime. We observed through a Gunn $r$ filter (ESO filter \#784) with heavy defocussing and exposure times of 150s.

The data were reduced using the {\sc defot} pipeline \citep{Southworth:2009vn}, which utilises an aperture-photometry routine {\sc aper.pro} ported from {\sc daophot} \citep{Stetson:1987kl}. The radius of the inner aperture was 45 pixels and the sky annulus extended from 60 to 100 pixels. Debiassing and flat-fielding the data did not make a significant difference to the results, so we neglected these calibrations.

A differential-photometry light curve was obtained for WASP-53 versus an ensemble comparison star. Due to the small field of view of NTT/EFOSC2, we were able to use only three comparison stars, all of which were at least two magnitudes fainter than WASP-53 in the $r$-band. The weights of the comparison stars, used in summing their fluxes to create the ensemble comparison star, were optimised to minimise the scatter in the data outside transit. Finally, the timestamps were moved to the BJD(TDB) timescale using routines from \citet{Eastman:2010ys}.

\section{Spectroscopic observations}\label{sec:spec}

We collected 98 CORALIE spectra on WASP-53 between the dates of 2010-12-04 and { 2016-11-21}, as well as 83 HARPS spectra between the date of 2011-08-28 and 2014-09-28. On WASP-81, we acquired { 67} spectra with CORALIE from 2011-09-28 to 2015-07-08, and 32 using HARPS between 2013-04-20 and 2016-10-21.

\subsection{Spectral analysis}

The analysis was performed on the standard pipeline reduction products, using
the methods given in \citet{Doyle:2013lr}. The \halpha\ line was used to give an initial estimate of the effective temperature (\teff). The surface gravity (\logg) was
determined from the Ca~{\sc i} line at 6439{\AA}, along with the Na~{\sc i} D lines. Additional \teff\ and \logg\ diagnotics were performed using the Fe lines. An ionisation balance between Fe~{\sc i} and Fe~{\sc ii} was required, along with a null dependence of the abundance on either equivalent width or excitation potential. This null dependence was also required to determine the microturbulence (\mictrb). The parameters obtained from the analysis are listed in Table~\ref{tab:stellarparams}. Some of those parameters will later be employed as priors to compute the most likely physical parameters of each system. The elemental abundances were determined from equivalent width measurements of several clean and unblended lines, and additional least squares fitting of lines was performed when required. The quoted error estimates include that given by the uncertainties in \teff, \logg, and \mictrb, as well as the scatter due to measurement and atomic data uncertainties. 

The individual HARPS spectra of WASP-53 were co-added to produce a single spectrum with an average S/N in excess of 100:1. 
The macroturbulence was assumed to be zero, since for mid-K stars it is expected to be lower than that of thermal broadening \citep{Gray:2008fj}. The projected stellar rotation velocity ($v\,\sin\,i_\star$) was determined by fitting the profiles of several unblended lines, yielding an upper limit of 2.7 $\pm$ 0.3 ~\kms\ for WASP-53. .   There is no significant detection of lithium in WASP-53. The equivalent width upper limit of 13m\AA, correspond to an abundance upper limit of $\log A$(Li) $<$ 0.32 $\pm$ 0.16. This implies an age of at least several hundreds Myr \citep{Sestito:2005ys}.

Similarly, the combination of the HARPS spectra obtained on WASP-81 produced a combined spectrum with an average S/N of 60:1. Here, we used a macroturbulent value of $3.84\pm0.73$~\kms\ after a relation from \citet{Doyle:2014qf}. $v\,\sin\,i_\star$  was found to be $1.20\pm0.69$~\kms .

\begin{table}
\caption{Stellar parameters of WASP-53A and WASP-81A}
\begin{tabular}{lrr} \hline
Parameter  & WASP-53A & WASP-81A  \\ 
\hline
$\alpha$ 	&	$02^{\rm h}07'38.22''$	& $20^{\rm h}16'49.89''$			\\
$\delta$ 	&	$-20^\circ39'43.0	''$	& $+03^\circ17'38.7''$			\\
\hline
$m_{ B}$		& $17.46\pm0.30$ $^{\rm a} $	&$13.14\pm0.30$ $^{\rm a} $\\
$m_{ V}$		& $12.19\pm0.30$ $^{\rm b} $	&$12.29\pm0.10$ $^{\rm b} $\\	
$m_{ R}$		& $11.85\pm0.30$ $^{\rm a} $	&$12.67\pm0.30$ $^{\rm a} $\\	
$m_{ r'}$		& $12.29\pm0.30$ $^{\rm c} $	&$12.36\pm0.30$ $^{\rm c} $\\	
$m_{ I}$		& $11.653\pm0.020$ $^{\rm d}$	&$11.326\pm0.053$ $^{\rm b} $\\	
$m_{ J}$		& $10.959\pm0.026$ $^{\rm e}$	&$11.263\pm0.027$ $^{\rm e} $	\\
$m_{ H}$		& $10.474\pm0.022$ $^{\rm e}$	&$10.913\pm0.024$ $^{\rm e} $	\\
$m_{ K}$		& $10.390\pm0.023$ $^{\rm e}$	&$10.892\pm0.026$ $^{\rm e} $	\\
\hline
\teff 	(K)			&	$4950 \pm 60$		&	$5890 \pm 120$	\\
\logg\	(\kms)	&	$4.40 \pm 0.20$	&	$4.27 \pm 0.09$	\\
\mictrb	(\kms)	&	$0.60 \pm 0.25$	&	$0.94 \pm 0.15$	\\
$v\,\sin\,i_\star$	 (\kms) &	$< 2.7 \pm 0.3$		&	$1.20 \pm 0.73$	\\
\hline
{[Fe/H]}			&	$0.22 \pm 0.11$	&	$-0.36 \pm 0.14$	\\
{[Ca/H]}			&	$0.16 \pm 0.15$	&	$-0.25\pm0.09$	\\
{[Sc/H]}			&	$0.19 \pm 0.11$	&	$-0.18\pm0.18$	\\
{[Ti/H]}			&	$0.26 \pm 0.15$	&	$-0.14\pm0.11$	\\
{[V/H]}			&	$0.44 \pm 0.20$	&	$-0.29\pm0.12$	\\
{[Cr/H]}			&	$0.22 \pm 0.11$	&	$-0.40\pm0.11$	\\
{[Mn/H]}			&	$0.29 \pm 0.20$	&	$-0.55\pm0.09$	\\
{[Co/H]}			&	$0.24 \pm 0.11$	&	$-0.36\pm0.14$	\\
{[Ni/H]}			&	$0.20 \pm 0.12$	&	$-0.34\pm0.15$	\\
$\log A$(Li)		&	$< 0.32 \pm 0.16$ 	&	$1.21 \pm 0.10$	\\
\hline
Mass ($M_{\sun}$)	&	$0.87 \pm 0.08$	&	$1.04\pm0.09$		\\
Radius ($R_{\sun}$)	&	$0.96 \pm 0.24$	&	$1.24\pm0.15$		\\
Spectral Type		&	K3 				&	G1				\\
Distance	(pc)		&	$235 \pm 55$		&	 $410\pm70$		\\ 
\hline 
\\
\end{tabular}
\label{tab:stellarparams}
\newline { Note:} Mass and Radius estimate using the
\citet{Torres:2010uq} calibration. Spectral Type estimated from \teff\
using the table in \citet{Gray:2008fj}. Abundances are relative to the solar values obtained by \citet{Asplund:2009vn}.
\newline { references} a) NOMAD; \citep{Zacharias:2004fk}  b) TASS; \citep{Droege:2006yq} c) CMC14; ViZier I/304/out d) DENIS; \citep{DENIS-Consortium:2005fj} e) 2MASS; \citet{Skrutskie:2006kx}
\end{table}

\subsection{Radial velocities}\label{subsec:rv}

\begin{figure*}  
\center
\begin{subfigure}[b]{0.49\textwidth}
	\caption{WASP-53}\label{fig:W53obs}
	\includegraphics[width= \textwidth]{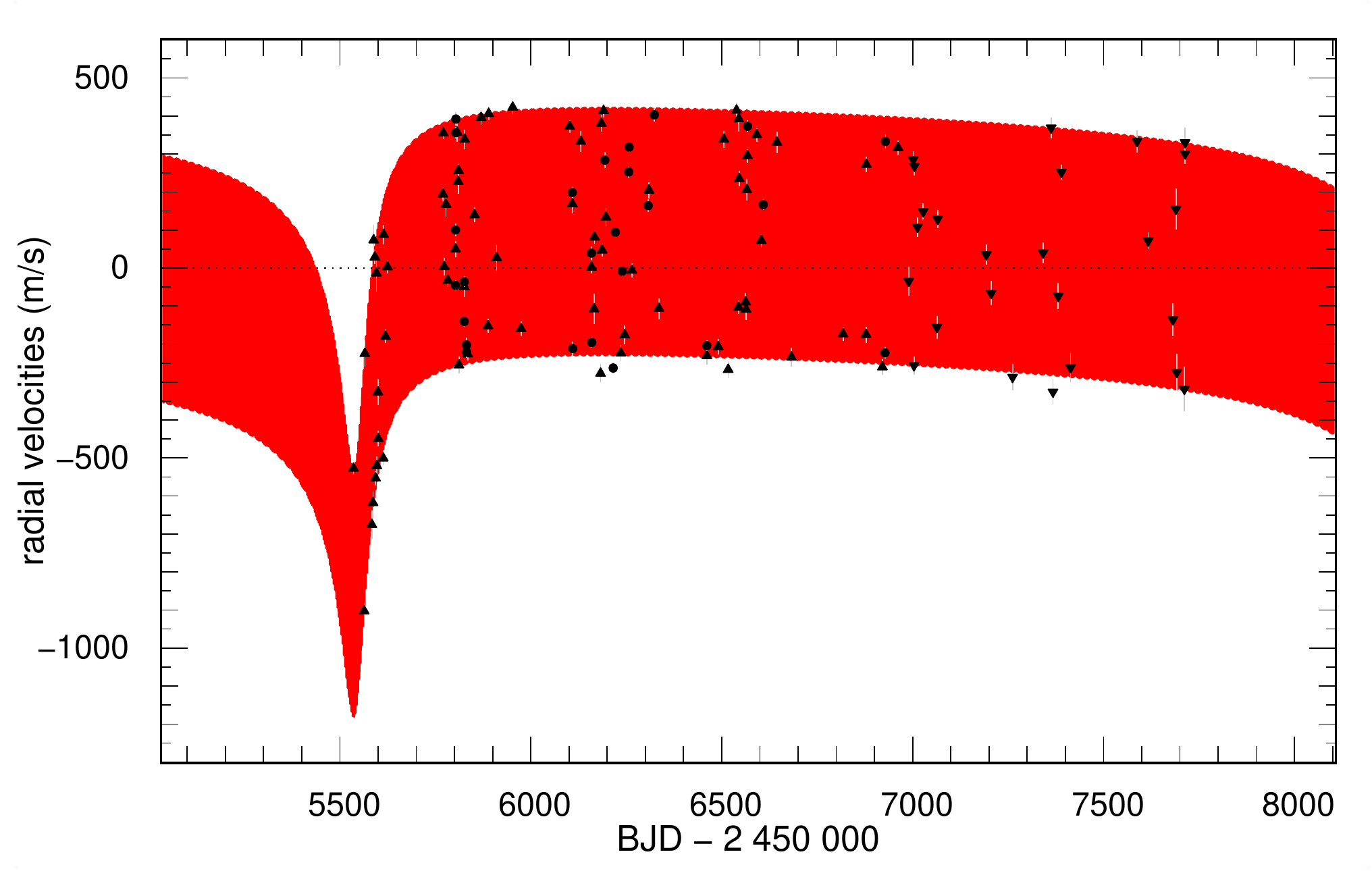}
\end{subfigure}
\begin{subfigure}[b]{0.49\textwidth}
	\caption{WASP-81}\label{fig:W81obs}
	\includegraphics[width= \textwidth]{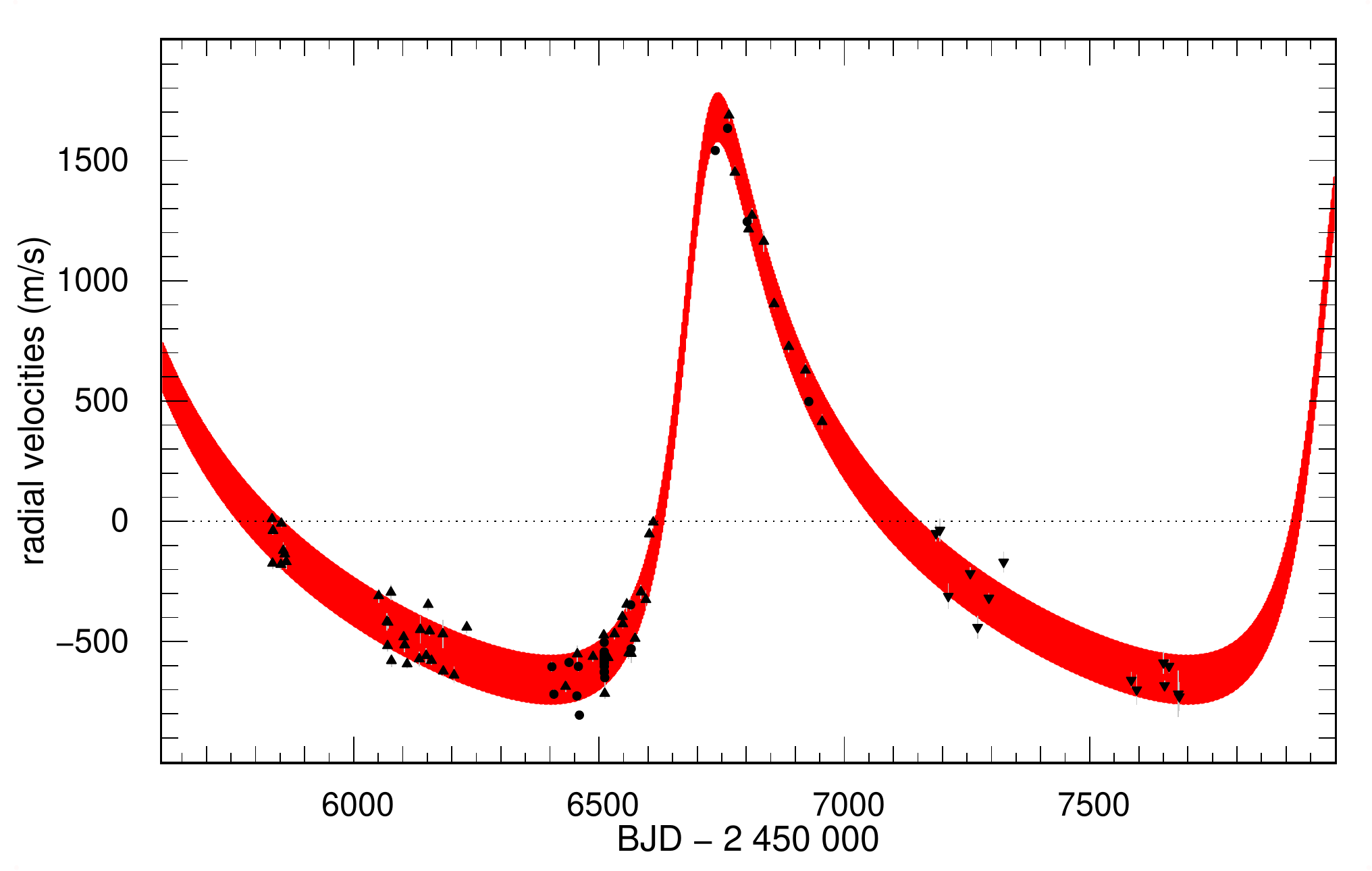}
\end{subfigure}
\begin{subfigure}[b]{0.49\textwidth}
	\includegraphics[width= \textwidth]{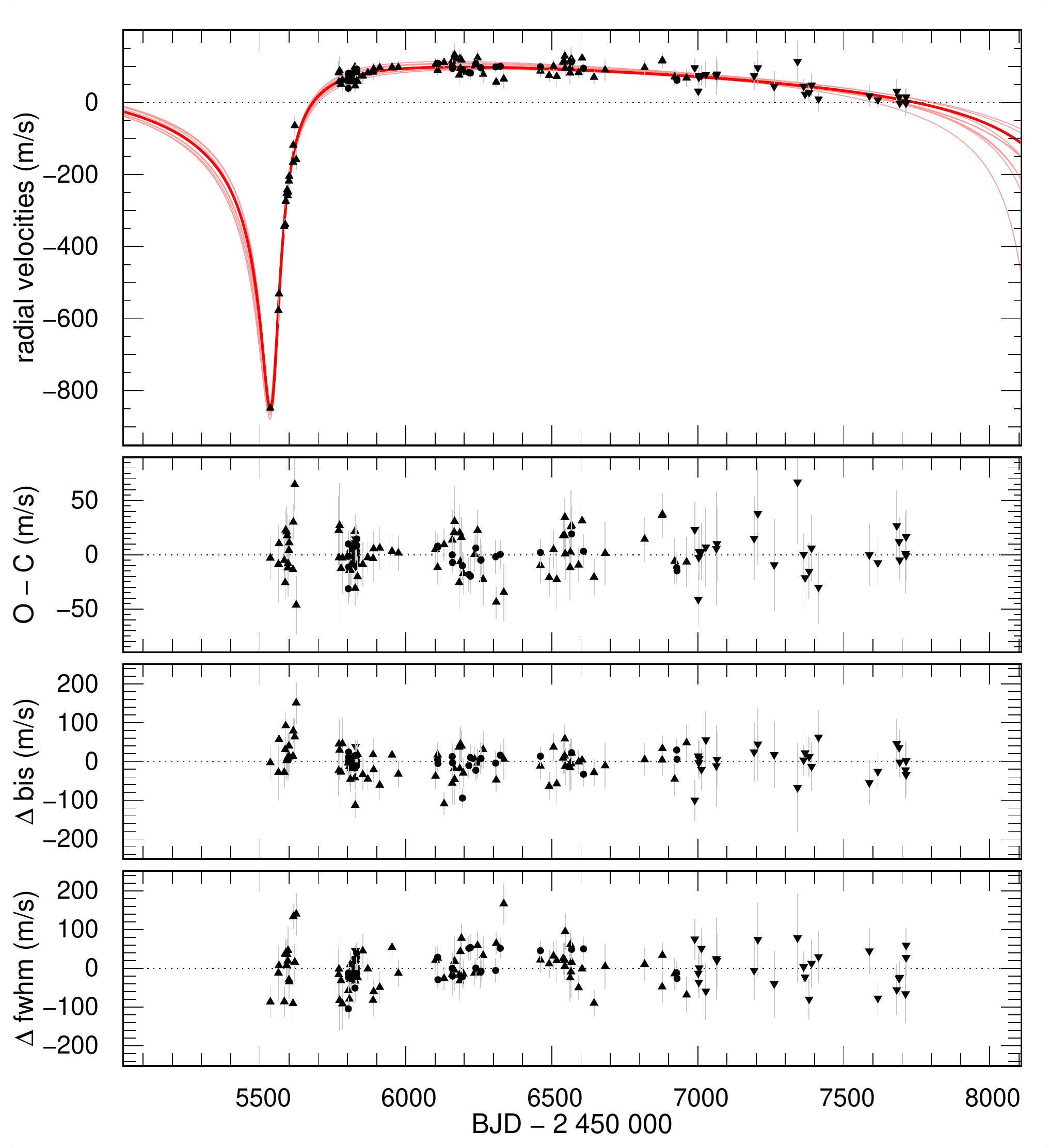}
\end{subfigure}
\begin{subfigure}[b]{0.49\textwidth}
	\includegraphics[width= \textwidth]{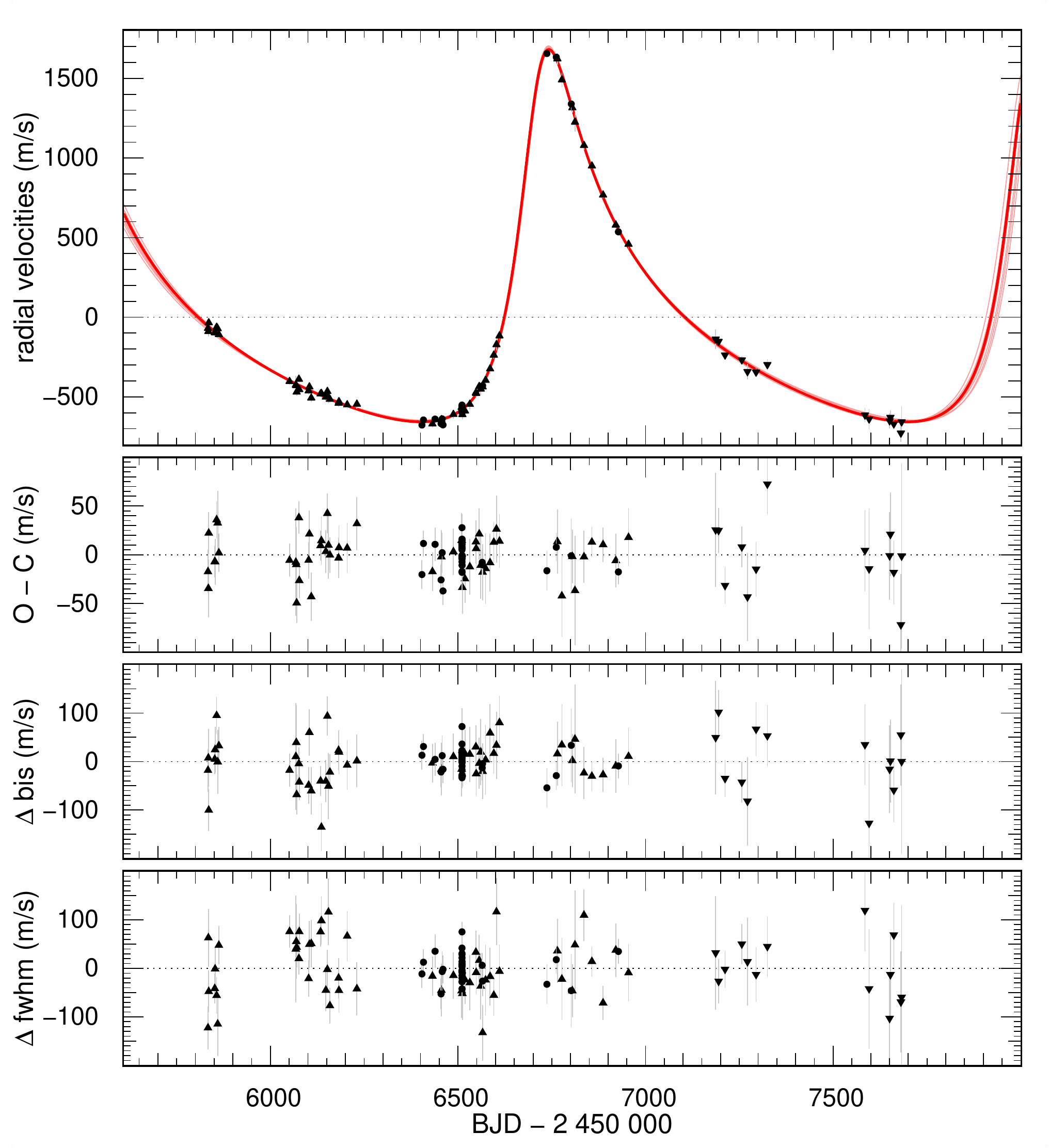}
\end{subfigure}
\caption{Radial velocities and models for WASP-53 ({\it left}) and WASP-81 ({\it right}). HARPS points are represented by discs, CORALIE is shown as triangles, pointed upwards prior to the upgrade and downwards for data acquired since. {\it top:} Radial velocity timeseries with the preferred two-planet model adjusted to the data. {\it middle:} same as on top, minus the inner planet. Ten alternative models, sampled randomly from the posterior are displayed in pink. Rossiter--McLaughlin sequences were removed from these plots; they can be found in Fig.~\ref{fig:ros}. Residuals are shown below the main plots. Further down, in order, we have the variation in the slope of the bisector span, and the variation in the FWHM of the cross-correlation function. }\label{fig:w53and81}  
\end{figure*}

The spectra were reduced using the standard CORALIE and HARPS reduction software. They have been shown to reach remarkable precision and accuracy, reaching below  1~m~s$^{-1}$ (e.g. \citet{Lovis:2006uq,Marmier:2013lr,Lopez-Morales:2014qv}). We extracted the radial velocities for both stars  by cross-correlating each spectrum with a binary mask. For WASP-53A we used a mask corresponding to a K5 spectral type, and for WASP-81A we employed a G2 mask. We fit the corresponding cross-correlation function with a Gaussian, whose mean provides us with the radial velocity \citep{Baranne:1996qa}. The corresponding values are displayed in the Journal of Observations (Appendix~\ref{app:obs}), along with observational data such as the individual exposure times. The observations are graphically presented in Fig.~\ref{fig:W53obs} for WASP-53 and in Fig.~\ref{fig:W81obs} for WASP-81.

\subsubsection{WASP-53}

\begin{figure*}  
\begin{center}
\begin{subfigure}[b]{0.45\textwidth}
	\caption{WASP-53b}\label{fig:W53b}
	\includegraphics[width=\textwidth]{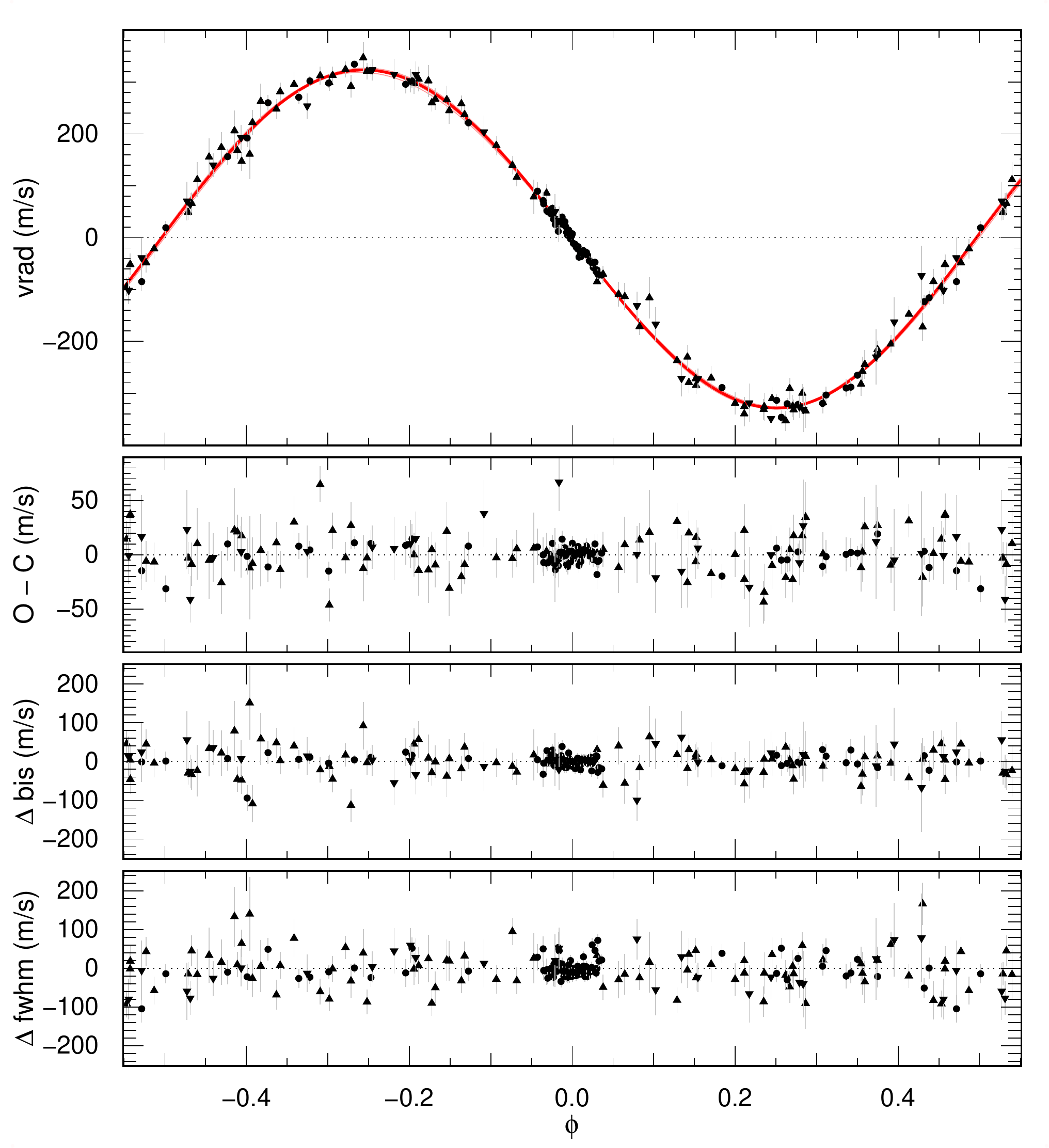}
\end{subfigure}
\begin{subfigure}[b]{0.45\textwidth}
	\caption{WASP-81b}\label{fig:W81b}
	\includegraphics[width=\textwidth]{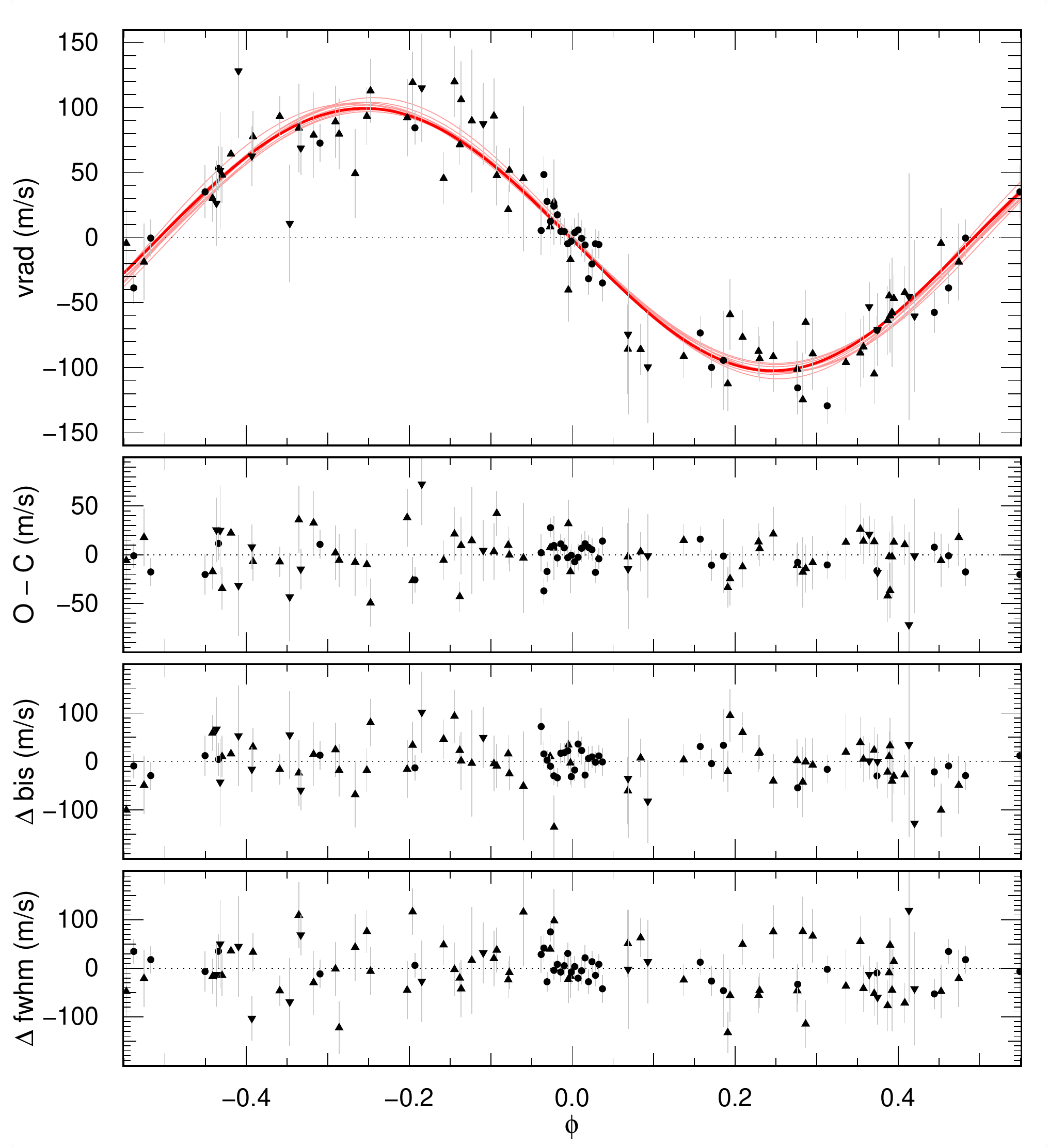}
\end{subfigure}
\caption{Radial velocity measurements phased with the orbit of the inner planet, after subtracting the variation due to the outer object. The preferred model is drawn in thick red, with another ten alternate models randomly picked from the posterior shown in pink.  Residuals are shown below the main plots. Further down we have the variation in the slope of the bisector span, and the variation in the FWHM of the cross-correlation function. The symbols are as in Fig.~\ref{fig:w53and81}. 
}\label{fig:phase}  
\end{center}
\end{figure*}

We first started to monitor WASP-53A with CORALIE, on 2010-12-05 targeting the orbital phase 0.75. The second spectrum was obtained 28 nights later, close to phase  0.25, and revealed a blue-shifted movement of nearly 400 m s$^{-1}$, compatible with a planetary object. The star was immediately flagged for intense follow-up, with the third spectrum acquired two nights after the second. The star was moving rapidly, and there was no sign of a change in the line width (FWHM), nor of its shape (span of the bisector slope), as can be visually inspected in Figs.~\ref{fig:w53and81} and \ref{fig:phase}. However, despite being obtained at phase 0.81, the star's velocity was puzzling,  being red-shifted by 300 m s$^{-1}$ compared to the first epoch. 
Our strategy has always been to follow any radial-velocity movement and identify its origin, planetary or not. Observations were continued. 

Before the observing season was over, we had confirmed a radial-velocity oscillation at a period matching the WASP signal (caused by WASP-53b), plus a rapid rise in the radial velocity. The following season we observed with both HARPS and CORALIE. WASP-53b's motion was quickly recovered and found to be in phase. Our monitoring continued and we observed the velocity of the star rise until it plateaued, indicating that an additional, massive, highly eccentric object had just finished passing through periastron. Our observations are on-going and, to this day, the velocity of the star has yet to decrease to the level observed at our first spectrum. In total we present { 98} spectra with CORALIE, including { 25} since an upgrade that saw the installation of new, octagonal, fibres (Nov 2014), and of a Fabry-P\'erot (Apr 2015) for the wavelength calibration throughout the night. We also gathered 83 spectra with HARPS, which include three timeseries obtained during transit, in order to capture the Rossiter--McLaughlin effect. 

The CORALIE data were divided into two independent datasets in order to account for any offset between before and after the upgrade. The HARPS set was divided into four sets, one for each Rossiter--McLaughlin effect (plus the measurement obtained the night before and after transit), and one set containing all the rest of the data. We obtained similar results by analysing the HARPS data as one set, however the division mitigates against any activity effect, such as done in \citet{Triaud:2009qy}. The Journal of Observations (Appendix~\ref{app:obs}) is separated in several tables according to the various sub-samples of radial-velocities.

\subsubsection{WASP-81}

CORALIE acquired our first spectrum on the WASP-81 system on 2011-09-29. The following night another was observed at the opposite phase. Its radial velocity revealed a variation and the target was placed in high priority. Within a month we had confirmed a variation at the WASP photometric period, and the star had set. The following season, we intended to monitor just enough to confirm that the oscillation was still in phase and to start routine long-term monitoring, as in the case of WASP-47 \citep{Neveu-VanMalle:2016xy}. The first measurement had a value clearly below expectations so we resumed an intense follow-up. As with WASP-53 we requested observing time on HARPS and monitored the system in parallel with CORALIE.
On the third season, as we had predicted, the velocity reached a minimum and started rising. HARPS was the first instrument on sky for the fourth season. We had anticipated that the system would have returned to a similar velocity as in the first points. We were surprised to find that the star was nearly 2~km~s$^{-1}$ higher than in the previous season. Shortly after, velocities started to drop and the outer orbit closed earlier this year. In total { 67} spectra were collected with CORALIE, { 14} of which were after the upgrade. With HARPS we gathered 32 measurements, of which 19 were obtained during a single night as an attempt to detect the Rossiter--McLaughlin effect. 

As with WASP-53 we separate the CORALIE data into two sets, before and after upgrade. Since we do not detect the R--M effect, all of the HARPS points are analysed as part of the same set. Refer to the Journal of Observations (Appendix~\ref{app:obs}) for further details.

\begin{figure*}  
\begin{center}
\begin{subfigure}[b]{0.45\textwidth}
	\caption{WASP-53b}\label{fig:W53ros}
	\includegraphics[width=\textwidth]{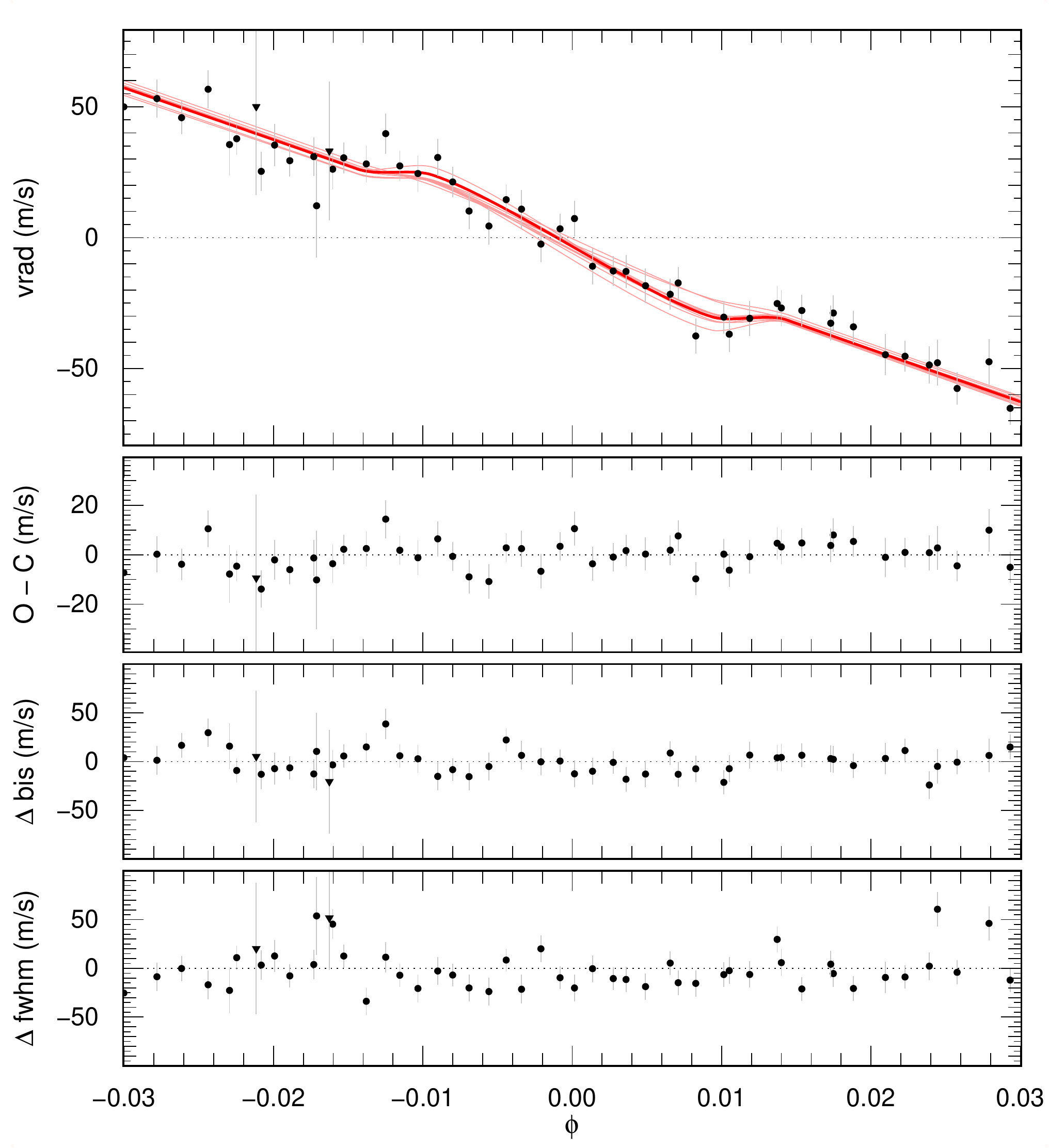}
\end{subfigure}
\begin{subfigure}[b]{0.45\textwidth}
	\caption{WASP-81b}\label{fig:W81ros}
	\includegraphics[width=\textwidth]{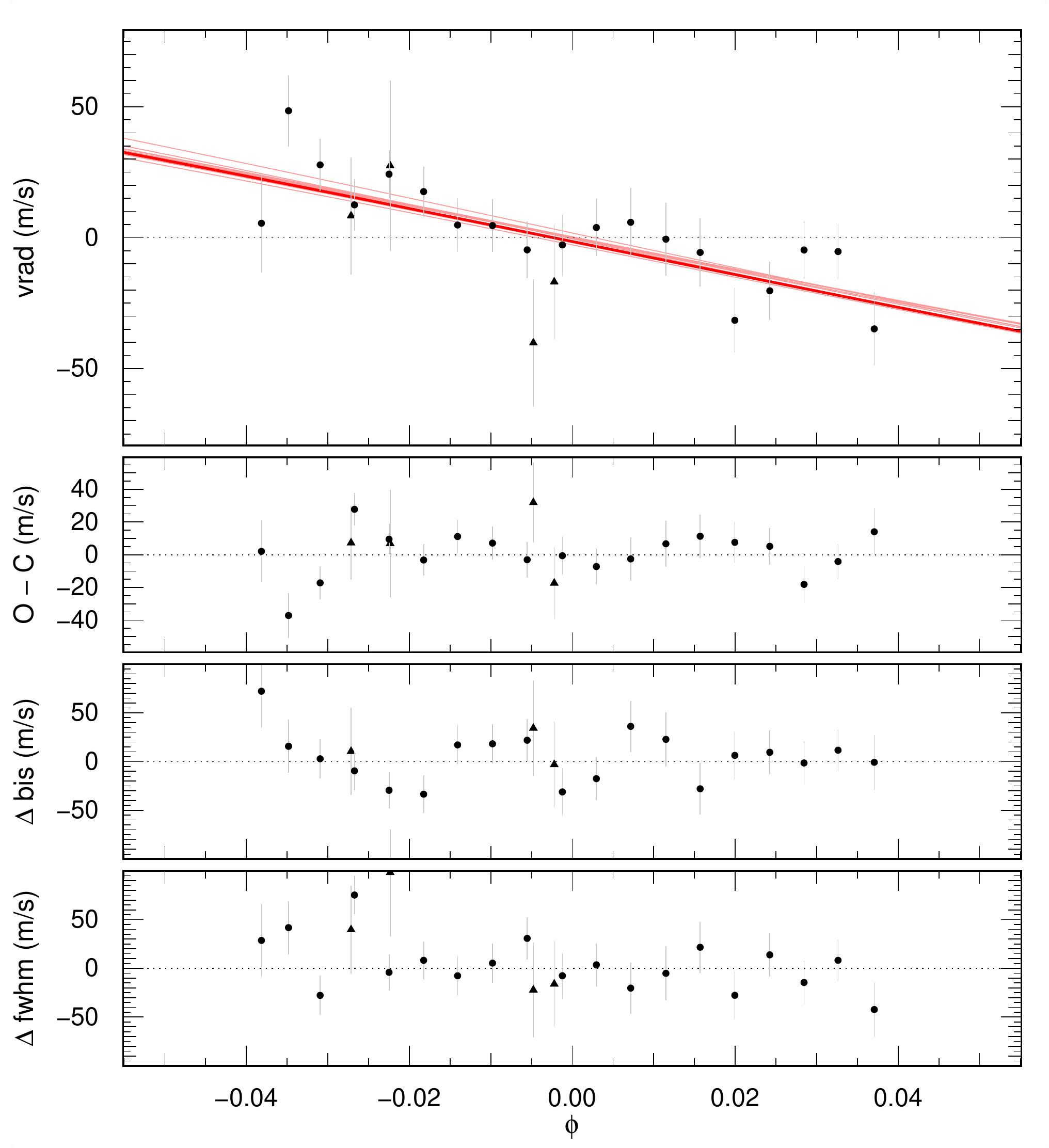}
\end{subfigure}
\caption{As for Fig.~\ref{fig:phase} but zoomed around transit time to show the Rossiter--McLaughlin effect. 
}\label{fig:ros}  
\end{center}
\end{figure*}

\section{Global analysis}\label{sec:analysis}

We analysed all of the photometric data and all of the radial velocity data together. We estimated the observed and physical parameters of the system using an MCMC algorithm, as detailed in \citet{Gillon:2012fj}. Our modus operandi is similar with the difference described below.

The photometric lightcurves were modelled using the formalism of \citet{Mandel:2002kx} for a transiting planet. The TRAPPIST and EulerCam lightcurves timestamps were transformed into BJD(TDB) from JD(UTC). The limb darkening was included in the form of a quadratic law, and its parameters were allowed to float, within the constraints of priors. The priors were computed by interpolating the data tabulated by \citet{Claret:2004fk}, consistent with the stellar parameters of Table~\ref{tab:stellarparams}. In addition, on each photometric timeseries, we allowed for a quadratic polynomial as a function of time in order to adjust for differential extinction relative to the ensemble of comparison stars. After this treatment some lightcurves still contained a significant residual scatter. For these we tried a number of other detrending functions, selecting them on the basis of a reduction in the global Bayesian Information Criterion (BIC thereafter; \citet{Schwarz:1978zz}). Those comparisons were performed by systematically using the same starting seed for the random elements. Details about which functions were selected and applied can be found in Appendix~\ref{app:models}. For those lightcurves where any new degree of complexity led to a worsened BIC, but where nevertheless the overall $\chi^2$ implied a poor fit, we { scaled\footnote{ we realise that this approach is not self-consistent, but our method is sufficient for small re-adjustements of the uncertainties and unambiguous detections.}.  our} error bars so as to approach a general reduced $\chi^2_{\rm r} = 1$. { The uncertainty increase ensures that our confidence intervals are not under-estimated. Those correction factors are available in Table~\ref{tab:phot}. A similar approach is used for the radial-velocities, by quadratically adding a {\it jitter} term to some sequences. Two of the nine sequences required a jitter of order 1.5-2 m s$^{-1}$}.

Individual detrended lightcurves and their transit model can be visually inspected in Fig.~\ref{fig:phot53}~\&~\ref{fig:phot81}. Residuals and fully modelled lightcurves are presented in Appendix~\ref{app:models}.


The radial velocities around the orbit were adjusted with two eccentric Keplerian functions (as in \citet{Hilditch:2001uq}), neglecting Newtonian effects. Since we cover less than an orbital period for WASP-53c and barely one for WASP-81c, we did not include a long term trend. Its inclusion leads to higher BIC values and its slope is mostly unconstrained. The Rossiter--McLaughlin effect was computed using the code written by \citet{Gimenez:2006kx}, following the formalism of \citet{Kopal:1942ai}. 
For WASP-81b the effect was not detected and the final fit does not include it.
For both systems we find the inner planets' orbits to be consistent with circular. Although allowing for eccentricity adds two parameters and increases the BIC, we nevertheless let these parameters float so as to include their uncertainties when marginalising the other parameters.

The MCMC's jump parameters are mostly set to match observables, which are then converted to physical parameters to compute the relevant models. $D$ is the transit depth, $b$ the impact parameter, $W$ the transit width. $T_0$ is the mid-transit time and $P$ the orbital period. 
We combine, the eccentricity $e$ and the angle of periastron $\omega$ into the pair $\sqrt{e}\cos \omega$, $\sqrt{e}\sin \omega$ which helps when exploring small eccentricities \citep{Triaud:2011vn}. Similarly we also construct the pair $\sqrt{v \sin I_\star}\cos \beta$, $\sqrt{v \sin I_\star}\sin \beta$ to model the Rossiter--McLaughlin effect. $\beta$ is the projected spin--orbit angle, and $v \sin I_\star$ the measure of the projected rotation velocity of the star, which in principle should match $v \sin i_\star$ from the spectral analysis. 
Instead of the semi-amplitude $K$ we use the jump parameter $K_2$ such that $K_2 = K \sqrt{1 - e^2} P^{1/3}$. This helps reduce some correlation between the parameters \citep{Ford:2006yq} helping with the exploration of parameter space.
For similar reasons we combine the limb-darkening coefficients into $c_1 = 2u_1 + u_2$, and $c_2 = u_1 - 2u_2$ following the recommendation of \citet{Holman:2006qy}. In the case of two {\it Euler} lightcurves for WASP-53 we also fit a sine function through the data, with a period $P$ and a $T_0$. Additional subscript indicate which lightcurves those are for.


All jump parameters are well constrained within one scale. Except for $P_{\rm c}$ and $K_{2{\rm ,c}}$, we sample our parameters' posteriors using Gaussian priors, whose variance is set with an initial Gibbs sampler. As an improvement over \citet{Gillon:2012fj} $P_{\rm c}$ and $K_{2{\rm ,c}}$ were sampled using non-informative priors in $\log$ space, otherwise known as {\it Jeffrey} priors. However their values remained largely within one error bar, and the application of Gaussian steps does not lead to qualitatively different results. 
The metallicity [Fe/H], and effective temperature $T_{\rm eff}$ complete this list of jump parameters. They are controlled by priors obtained from Table~\ref{tab:stellarparams} and used to compute at every MCMC step a stellar mass and a stellar radius produced in a fashion similar to \citet{Torres:2010uq}.

\begin{table}
  \caption{Median and $1\sigma$ confidence regions for the jump parameters that evolve in our MCMC chains. Errors on the last two digits of each parameters are given in brackets. { Parameters not dependent on information contained within the adjusted data are highlighted with asterisks} 
   }\label{tab:jump}
  \begin{tabular}{lrrl}
  \hline
  \hline
  Parameters (Units) & WASP-53 & WASP-81 &\\
  \hline
  {\it the star} &&\\
${\rm [Fe/H]}$	(dex)		&$0.22_{(-0.11)}^{(+0.11)}$	&$-0.36_{(-0.14)}^{(+0.14)}$		&*	\\
$T_{\rm eff}$	(K)		&$4953_{(-60)}^{(+59)}$		&$5870_{(-120)}^{(+120)}$	&*	\\
$c_{1,JR}$			&-- 						&$0.918_{(-33)}^{(+33)}$		&*	\\
$c_{2,JR}$			&-- 						&$-0.280_{(-23)}^{(+23)}$	&*	\\
$c_{1,BB}$			&$1.069_{(-38)}^{(+37)}$ 		&$0.772_{(-57)}^{(+57)}$		&*	\\
$c_{2,BB}$			&$0.031_{(-30)}^{(+31)}$ 		&$-0.321_{(-38)}^{(+37)}$	&*	\\
$c_{1,I+z}$			&$0.995_{(-61)}^{(+62)}$ 		&$0.807_{(-57)}^{(+55)}$		&*	\\
$c_{2,I+z}$			&$-0.020_{(-37)}^{(+37)}$ 	&$-0.297_{(-32)}^{(+30)}$	&*	\\
$c_{1,r'}$				&$1.316_{(-29)}^{(+29)}$ 		&--	&*	\\
$c_{2,r'}$				&$0.272_{(-28)}^{(+28)}$ 		&--	&*	\\
\hline
  {\it inner planet}\\
$D_{\rm b}$						&$0.01831_{(-34)}^{(+33)}$		&$0.01254_{(-26)}^{(+27)}$		&	\\
$b_{\rm b}$		($R_\odot$)		&$0.562_{(-22)}^{(+20)}$	 		&$0.15_{(-0.10)}^{(+0.10)}$			&	\\
$W_{\rm b}$		(d)				&$0.09469_{(-64)}^{(+65)}$		&$0.14501_{(-64)}^{(+73)}$		&	\\
$T_{\rm 0,b}$ (BJD -2\,450\,000)		&$5943.56695_{(-12)}^{(+11)}$	&$6195.57462_{(-20)}^{(+20)}$	&	\\
$P_{\rm b}$		(d)				&$3.3098443_{(-20)}^{(+20)}$		&$2.7164762_{(-23)}^{(+23)}$		&	\\
$\sqrt{e_{\rm b}}\cos \omega_{\rm b}$	&$-0.070_{(-21)}^{(+24)}$		&$-0.026_{(-76)}^{(+83)}$		&	\\
$\sqrt{e_{\rm b}}\sin \omega_{\rm b}$	&$-0.085_{(-33)}^{(+46)}$		&$-0.00_{(-0.12)}^{(+0.12)}$			&	\\
$K_{\rm2, b}$		($m s^{-1} d^{1/3}$)	&$485.9_{(-2.7)}^{(+2.7)}$		&$140.7_{(-4.6)}^{(+4.7)}$		&	\\
$\sqrt{v \sin I_\star}\cos \beta_{\rm b}$	&$0.91_{(-0.13)}^{(+0.11)}$			&--	&	\\
$\sqrt{v \sin I_\star}\sin \beta_{\rm b}$	&$-0.07_{(-0.19)}^{(+0.19)}$			&--	&	\\
\hline
  {\it outer planet}\\
$T_{\rm 0,c}$ (BJD -2\,450\,000)		&$5456_{(-13)}^{(+11)}$	&$6936.5_{(-2.5)}^{(+2.6)}$	&	\\
$P_{\rm c}$		(day)				&$2840_{(-130)}^{(+170)}$	&$1297.2_{(-7.8)}^{(+8.1)}$	&	\\
$\sqrt{e_{\rm c}}\cos \omega_{\rm c}$	&$-0.867_{(-12)}^{(+12)}$	&$0.5871_{(-42)}^{(+40)}$	&	\\
$\sqrt{e_{\rm c}}\sin \omega_{\rm c}$		&$-0.292_{(-30)}^{(+30)}$	&$-0.4609_{(-61)}^{(+61)}$	&	\\
$K_{\rm 2,c}$		($m s^{-1} d^{1/3}$)	&$3685_{(-89)}^{(+110)}$		&$10\,557_{(-50)}^{(+51)}$	&	\\
\hline
jump parameters: & 23 & 21 &\\
\hline
\end{tabular}
\end{table}

\begin{table}
  \caption{{ Final estimates for the median and $1\sigma$ confidence regions}, for various interesting parameters of the WASP-53 and WASP-81 systems. They were estimated from the posteriors of the jump parameters outlined in Table~\ref{tab:jump}. Errors on the last two digits of each parameters are given in brackets. Upper limits are for $3\sigma$ confidence. { Parameters not dependent on information contained within the adjusted data are highlighted with asterisks}}\label{tab:res}
  \begin{tabular}{lrrc}
  \hline
  \hline
  Physical Parameters & WASP-53 & WASP-81 \\
  \hline
  {\it the star}\\
$M_\star$			($M_\odot$)		&$0.839_{(-54)}^{(+54)}$	& $1.080_{(-58)}^{(+59)}$	&*\\
$R_\star$			($R_\odot$)		&$0.798_{(-23)}^{(+23)}$	& $1.283_{(-37)}^{(+40)}$	&\\
$\rho_\star$		($\rho_\odot$)		&$1.648_{(-85)}^{(+91)}$	& $0.513_{(-37)}^{(+30)}$	&\\
$L_\star$			($L_\odot$)		&$0.344_{(-23)}^{(+23)}$	& $1.76_{(-0.18)}^{(+0.20)}$	&\\
\\
$T_{\rm eff}$	(K)			&$4953_{(-60)}^{(+60)}$		&$5870_{(-120)}^{(+120)}$	&*\\
$\log g_\star$	(cgs)			&$4.553_{(-20)}^{(+19)}$		& $4.258_{(-27)}^{(+22)}$	 &*\\
${\rm [Fe/H]}$	(dex)			&$0.22_{(-0.11)}^{(+0.11)}$	&$-0.36_{(-0.14)}^{(+0.14)}$	&*	\\
$ v \sin I_\star$	(km s$^{-1}$)	&$0.86_{(-0.21)}^{(+0.21)}$		& --	 &\\
\hline
{\it inner planet}\\
$P_{\rm b}$		(day)				&$3.3098443_{(-20)}^{(+20)}$		&$2.7164762_{(-23)}^{(+23)}$		\\
$T_{\rm 0,b}$ (BJD -2\,450\,000)		&$5943.56695_{(-12)}^{(+11)}$	&$6195.57462_{(-20)}^{(+23)}$	\\
$ K_{\rm b}$		(m s$^{-1}$)		&$326.1_{(-1.8)}^{(+1.8)}$		& $100.8_{(-3.3)}^{(+3.4)}$	\\
\\
$M_{\rm b}$		($M_{\rm Jup}$)	&$2.132_{(-94)}^{(+92)}$	& $0.729_{(-35)}^{(+36)}$	\\
$R_{\rm b}$ 		($R_{\rm Jup}$)	&$1.074_{(-37)}^{(+37)}$	& $1.429_{(-46)}^{(+51)}$	\\
$\rho_{\rm b}$		($\rho_{\rm Jup}$)	&$1.72_{(-0.13)}^{(+0.15)}$	& $0.250_{(-23)}^{(+23)}$	\\
$\log g_{\rm b}$	(cgs)				&$3.680_{(-22)}^{(+22)}$	& $2.967_{(-30)}^{(+27)}$	\\
$ a_{\rm b}/R_\star$					&$11.05_{(-0.19)}^{(+0.20)}$	& $6.56_{(-0.16)}^{(+0.13)}$	\\
$T_{\rm b,eq}$		(K)				&$1053_{(-16)}^{(+16)}$	& $1623_{(-37)}^{(+38)}$	\\
\\
$ a_{\rm b}$		(AU)				&$0.04101_{(-91)}^{(+83)}$	& $0.03908_{(-72)}^{(+70)}$	\\
$ i_{\rm b}$		(deg)			&$87.08_{(-0.15)}^{(+0.16)}$		& $88.69_{(-0.92)}^{(+0.88)}$	\\
$ \beta_{\rm b}$	(deg)			&$-4_{(-12)}^{(+12)}$		& --	\\
$ e_{\rm b}$						&$<0.030$				& $<0.066$	\\
\hline
{\it outer planet}\\
$P_{\rm c}$		(day)				&$> 2840_{(-130)}^{(+170)}$	&$1297.2_{(-7.8)}^{(+8.1)}$	\\
$T_{\rm 0,c}$ (BJD -2\,450\,000)		&$5456_{(-13)}^{(+11)}$		&$6936.5_{(-2.5)}^{(+2.6)}$	\\
$ K_{\rm c}$		(m s$^{-1}$)		&$> 475.6_{(-8.0)}^{(+8.2)}$	& $1169.3_{(-6.6)}^{(+6.9)}$	\\
\\
$M_{\rm c} \sin i_{\rm c}$	($M_{\rm Jup}$)	&$> 16.35_{(-0.82)}^{(+0.85)}$		& $56.6_{(-2.0)}^{(+2.0)}$	\\
$ a_{\rm c}$		(AU)					&$> 3.73_{(-0.14)}^{(+0.16)}$		& $2.426_{(-45)}^{(+44)}$	\\
$ e_{\rm c}$							&$0.8369_{(-70)}^{(+69)}$		& $0.5570_{(-44)}^{(+44)}$	\\
$ \omega_{\rm c}$	(deg)					&$198.6_{(-2.0)}^{(+2.0)}$	& $321.86_{(-0.51)}^{(+0.52)}$	\\
\hline
\end{tabular}
\end{table}

For our final analysis we set 10 chains of 100\,000 steps, starting from different seeds. All converged to similar BIC values. The first 20\,000 steps were systematically removed (to allow for burn-in), and the remainder were analysed leading to our results. 
For each of our 10 chains we extract the median value for each parameter and compare them to one another. They are usually of order 0.1\% different from one another, except for the jump parameters responsible for modelling WASP-53c, which can vary as much as 60\%. This is because the orbit is not closed. We discuss this further in the next section.

The posterior probability distributions have been stored and can be requested by email to the lead author. We present the median values and $1\sigma$ region of our posteriors in Table~\ref{tab:jump} for the jump parameters, and in Table~\ref{tab:res} for the physical parameters. The results are discussed in the following section.


\section{Results}\label{sec:res}

\begin{figure*}  
\center
\includegraphics[width= 0.65\textwidth]{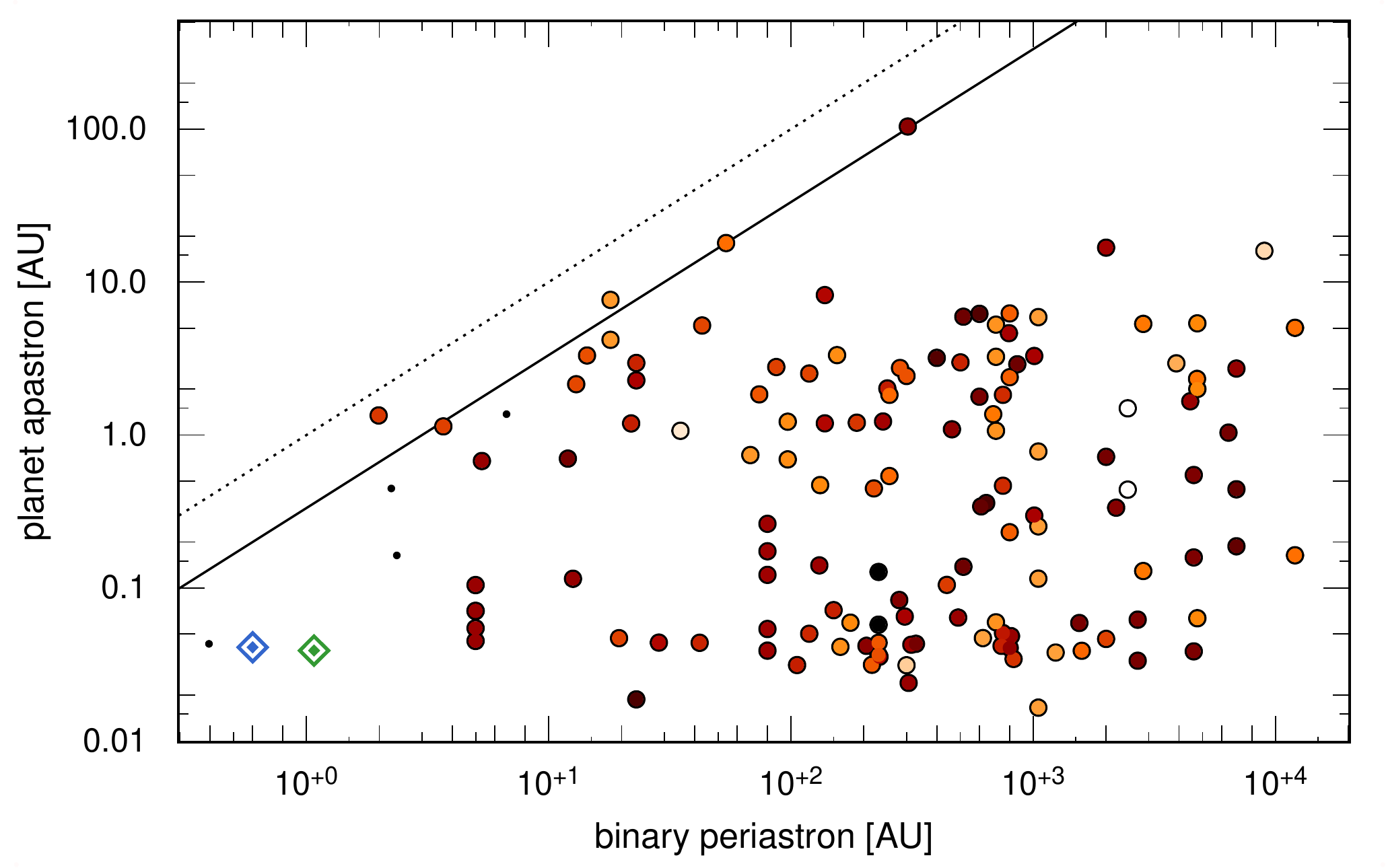}
\caption{Planetary apastron versus binary periastron (or separation if eccentricity is unknown) in astronomical units for known S-type planetary systems. The colour of the dots reflects the logarithm of the ratio of the planet-hosting star mass to the mass of its stellar companion(s) (white = 0.2, black = 18). WASP-53 and WASP-81 are highlighted as a blue and green diamond, respectively. The small black dots represent four systems containing a gas-giant and a brown dwarf. The dotted line is a 1:1 line, and the plain line is a 3:1 contour. Above that line systems are usually unstable \citep{Dvorak:1986fk,Holman:1999lr}. Data collected from \href{http://openexoplanetcatalogue.com}{openexoplanetcatalogue.com}, from \href{http://www.univie.ac.at/adg/schwarz/multiple.html}{www.univie.ac.at/adg/schwarz/multiple.html}, and from \href{http://www.exoplanet.eu}{exoplanet.eu}.
}\label{fig:binaryplanet}  
\end{figure*}

\subsection{WASP-53}\label{sec:wasp53}

WASP-53 is a system composed of a star and two orbiting objects. WASP-53b is a hot Jupiter with a mass $M_{\rm b} = 2.1\pm 0.1 M_{\rm Jup}$, with a radius $R_{\rm b}  = 1.07\pm 0.04 R_{\rm Jup}$. We find the inner orbit to be consistent with zero eccentricity, placing a 99\% confidence limit of 0.03. The Rossiter--McLaughlin effect is weakly detected. We find a lower amplitude than we anticipated, likely because $v \sin i_\star$ is over-estimated owing to an under-estimation of macroturbulence, as already noted for a number of late-type dwarfs (e.g. \citet{Triaud:2011vn,Triaud:2015fk}). We find the spin--orbit angle $\beta = -4^\circ \pm 12$. The planet appears to be coplanar.

The orbit of WASP-53c does not close within the span of our observations, which affects the various chains we launched. The median values on individual jump parameters values can vary by as much as 50\%, which is reflected in the large errors in Table~\ref{tab:jump} and \ref{tab:res}. This means that while we provide median values and their $1\sigma$ confidence ranges, those are in fact more akin to lower limits on $M_{\rm c}$, $e_{\rm c}$, $P_{\rm c}$ etc. WASP-53c is at least $16 M_{\rm Jup}$, with a period likely longer than 2500 days. The eccentricity of its orbit is high with our data being most consistent with $0.84\pm 0.01$. 

We were lucky to observe the system during the final phases of WASP-53c's periastron passage (but unlucky to miss the first half). If removing the first season of CORALIE data, we only detect a quadratic drift with a weak curvature and would never have guessed the presence of such a massive companion within the system.

\subsection{WASP-81}\label{sec:wasp81}

WASP-81 is a system composed of a star and two orbiting objects. WASP-81b is a hot Jupiter whose mass is $M_{\rm b} = 0.73\pm 0.04 M_{\rm Jup}$ and radius $R_{\rm b}  = 1.43\pm 0.05 R_{\rm Jup}$. Its orbit is consistent with being circular and we place a 99\%-confidence upper limit at 0.07. The Rossiter--McLaughlin effect is not detected. Its amplitude is projected to be less than 10 m s$^{-1}$. The low impact parameter means that the spin--orbit angle will be degenerate with $v \sin I_\star$ as in \citet{Triaud:2011vn}.

WASP-81c has a minimum mass $M_{\rm c} = 57\pm 2 M_{\rm Jup}$, a period $P_{\rm c} = 1297\pm 8$ days and an eccentricity of order 0.56.

\begin{table}
  \caption{Dates on which WASP-53c and WASP-81c may transit. Numbers are calendar dates and Barycentric Julian Dates (BJD) -- 2\,450\,000. }\label{tab:dates}
  \begin{tabular}{ll}
  \hline
  \hline
WASP-53c & WASP-81c \\
  \hline

  {\it passed dates}\\
--								& $3044.9_{-24}^{+24}$	(2004-02-09)\\
--								& $4342.1_{-16}^{+16}$	(2007-08-29)\\
$2616.0_{-130}^{+170}$ 	(2002-12-07)	& $5639.3_{-8.2}^{+8.5}$	(2011-03-18)\\
$5455.5_{-13}^{+11}$	(2010-09-16)	& $6936.5_{-2.5}^{+2.6}$	(2014-10-06)\\ \hline
  {\it future dates}\\
$8295.1_{-130}^{+170}$	(2018-06-25)	& $8233.8_{-8.2}^{+8.5}$	(2018-04-25)\\
--								& $9531.0_{-16}^{+16}$	(2021-11-12)\\
\hline
\end{tabular}
\end{table}

\begin{figure*}  
\begin{center}
\begin{subfigure}[b]{0.45\textwidth}
	\caption{WASP-53}\label{fig:megno53}
	\includegraphics[width=\textwidth]{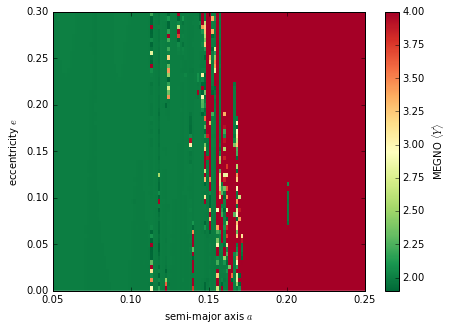}
\end{subfigure}
\begin{subfigure}[b]{0.45\textwidth}
	\caption{WASP-81}\label{fig:megno81}
	\includegraphics[width=\textwidth]{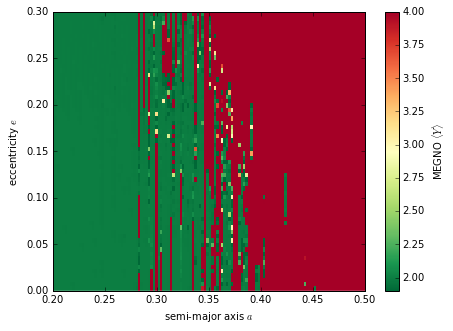}
\end{subfigure}
\caption{MEGNO maps showing regions of stability in green, and { chaos} in red, showing where planets with the masses of WASP-53b and WASP-81b could exist. Any value in excess of 4 has the same colour. All regions on the left-hand side of the graphs are stable, and all regions on the right-hand side are unstable (within the outer orbit). Maps computed by integrating for $5\times 10^4$ years
}\label{fig:megno}  
\end{center}
\end{figure*} 

\begin{figure*}  
\begin{center}
\begin{subfigure}[b]{0.45\textwidth}
	\caption{WASP-53}\label{fig:megno53-2}
	\includegraphics[width=\textwidth]{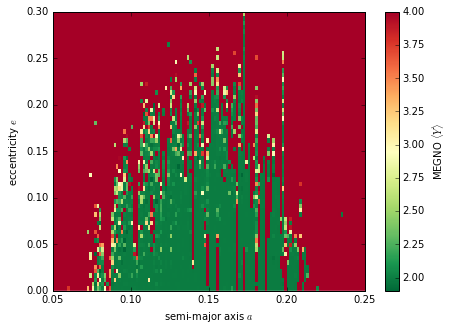}
\end{subfigure}
\begin{subfigure}[b]{0.45\textwidth}
	\caption{WASP-81}\label{fig:megno81-2}
	\includegraphics[width=\textwidth]{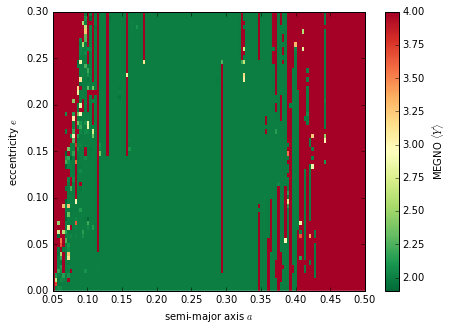}
\end{subfigure}
\caption{MEGNO maps showing regions of stability in green, and { chaos} in red, for mass-less particles between the positions of the inner and outer objects in the WASP-53 and WASP-81 systems. Any value in excess of 4 has the same colour. Maps computed by integrating for $5\times 10^4$ years
}\label{fig:megno2}  
\end{center}
\end{figure*} 

\section{Discussion}\label{sec:discus}

We have discovered two hot Jupiters orbiting the primary star of two tight binary systems. WASP-53b is super-Jupiter in mass, while WASP-81b is sub-Jupiter. Both occupy orbits that are typical for hot Jupiters \citep[eg.][]{Santerne:2016lr}. Those two planets are both accompanied by brown-dwarf-mass objects, on highly eccentric orbits of a few AU. In Fig.~\ref{fig:binaryplanet}, we plot other known planetary systems orbiting one star of a multiple stellar system, showing how atypical WASP-53 and WASP-81 are within the current exoplanet population. Only four other systems have a stellar companion with periastra closer than 10 AU: Kepler-444 \citep{Dupuy:2016rm}, KOI-1257 \citep{Santerne:2014oq}, HD\,59686 \citep{Ortiz:2016kq} and the astonishing, maybe retrograde, $\nu$ Oct \citep{Ramm:2015yu} (the leftmost red dot, above the plain line). In addition, four gas giants have outer brown-dwarf companions: HAT-P-13 \citep{Knutson:2014fg}, HIP\,5158 \citep{Feroz:2011it}, HD\,168443 \citep{Sahlmann:2011qy}, and HD\,38529 \citep{Benedict:2010gd}, in architectures similar to WASP-53 and WASP-81.

We now review the elements that make those systems stand out. We speculate about their origin, and propose some observational tests to verify some of our scenarios.

We use the stability criterion numerically determined by \citet{Holman:1999lr} to compute the widest orbital separation that each of the hot Jupiters could have occupied. In the case of WASP-53 we obtain a critical semi-major axis $a_{\rm crit} = 0.16\pm0.15$ AU, and for WASP-81, $a_{\rm crit} = 0.38 \pm 0.06$ AU. This criterion was numerically determined for a mass ratio $\mu = m_2 / (m_1 + m_2) > 0.1$, which is not satisfied for either of our systems, and likely explains the large uncertainty for WASP-53. We therefore proceed with a stability criterion devised by \citet{Petrovich:2015vn} for hierarchical planetary system. { We find that unstable orbit start emerging for $0.14 < a_{\rm b}<0.17$ for WASP-53, and $0.30 < a_{\rm b}<0.38$ for WASP-81, in good agreement with the previous approach.}

We further verify this with the method proposed by \citet{Cincotta:2003jk}, which uses a marker called the MEGNO { (the Mean Exponential Growth factor of Nearby Orbits), as implemented in {\sc rebound} by \citet{Rein:2015qf}. The MEGNO is a good tracker of orbital chaos, meaning that  infinitesimal changes in initial parameters lead to diverging solutions. For quasi-periodic orbits, thus those showing no chaotic behaviour, the MEGNO reaches a value of 2 \citep{Hinse:2010xy}. Larger values of the MEGNO, typically $> 4$, indicate significant changes in the orbital parameters, a sign of chaos. Although this does not necessarily translate by unstable orbits \citep[e.g.][]{Deck:2012kx}, it often tracks them as we saw above. For our case the MEGNO outlines where nearly closed orbits exist and therefore informs us on where any disc material may have been stable, or where additional planets may exist.}

We used the parameters provided in Tab.~\ref{tab:res} for the outer companions, assumed coplanarity between the inner and outer orbits, and computed the MEGNO for a particle with a mass and separation for WASP-53b and WASP-81b. We integrated each system for $5\times 10^6$ years and obtained values of 2.0062 and 1.9999 respectively, indicating stability. We expanded those simulation and explore the $(a,e)$ parameter space. To compute the maps presented in Fig.~\ref{fig:megno} we integrated each of the pixels  over $5\times 10^4$ years. We observe from those results that wide regions of { chaos} exist, which is  consistent with the work of \citet{Holman:1999lr} and \citet{Petrovich:2015vn}. Both WASP-53 and WASP-81 appear stable over long periods of time. In the case of WASP-53 we observe that only regions closer in than 0.15 AU retain stable orbits. For WASP-81 there is slightly more space with orbits within 0.3 AU being generally stable.

We also investigated whether other planets could exist between objects b and c by adding a third massless particle. The results are displayed in Fig.~\ref{fig:megno2}, which show a range of stable orbits (in green). While this suggests that other planets could still be identified within our systems, we think it unlikely since hot Jupiters are usually found to be isolated. Only one hot Jupiter is known to have other planetary companions within an astronomical unit \citep{Becker:2015fk,Neveu-VanMalle:2016xy}, and a recent analysis of the {\it Kepler} data shows that a lack of nearby companions is an important aspect that sets hot Jupiters apart from other gas giants \citep{Huang:2016yq}.

\subsection{Determination of the mass and orbital inclination of WASP-53c and WASP-81c}

{\it Gaia} \citep{Perryman:2001lr} is an ESA mission launched in 2013 currently scanning the sky and measuring stellar positions with a precision of order 30 micro-arcseconds for stars brighter than optical magnitude 12 \citep{de-Bruijne:2012qy}. Several studies have investigated {\it Gaia}'s potential for detecting gas giants using astrometry \citep{Casertano:2008zp, Sozzetti:2014qy,Perryman:2014rf,Sahlmann:2015xy}, with \citet{Neveu:2012lr} looking into the  combination of radial-velocities with astrometric measurements. 

From \citet{Perryman:2014rf}, WASP-53c and WASP-81c will produce an astrometric displacement of their host stars, $\alpha$, defined as:
\begin{equation}
\alpha = \left({M_{\rm p} \over M_\star}\right) \left({a \over 1 {\rm AU}}\right)  \left({d \over 1 {\rm pc}}\right)^{-1} {\rm arcsec}
\end{equation}

WASP-53 and WASP-81 will move on the sky by $\alpha = 260$, and $300\,\mu$as respectively, caused by their outer companions, assuming an orbital inclination $i_{\rm c} = 90^\circ$, which is the poorest scenario possible. We can expect of { order $N_{\rm obs} = 70$ astrometric} measurements\footnote{78 expected measurements for WASP-53, and 60 for WASP-81 according to the following tool: \href{http://gaia.esac.esa.int/gost/}{http://gaia.esac.esa.int/gost/}} { with typical uncertainties $\sigma = 40 \mu$as} \citep{de-Bruijne:2012qy} to be collected on our two targets. This translates to an astronometric signal-to-noise, { where $S/N = 
\alpha \sqrt{N_{\rm obs}} / \sigma$},  in excess of 50 for both systems. If instead $i_{\rm c} = 10^\circ$, we obtain $\alpha = 1500$, and $1700\mu$as { respectively}\footnote{we do not detect a secondary set of lines in either of our spectra. This is equivalent to a limit of $i_{\rm c}> 2^\circ$, and $i_{\rm c}> 5^\circ$ respectively.}. The amplitude of the orbital motion alone should inform us of the mutual inclination between the inner and outer planet. For { astrometric} signal-to-noise values of $ > 20$ the orbital inclination will typically be estimated with a precision of $<10^\circ$ \citep{Sahlmann:2015xy}.

\subsection{Transit timing variations}

Using {\sc rebound} \citep{Rein:2015yu} we integrated the system over a few orbital periods of the outer companion and recorded when transits of the inner planet happened. For WASP-53b we expect total transit-timing variations, caused by the perturbing effect of the outer companion, { to be of order 35s}. However, most of the variation happens during periastron, { which means that during nearly 100 transit epochs we covered for WASP-53b, we expect no variation to be measurable)}. We expect a detectable signal to appear within the next two years if our solution for the outer companion is correct. { After WASP-53c swings again via periastron, the ephemeris for WASP-53b will become offset by approximately 35s. This offset remains constant until the following periastron when it will offset again by the same amount}. We repeated the procedure for WASP-81 and find { a similar TTV behaviour for WASP-81b with offset of approximately 30s compared to the value we produce here.}

\subsection{Estimating $k_2$}

Secular interactions between pairs of orbiting planets usually excites their orbital eccentricities. In the case of WASP-53 and WASP-81, the innermost planetary orbit will be affected by tidal forces which tend to damp eccentricity, while the outer, massive companion occupies a highly eccentric orbit which will excite the inner planet's eccentricity. \citet{Mardling:2007lr} investigated this secular problem and found that the inner planet will reach an equilibrium eccentricity, called a {\it fixed-point}, with a value dependent on the planet's internal density profile, and parametrised by $k_2$, the tidal Love number \citep{Sterne:1939fj}. The more mass that gets included into the core of the planet, the larger will be the fixed-point eccentricity \citep{Batygin:2009qy}. As such a measure of the eccentricity of the tidally damped, inner orbit can yield the core-mass fraction of exoplanets. 
Recently, \citet{Buhler:2016yq} investigated the case of the HAT-P-13 system that presents an architecture similar to WASP-53 and WASP-81, and managed to constrain the core mass of the transiting hot Jupiter to 11 M$_\oplus$ by measuring an eccentricity of $0.007 \pm 0.001$. We see here how our two systems compare.

We estimated the inner eccentricity, first assuming no tidal damping, using eq.~36 in \cite{Mardling:2007lr}, and found fixed-point eccentricities of 0.00053 for WASP-53b, and 0.0027 for WASP-81b (HAT-P-13's parameters yield 0.0087). Any tidal dissipation will reduce these values by an amount dependent on the internal composition. Similarly, any mutual inclination between the outer and inner orbits will reduce these values. Sadly, because the eccentricities we expect of WASP-53b and WASP-81b are so small, we find it unlikely that their internal composition will be determined soon. If however, the eccentricity is one day measured, we should expect apsidal alignment or anti-alignment between the inner and outer orbits. The Love number can also be extracted from the way a transit lightcurve gets affected by apsidal precession \citep{Ragozzine:2009vn}.

\subsection{Possible scenarios for the planets' formation and orbital evolution}

We here speculate about the sequence of events leading to systems like WASP-53 and WASP-81, and also to HAT-P-13b \citep{Bakos:2009fj}, HD\,59686Ab \citep{Ortiz:2016kq} and others.
The presence of both a planet and a brown-dwarf mass object  \citep[one on the planetary side, the other on the stellar side of the brown-dwarf desert;][]{Grether:2006kx,Sahlmann:2011qy} within the same system possibly suggests that core accretion and gravitational collapse can both operate within the same disc environment. They may also be smaller fragments from the nebula that created WASP-53A and WASP-81A.

Both WASP-53b and WASP-81b could initially have formed on circumbinary orbits, to be captured by the primary following a dynamical instability. This sort of scenario has been investigated by \citet{Sutherland:2016vn}, and shown to be an unlikely outcome, which, when it does happen, favours capture of the planet by the secondary instead of the primary. Therefore, WASP-53b and WASP-81b most likely formed within a disc surrounding WASP-53A and WASP-81A.

We can think of two alternate scenarios and make an appeal for theorists to investigate them since they could teach us much about how gas giants form. 

WASP-53c and WASP-81c are too massive compared to the protoplanetary disc to have followed a type-II migration \citep{Duffell:2014rz,Durmann:2015rm}. If no or little orbital evolution has happened within the two systems that we studied, then WASP-53b and WASP-81b must have formed well within the snow line, on orbits shorter than 0.15 and 0.3 AU respectively. Theoretical work by \citet{Batygin:2015lr} and \citet{Lee:2015kq} suggests that the formation of gas giants within one astronomical unit is feasible. According to \citet{Fung:2014fv} gas can flow through the gap carved by a planet, particularly so if it reaches masses near that of a brown dwarf. The more massive the object, the more perturbations are produced at the outer gap's edge, launching streams of gas that replenish the inner disc to provide enough mass to allow the formation of a gas giant. The large eccentricities of WASP-53c and WASP-81c might have enhanced this effect. However, \citet{Lambrechts:2014lq} and \citet{Rosotti:2016lq} find that planets more massive than 20--30 Earth masses prevent the flow of dust grains across the same gap. Accordingly, if WASP-53b and WASP-81b formed in-situ via core-accretion, they could only have used solids within about 0.2 AU, before being able to accrete gas. If this scenario is correct, systems like WASP-53 and WASP-81 can inform us about the efficiency of core accretion, as well the minimum core mass necessary to accrete significant gas envelopes. This would also leave the planets poorer in metals than otherwise, something which is possible to determine via transmission spectroscopy \citep{Seager:2010kx,Madhusudhan:2014jk}

WASP-53c and WASP-81c are both eccentric. { It has been argued that disc-planet interactions can excite the eccentricity of gap-opening planets, but not to the values that we observe for WASP-53c and WASP-81c} \citep[e.g.][]{Goldreich:2003uq,DAngelo:2006fj,Rosotti:2016qy,Teyssandier:2016kx}. This might imply that WASP-53c and WASP-81c reached their current orbital parameters after disc dispersal, possibly due to dynamical interactions with third, yet unseen companions to WASP-53 and WASP-81. If this is the case then either WASP-53b and WASP-81b disc-migrated well before WASP-53c and WASP-81c reached their current orbits, or, WASP-53b and WASP-81b reached their current orbit following a high-eccentricity migration produced by the same instability that left their outer companions on eccentric orbital paths. In either of those cases, we expect a significant mutual inclination between the inner and outer orbits, which {\it Gaia} should in principle be able to measure. Coplanarity would favour the scenario outlined in the previous paragraph.

\section{Concluding words}\label{sec:concl}

WASP-53 \& WASP-81 are peculiar systems composed of both a planet and a brown dwarf. This orbital set-up is a relic of its past formation. Investigating them in further studies, notably with the help of {\it Gaia}, will prove invaluable for  understanding planet formation and the subsequent orbital evolution, but also the relation between planet formation, brown dwarf formation and stellar formation.


\section*{Nota Bene}
Dates are given in the BJD-TDB standard. The radii we used for Jupiter and the Sun are the volumetric mean radii.

For clarity, we used the subscripts $\star$ for the star, $b$ for the inner planet, and $c$ for the outer object, all throughout.

\section*{Acknowledgments}

AT would like to acknowledge inspiring discussions with the following people: Rosemary Mardling, Hanno Rein, Simon Hodgkin, Daniel Tamayo, Richard Booth, Cristobal Petrovich, Yanqin Wu, Kaitlin Kratter, Jean Teyssandier, Ari Silburt and David Martin.

TRAPPIST is a project funded by the Belgian Fund for Scientific Research (Fond National de la Recherche Scientifique, 
F.R.S-FNRS) under grant FRFC 2.5.594.09.F, with the participation of the Swiss National Science Fundation (SNF). 
The {\it Euler} Swiss telescope is supported by the Swiss National Science Foundation.
We are all very grateful to ESO and its La Silla  staff for their continuous support.   WASP-South is hosted by the South African Astronomical Observatory while SuperWASP-North is hosted by the Issac Newton Group at the Observatorio del Roque de los Muchachos on  La Palma. We are grateful for the ongoing support and assistance of these observatories.   

M. Gillon and E. Jehin are Research Associates at the F.R.S-FNRS; L. Delrez received the support of the F.R.I.A. fund of the FNRS.

This publication makes use of data products from 2MASS, whose data was obtained through \href{http://simbad.u-strasbg.fr/simbad/}{Simbad} and \href{http://vizier.u-strasbg.fr/viz-bin/VizieR}{VizieR} services hosted at the \href{http://cds.u-strasbg.fr}{CDS-Strasbourg}.
The Two Micron All Sky Survey is a joint project of the University of Massachusetts and the Infrared Processing and Analysis Center/California Institute of Technology, funded by the National Aeronautics and Space Administration and the National Science Foundation.

References to exoplanetary systems were obtained through the use of the paper repositories, \href{http://adsabs.harvard.edu/abstract_service.html}{ADS} and \href{http://arxiv.org/archive/astro-ph}{arXiv}, but also through frequent visits to the \href{http://exoplanet.eu}{exoplanet.eu} \citep{Schneider:2011lr} and \href{http://exoplanets.org}{exoplanets.org} \citep{Wright:2011fj} websites.

\bibliographystyle{mn2e}
\bibliography{../1Mybib.bib}

\begin{thebibliography}{124}
\expandafter\ifx\csname natexlab\endcsname\relax\def\natexlab#1{#1}\fi

\bibitem[{{Albrecht} {et~al}\mbox{.}(2012{\natexlab{a}}){Albrecht}, {Winn},
  {Butler}, {Crane}, {Shectman}, {Thompson}, {Hirano}, \&
  {Wittenmyer}}]{Albrecht:2012lr}
{Albrecht} S., {Winn} J.~N., {Butler} R.~P., {Crane} J.~D., {Shectman} S.~A.,
  {Thompson} I.~B., {Hirano} T., {Wittenmyer} R.~A., 2012{\natexlab{a}}, \apj,
  744, 189

\bibitem[{{Albrecht} {et~al}\mbox{.}(2012{\natexlab{b}}){Albrecht}, {Winn},
  {Johnson}, {Howard}, {Marcy}, {Butler}, {Arriagada}, {Crane}, {Shectman},
  {Thompson}, {Hirano}, {Bakos}, \& {Hartman}}]{Albrecht:2012lp}
{Albrecht} S. {et~al.}, 2012{\natexlab{b}}, \apj, 757, 18

\bibitem[{{Alibert} {et~al}\mbox{.}(2005){Alibert}, {Mousis}, {Mordasini}, \&
  {Benz}}]{Alibert:2005fk}
{Alibert} Y., {Mousis} O., {Mordasini} C., {Benz} W., 2005, \apjl, 626, L57

\bibitem[{{Anderson} {et~al}\mbox{.}(2010){Anderson}, {Hellier}, {Gillon},
  {Triaud}, {Smalley}, {Hebb}, {Collier Cameron}, {Maxted}, {Queloz}, {West},
  {Bentley}, {Enoch}, {Horne}, {Lister}, {Mayor}, {Parley}, {Pepe}, {Pollacco},
  {S{\'e}gransan}, {Udry}, \& {Wilson}}]{Anderson:2010fj}
{Anderson} D.~R. {et~al.}, 2010, \apj, 709, 159

\bibitem[{{Anderson} {et~al}\mbox{.}(2015){Anderson}, {Triaud}, {Turner},
  {Brown}, {Clark}, {Smalley}, {Collier Cameron}, {Doyle}, {Gillon}, {Hellier},
  {Lovis}, {Maxted}, {Pollacco}, {Queloz}, \& {Smith}}]{Anderson:2015lr}
---, 2015, \apjl, 800, L9

\bibitem[{{Asplund} {et~al}\mbox{.}(2009){Asplund}, {Grevesse}, {Sauval}, \&
  {Scott}}]{Asplund:2009vn}
{Asplund} M., {Grevesse} N., {Sauval} A.~J., {Scott} P., 2009, \araa, 47, 481

\bibitem[{{Bakos} {et~al}\mbox{.}(2009){Bakos}, {Howard}, {Noyes}, {Hartman},
  {Torres}, {Kov{\'a}cs}, {Fischer}, {Latham}, {Johnson}, {Marcy}, {Sasselov},
  {Stefanik}, {Sip{\H o}cz}, {Kov{\'a}cs}, {Esquerdo}, {P{\'a}l},
  {L{\'a}z{\'a}r}, {Papp}, \& {S{\'a}ri}}]{Bakos:2009fj}
{Bakos} G.~{\'A}. {et~al.}, 2009, \apj, 707, 446

\bibitem[{{Baranne} {et~al}\mbox{.}(1996){Baranne}, {Queloz}, {Mayor},
  {Adrianzyk}, {Knispel}, {Kohler}, {Lacroix}, {Meunier}, {Rimbaud}, \&
  {Vin}}]{Baranne:1996qa}
{Baranne} A. {et~al.}, 1996, \aaps, 119, 373

\bibitem[{{Baruteau} {et~al}\mbox{.}(2014){Baruteau}, {Crida}, {Paardekooper},
  {Masset}, {Guilet}, {Bitsch}, {Nelson}, {Kley}, \&
  {Papaloizou}}]{Baruteau:2014rp}
{Baruteau} C. {et~al.}, 2014, Protostars and Planets VI, 667

\bibitem[{{Batygin}(2012)}]{Batygin:2012hl}
{Batygin} K., 2012, \nat, 491, 418

\bibitem[{{Batygin}, {Bodenheimer} \& {Laughlin}(2009){Batygin}, {Bodenheimer},
  \& {Laughlin}}]{Batygin:2009qy}
{Batygin} K., {Bodenheimer} P., {Laughlin} G., 2009, \apjl, 704, L49

\bibitem[{{Batygin}, {Bodenheimer} \& {Laughlin}(2015){Batygin}, {Bodenheimer},
  \& {Laughlin}}]{Batygin:2015lr}
{Batygin} K., {Bodenheimer} P.~H., {Laughlin} G.~P., 2015, ArXiv e-prints

\bibitem[{{Becker} {et~al}\mbox{.}(2015){Becker}, {Vanderburg}, {Adams},
  {Rappaport}, \& {Schwengeler}}]{Becker:2015fk}
{Becker} J.~C., {Vanderburg} A., {Adams} F.~C., {Rappaport} S.~A.,
  {Schwengeler} H.~M., 2015, \apjl, 812, L18

\bibitem[{{Benedict} {et~al}\mbox{.}(2010){Benedict}, {McArthur}, {Bean},
  {Barnes}, {Harrison}, {Hatzes}, {Martioli}, \& {Nelan}}]{Benedict:2010gd}
{Benedict} G.~F., {McArthur} B.~E., {Bean} J.~L., {Barnes} R., {Harrison}
  T.~E., {Hatzes} A., {Martioli} E., {Nelan} E.~P., 2010, \aj, 139, 1844

\bibitem[{{Bodenheimer}, {Hubickyj} \& {Lissauer}(2000){Bodenheimer},
  {Hubickyj}, \& {Lissauer}}]{Bodenheimer:2000lr}
{Bodenheimer} P., {Hubickyj} O., {Lissauer} J.~J., 2000, \icarus, 143, 2

\bibitem[{{Buhler} {et~al}\mbox{.}(2016){Buhler}, {Knutson}, {Batygin},
  {Fulton}, {Fortney}, {Burrows}, \& {Wong}}]{Buhler:2016yq}
{Buhler} P.~B., {Knutson} H.~A., {Batygin} K., {Fulton} B.~J., {Fortney} J.~J.,
  {Burrows} A., {Wong} I., 2016, \apj, 821, 26

\bibitem[{{Casertano} {et~al}\mbox{.}(2008){Casertano}, {Lattanzi}, {Sozzetti},
  {Spagna}, {Jancart}, {Morbidelli}, {Pannunzio}, {Pourbaix}, \&
  {Queloz}}]{Casertano:2008zp}
{Casertano} S. {et~al.}, 2008, \aap, 482, 699

\bibitem[{{C{\'e}bron} {et~al}\mbox{.}(2011){C{\'e}bron}, {Moutou}, {Le Bars},
  {Le Gal}, \& {Far{\`e}s}}]{Cebron:2011lr}
{C{\'e}bron} D., {Moutou} C., {Le Bars} M., {Le Gal} P., {Far{\`e}s} R., 2011,
  in European Physical Journal Web of Conferences, Vol.~11, European Physical
  Journal Web of Conferences, p. 3003

\bibitem[{{Cincotta}, {Giordano} \& {Sim{\'o}}(2003){Cincotta}, {Giordano}, \&
  {Sim{\'o}}}]{Cincotta:2003jk}
{Cincotta} P.~M., {Giordano} C.~M., {Sim{\'o}} C., 2003, Physica D Nonlinear
  Phenomena, 182, 151

\bibitem[{{Claret}(2004)}]{Claret:2004fk}
{Claret} A., 2004, \aap, 428, 1001

\bibitem[{{Collier Cameron} {et~al}\mbox{.}(2007){Collier Cameron}, {Wilson},
  {West}, {Hebb}, {Wang}, {Aigrain}, {Bouchy}, {Christian}, {Clarkson},
  {Enoch}, {Esposito}, {Guenther}, {Haswell}, {H{\'e}brard}, {Hellier},
  {Horne}, {Irwin}, {Kane}, {Loeillet}, {Lister}, {Maxted}, {Mayor}, {Moutou},
  {Parley}, {Pollacco}, {Pont}, {Queloz}, {Ryans}, {Skillen}, {Street}, {Udry},
  \& {Wheatley}}]{Collier-Cameron:2007pb}
{Collier Cameron} A. {et~al.}, 2007, \mnras, 380, 1230

\bibitem[{{D'Angelo}, {Lubow} \& {Bate}(2006){D'Angelo}, {Lubow}, \&
  {Bate}}]{DAngelo:2006fj}
{D'Angelo} G., {Lubow} S.~H., {Bate} M.~R., 2006, \apj, 652, 1698

\bibitem[{{Dawson}(2014)}]{Dawson:2014kx}
{Dawson} R.~I., 2014, \apjl, 790, L31

\bibitem[{{de Bruijne}(2012)}]{de-Bruijne:2012qy}
{de Bruijne} J.~H.~J., 2012, \apss, 341, 31

\bibitem[{{Deck} {et~al}\mbox{.}(2012){Deck}, {Holman}, {Agol}, {Carter},
  {Lissauer}, {Ragozzine}, \& {Winn}}]{Deck:2012kx}
{Deck} K.~M., {Holman} M.~J., {Agol} E., {Carter} J.~A., {Lissauer} J.~J.,
  {Ragozzine} D., {Winn} J.~N., 2012, \apjl, 755, L21

\bibitem[{{Delrez} {et~al}\mbox{.}(2014){Delrez}, {Van Grootel}, {Anderson},
  {Collier-Cameron}, {Doyle}, {Fumel}, {Gillon}, {Hellier}, {Jehin}, {Lendl},
  {Neveu-VanMalle}, {Maxted}, {Pepe}, {Pollacco}, {Queloz}, {S{\'e}gransan},
  {Smalley}, {Smith}, {Southworth}, {Triaud}, {Udry}, \&
  {West}}]{Delrez:2014qf}
{Delrez} L. {et~al.}, 2014, \aap, 563, A143

\bibitem[{{DENIS Consortium}(2005)}]{DENIS-Consortium:2005fj}
{DENIS Consortium}, 2005, VizieR Online Data Catalog, 2263, 0

\bibitem[{{Doyle} {et~al}\mbox{.}(2014){Doyle}, {Davies}, {Smalley}, {Chaplin},
  \& {Elsworth}}]{Doyle:2014qf}
{Doyle} A.~P., {Davies} G.~R., {Smalley} B., {Chaplin} W.~J., {Elsworth} Y.,
  2014, \mnras, 444, 3592

\bibitem[{{Doyle} {et~al}\mbox{.}(2013){Doyle}, {Smalley}, {Maxted},
  {Anderson}, {Cameron}, {Gillon}, {Hellier}, {Pollacco}, {Queloz}, {Triaud},
  \& {West}}]{Doyle:2013lr}
{Doyle} A.~P. {et~al.}, 2013, \mnras, 428, 3164

\bibitem[{{Droege} {et~al}\mbox{.}(2006){Droege}, {Richmond}, {Sallman}, \&
  {Creager}}]{Droege:2006yq}
{Droege} T.~F., {Richmond} M.~W., {Sallman} M.~P., {Creager} R.~P., 2006,
  \pasp, 118, 1666

\bibitem[{{Duffell} {et~al}\mbox{.}(2014){Duffell}, {Haiman}, {MacFadyen},
  {D'Orazio}, \& {Farris}}]{Duffell:2014rz}
{Duffell} P.~C., {Haiman} Z., {MacFadyen} A.~I., {D'Orazio} D.~J., {Farris}
  B.~D., 2014, \apjl, 792, L10

\bibitem[{{Dupuy} {et~al}\mbox{.}(2016){Dupuy}, {Kratter}, {Kraus}, {Isaacson},
  {Mann}, {Ireland}, {Howard}, \& {Huber}}]{Dupuy:2016rm}
{Dupuy} T.~J., {Kratter} K.~M., {Kraus} A.~L., {Isaacson} H., {Mann} A.~W.,
  {Ireland} M.~J., {Howard} A.~W., {Huber} D., 2016, \apj, 817, 80

\bibitem[{{D{\"u}rmann} \& {Kley}(2015)}]{Durmann:2015rm}
{D{\"u}rmann} C., {Kley} W., 2015, \aap, 574, A52

\bibitem[{{Dvorak}(1986)}]{Dvorak:1986fk}
{Dvorak} R., 1986, \aap, 167, 379

\bibitem[{{Eastman}, {Siverd} \& {Gaudi}(2010){Eastman}, {Siverd}, \&
  {Gaudi}}]{Eastman:2010ys}
{Eastman} J., {Siverd} R., {Gaudi} B.~S., 2010, \pasp, 122, 935

\bibitem[{{Feroz}, {Balan} \& {Hobson}(2011){Feroz}, {Balan}, \&
  {Hobson}}]{Feroz:2011it}
{Feroz} F., {Balan} S.~T., {Hobson} M.~P., 2011, \mnras, 416, L104

\bibitem[{{Ford}(2006)}]{Ford:2006yq}
{Ford} E.~B., 2006, \apj, 642, 505

\bibitem[{{Fung}, {Shi} \& {Chiang}(2014){Fung}, {Shi}, \&
  {Chiang}}]{Fung:2014fv}
{Fung} J., {Shi} J.-M., {Chiang} E., 2014, \apj, 782, 88

\bibitem[{{Gaudi} \& {Winn}(2007)}]{Gaudi:2007vn}
{Gaudi} B.~S., {Winn} J.~N., 2007, \apj, 655, 550

\bibitem[{{Gillon} {et~al}\mbox{.}(2013){Gillon}, {Anderson},
  {Collier-Cameron}, {Doyle}, {Fumel}, {Hellier}, {Jehin}, {Lendl}, {Maxted},
  {Montalb{\'a}n}, {Pepe}, {Pollacco}, {Queloz}, {S{\'e}gransan}, {Smith},
  {Smalley}, {Southworth}, {Triaud}, {Udry}, \& {West}}]{Gillon:2013yg}
{Gillon} M. {et~al.}, 2013, \aap, 552, A82

\bibitem[{{Gillon} {et~al}\mbox{.}(2011){Gillon}, {Jehin}, {Magain}, {Chantry},
  {Hutsem{\'e}kers}, {Manfroid}, {Queloz}, \& {Udry}}]{Gillon:2011qf}
{Gillon} M., {Jehin} E., {Magain} P., {Chantry} V., {Hutsem{\'e}kers} D.,
  {Manfroid} J., {Queloz} D., {Udry} S., 2011, in European Physical Journal Web
  of Conferences, Vol.~11, European Physical Journal Web of Conferences, p.
  6002

\bibitem[{{Gillon} {et~al}\mbox{.}(2012){Gillon}, {Triaud}, {Fortney},
  {Demory}, {Jehin}, {Lendl}, {Magain}, {Kabath}, {Queloz}, {Alonso},
  {Anderson}, {Collier Cameron}, {Fumel}, {Hebb}, {Hellier}, {Lanotte},
  {Maxted}, {Mowlavi}, \& {Smalley}}]{Gillon:2012fj}
{Gillon} M. {et~al.}, 2012, \aap, 542, A4

\bibitem[{{Gim{\'e}nez}(2006)}]{Gimenez:2006kx}
{Gim{\'e}nez} A., 2006, \apj, 650, 408

\bibitem[{{Goldreich} \& {Sari}(2003)}]{Goldreich:2003uq}
{Goldreich} P., {Sari} R., 2003, \apj, 585, 1024

\bibitem[{{Gray}(2008)}]{Gray:2008fj}
{Gray} D.~F., 2008, {The Observation and Analysis of Stellar Photospheres},
  {Gray, D.~F.}, ed.

\bibitem[{{Grether} \& {Lineweaver}(2006)}]{Grether:2006kx}
{Grether} D., {Lineweaver} C.~H., 2006, \apj, 640, 1051

\bibitem[{{Guillochon}, {Ramirez-Ruiz} \& {Lin}(2011){Guillochon},
  {Ramirez-Ruiz}, \& {Lin}}]{Guillochon:2011fk}
{Guillochon} J., {Ramirez-Ruiz} E., {Lin} D., 2011, \apj, 732, 74

\bibitem[{{Hatzes} \& {Wuchterl}(2005)}]{Hatzes:2005ys}
{Hatzes} A.~P., {Wuchterl} G., 2005, \nat, 436, 182

\bibitem[{{H{\'e}brard} {et~al}\mbox{.}(2008){H{\'e}brard}, {Bouchy}, {Pont},
  {Loeillet}, {Rabus}, {Bonfils}, {Moutou}, {Boisse}, {Delfosse}, {Desort},
  {Eggenberger}, {Ehrenreich}, {Forveille}, {Lagrange}, {Lovis}, {Mayor},
  {Pepe}, {Perrier}, {Queloz}, {Santos}, {S{\'e}gransan}, {Udry}, \&
  {Vidal-Madjar}}]{Hebrard:2008mz}
{H{\'e}brard} G. {et~al.}, 2008, \aap, 488, 763

\bibitem[{{Helled} {et~al}\mbox{.}(2014){Helled}, {Bodenheimer}, {Podolak},
  {Boley}, {Meru}, {Nayakshin}, {Fortney}, {Mayer}, {Alibert}, \&
  {Boss}}]{Helled:2014kq}
{Helled} R. {et~al.}, 2014, Protostars and Planets VI, 643

\bibitem[{{Hilditch}(2001)}]{Hilditch:2001uq}
{Hilditch} R.~W., 2001, {An Introduction to Close Binary Stars}, {Hilditch,
  R.~W.}, ed.

\bibitem[{{Hinse} {et~al}\mbox{.}(2010){Hinse}, {Christou}, {Alvarellos}, \&
  {Go{\'z}dziewski}}]{Hinse:2010xy}
{Hinse} T.~C., {Christou} A.~A., {Alvarellos} J.~L.~A., {Go{\'z}dziewski} K.,
  2010, \mnras, 404, 837

\bibitem[{{Holman} \& {Wiegert}(1999)}]{Holman:1999lr}
{Holman} M.~J., {Wiegert} P.~A., 1999, \aj, 117, 621

\bibitem[{{Holman} {et~al}\mbox{.}(2006){Holman}, {Winn}, {Latham},
  {O'Donovan}, {Charbonneau}, {Bakos}, {Esquerdo}, {Hergenrother}, {Everett},
  \& {P{\'a}l}}]{Holman:2006qy}
{Holman} M.~J. {et~al.}, 2006, \apj, 652, 1715

\bibitem[{{Huang}, {Wu} \& {Triaud}(2016){Huang}, {Wu}, \&
  {Triaud}}]{Huang:2016yq}
{Huang} C., {Wu} Y., {Triaud} A.~H.~M.~J., 2016, \apj, 825, 98

\bibitem[{{Jehin} {et~al}\mbox{.}(2011){Jehin}, {Gillon}, {Queloz}, {Magain},
  {Manfroid}, {Chantry}, {Lendl}, {Hutsem{\'e}kers}, \& {Udry}}]{Jehin:2011dk}
{Jehin} E. {et~al.}, 2011, The Messenger, 145, 2

\bibitem[{{Knutson} {et~al}\mbox{.}(2014){Knutson}, {Fulton}, {Montet}, {Kao},
  {Ngo}, {Howard}, {Crepp}, {Hinkley}, {Bakos}, {Batygin}, {Johnson}, {Morton},
  \& {Muirhead}}]{Knutson:2014fg}
{Knutson} H.~A. {et~al.}, 2014, \apj, 785, 126

\bibitem[{{Kopal}(1942)}]{Kopal:1942ai}
{Kopal} Z., 1942, Proceedings of the National Academy of Science, 28, 133

\bibitem[{{Lai}(2014)}]{Lai:2014nx}
{Lai} D., 2014, \mnras, 440, 3532

\bibitem[{{Lai}, {Foucart} \& {Lin}(2011){Lai}, {Foucart}, \&
  {Lin}}]{Lai:2011ul}
{Lai} D., {Foucart} F., {Lin} D.~N.~C., 2011, \mnras, 412, 2790

\bibitem[{{Lambrechts}, {Johansen} \& {Morbidelli}(2014){Lambrechts},
  {Johansen}, \& {Morbidelli}}]{Lambrechts:2014lq}
{Lambrechts} M., {Johansen} A., {Morbidelli} A., 2014, \aap, 572, A35

\bibitem[{{Lang} {et~al}\mbox{.}(2010){Lang}, {Hogg}, {Mierle}, {Blanton}, \&
  {Roweis}}]{Lang:2010yq}
{Lang} D., {Hogg} D.~W., {Mierle} K., {Blanton} M., {Roweis} S., 2010, \aj,
  139, 1782

\bibitem[{{Lee} \& {Chiang}(2015)}]{Lee:2015kq}
{Lee} E.~J., {Chiang} E., 2015, \apj, 811, 41

\bibitem[{{Lendl} {et~al}\mbox{.}(2012){Lendl}, {Anderson}, {Collier-Cameron},
  {Doyle}, {Gillon}, {Hellier}, {Jehin}, {Lister}, {Maxted}, {Pepe},
  {Pollacco}, {Queloz}, {Smalley}, {S{\'e}gransan}, {Smith}, {Triaud}, {Udry},
  {West}, \& {Wheatley}}]{Lendl:2012qy}
{Lendl} M. {et~al.}, 2012, \aap, 544, A72

\bibitem[{{Lendl} {et~al}\mbox{.}(2014){Lendl}, {Triaud}, {Anderson}, {Collier
  Cameron}, {Delrez}, {Doyle}, {Gillon}, {Hellier}, {Jehin}, {Maxted},
  {Neveu-VanMalle}, {Pepe}, {Pollacco}, {Queloz}, {S{\'e}gransan}, {Smalley},
  {Smith}, {Udry}, {Van Grootel}, \& {West}}]{Lendl:2014yu}
---, 2014, \aap, 568, A81

\bibitem[{{Lin}, {Bodenheimer} \& {Richardson}(1996){Lin}, {Bodenheimer}, \&
  {Richardson}}]{Lin:1996yq}
{Lin} D.~N.~C., {Bodenheimer} P., {Richardson} D.~C., 1996, \nat, 380, 606

\bibitem[{{L{\'o}pez-Morales} {et~al}\mbox{.}(2014){L{\'o}pez-Morales},
  {Triaud}, {Rodler}, {Dumusque}, {Buchhave}, {Harutyunyan}, {Hoyer}, {Alonso},
  {Gillon}, {Kaib}, {Latham}, {Lovis}, {Pepe}, {Queloz}, {Raymond},
  {S{\'e}gransan}, {Waldmann}, \& {Udry}}]{Lopez-Morales:2014qv}
{L{\'o}pez-Morales} M. {et~al.}, 2014, \apjl, 792, L31

\bibitem[{{Lovis} {et~al}\mbox{.}(2006){Lovis}, {Mayor}, {Pepe}, {Alibert},
  {Benz}, {Bouchy}, {Correia}, {Laskar}, {Mordasini}, {Queloz}, {Santos},
  {Udry}, {Bertaux}, \& {Sivan}}]{Lovis:2006uq}
{Lovis} C. {et~al.}, 2006, \nat, 441, 305

\bibitem[{{Madhusudhan} {et~al}\mbox{.}(2014){Madhusudhan}, {Crouzet},
  {McCullough}, {Deming}, \& {Hedges}}]{Madhusudhan:2014jk}
{Madhusudhan} N., {Crouzet} N., {McCullough} P.~R., {Deming} D., {Hedges} C.,
  2014, \apjl, 791, L9

\bibitem[{{Mandel} \& {Agol}(2002)}]{Mandel:2002kx}
{Mandel} K., {Agol} E., 2002, \apjl, 580, L171

\bibitem[{{Mardling}(2007)}]{Mardling:2007lr}
{Mardling} R.~A., 2007, \mnras, 382, 1768

\bibitem[{{Marigo} {et~al}\mbox{.}(2008){Marigo}, {Girardi}, {Bressan},
  {Groenewegen}, {Silva}, \& {Granato}}]{Marigo:2008lr}
{Marigo} P., {Girardi} L., {Bressan} A., {Groenewegen} M.~A.~T., {Silva} L.,
  {Granato} G.~L., 2008, \aap, 482, 883

\bibitem[{{Marmier} {et~al}\mbox{.}(2013){Marmier}, {S{\'e}gransan}, {Udry},
  {Mayor}, {Pepe}, {Queloz}, {Lovis}, {Naef}, {Santos}, {Alonso}, {Alves},
  {Berthet}, {Chazelas}, {Demory}, {Dumusque}, {Eggenberger}, {Figueira},
  {Gillon}, {Hagelberg}, {Lendl}, {Mardling}, {M{\'e}gevand}, {Neveu},
  {Sahlmann}, {Sosnowska}, {Tewes}, \& {Triaud}}]{Marmier:2013lr}
{Marmier} M. {et~al.}, 2013, \aap, 551, A90

\bibitem[{{Mayor} \& {Queloz}(1995)}]{Mayor:1995uq}
{Mayor} M., {Queloz} D., 1995, \nat, 378, 355

\bibitem[{{Naoz} {et~al}\mbox{.}(2011){Naoz}, {Farr}, {Lithwick}, {Rasio}, \&
  {Teyssandier}}]{Naoz:2011lr}
{Naoz} S., {Farr} W.~M., {Lithwick} Y., {Rasio} F.~A., {Teyssandier} J., 2011,
  \nat, 473, 187

\bibitem[{{Neveu} {et~al}\mbox{.}(2012){Neveu}, {Sahlmann}, {Queloz}, \&
  {S{\'e}gransan}}]{Neveu:2012lr}
{Neveu} M., {Sahlmann} J., {Queloz} D., {S{\'e}gransan} D., 2012, in Orbital
  Couples: Pas de Deux in the Solar System and the Milky Way, {Arenou} F.,
  {Hestroffer} D., eds., pp. 81--85

\bibitem[{{Neveu-VanMalle} {et~al}\mbox{.}(2016){Neveu-VanMalle}, {Queloz},
  {Anderson}, {Brown}, {Collier Cameron}, {Delrez}, {D{\'{\i}}az}, {Gillon},
  {Hellier}, {Jehin}, {Lister}, {Pepe}, {Rojo}, {S{\'e}gransan}, {Triaud},
  {Turner}, \& {Udry}}]{Neveu-VanMalle:2016xy}
{Neveu-VanMalle} M. {et~al.}, 2016, \aap, 586, A93

\bibitem[{{Ortiz} {et~al}\mbox{.}(2016){Ortiz}, {Reffert}, {Trifonov},
  {Quirrenbach}, {Mitchell}, {Nowak}, {Buenzli}, {Zimmerman}, {Bonnefoy},
  {Skemer}, {Defr{\`e}re}, {Lee}, {Fischer}, \& {Hinz}}]{Ortiz:2016kq}
{Ortiz} M. {et~al.}, 2016, ArXiv e-prints

\bibitem[{{Perryman} {et~al}\mbox{.}(2014){Perryman}, {Hartman}, {Bakos}, \&
  {Lindegren}}]{Perryman:2014rf}
{Perryman} M., {Hartman} J., {Bakos} G.~{\'A}., {Lindegren} L., 2014, \apj,
  797, 14

\bibitem[{{Perryman} {et~al}\mbox{.}(2001){Perryman}, {de Boer}, {Gilmore},
  {H{\o}g}, {Lattanzi}, {Lindegren}, {Luri}, {Mignard}, {Pace}, \& {de
  Zeeuw}}]{Perryman:2001lr}
{Perryman} M.~A.~C. {et~al.}, 2001, \aap, 369, 339

\bibitem[{{Petrovich}(2015{\natexlab{a}})}]{Petrovich:2015qf}
{Petrovich} C., 2015{\natexlab{a}}, \apj, 799, 27

\bibitem[{{Petrovich}(2015{\natexlab{b}})}]{Petrovich:2015vn}
---, 2015{\natexlab{b}}, \apj, 808, 120

\bibitem[{{Pollacco} {et~al}\mbox{.}(2006){Pollacco}, {Skillen}, {Collier
  Cameron}, {Christian}, {Hellier}, {Irwin}, {Lister}, {Street}, {West},
  {Anderson}, {Clarkson}, {Deeg}, {Enoch}, {Evans}, {Fitzsimmons}, {Haswell},
  {Hodgkin}, {Horne}, {Kane}, {Keenan}, {Maxted}, {Norton}, {Osborne},
  {Parley}, {Ryans}, {Smalley}, {Wheatley}, \& {Wilson}}]{Pollacco:2006fj}
{Pollacco} D.~L. {et~al.}, 2006, \pasp, 118, 1407

\bibitem[{{Pollack} {et~al}\mbox{.}(1996){Pollack}, {Hubickyj}, {Bodenheimer},
  {Lissauer}, {Podolak}, \& {Greenzweig}}]{Pollack:1996uq}
{Pollack} J.~B., {Hubickyj} O., {Bodenheimer} P., {Lissauer} J.~J., {Podolak}
  M., {Greenzweig} Y., 1996, \icarus, 124, 62

\bibitem[{{Queloz} {et~al}\mbox{.}(2000){Queloz}, {Eggenberger}, {Mayor},
  {Perrier}, {Beuzit}, {Naef}, {Sivan}, \& {Udry}}]{Queloz:2000rt}
{Queloz} D., {Eggenberger} A., {Mayor} M., {Perrier} C., {Beuzit} J.~L., {Naef}
  D., {Sivan} J.~P., {Udry} S., 2000, \aap, 359, L13

\bibitem[{{Ragozzine} \& {Wolf}(2009)}]{Ragozzine:2009vn}
{Ragozzine} D., {Wolf} A.~S., 2009, \apj, 698, 1778

\bibitem[{{Ramm}(2015)}]{Ramm:2015yu}
{Ramm} D.~J., 2015, \mnras, 449, 4428

\bibitem[{{Rasio} \& {Ford}(1996)}]{Rasio:1996ly}
{Rasio} F.~A., {Ford} E.~B., 1996, Science, 274, 954

\bibitem[{{Rein} \& {Spiegel}(2015)}]{Rein:2015yu}
{Rein} H., {Spiegel} D.~S., 2015, \mnras, 446, 1424

\bibitem[{{Rein} \& {Tamayo}(2015)}]{Rein:2015qf}
{Rein} H., {Tamayo} D., 2015, \mnras, 452, 376

\bibitem[{{Rogers} {et~al}\mbox{.}(2013){Rogers}, {Lin}, {McElwaine}, \&
  {Lau}}]{Rogers:2013la}
{Rogers} T.~M., {Lin} D.~N.~C., {McElwaine} J.~N., {Lau} H.~H.~B., 2013, \apj,
  772, 21

\bibitem[{{Rosotti} {et~al}\mbox{.}(2016{\natexlab{a}}){Rosotti}, {Booth},
  {Clarke}, {Teyssandier}, {Facchini}, \& {Mustill}}]{Rosotti:2016qy}
{Rosotti} G.~P., {Booth} R.~A., {Clarke} C.~J., {Teyssandier} J., {Facchini}
  S., {Mustill} A.~J., 2016{\natexlab{a}}, ArXiv e-prints

\bibitem[{{Rosotti} {et~al}\mbox{.}(2016{\natexlab{b}}){Rosotti}, {Juhasz},
  {Booth}, \& {Clarke}}]{Rosotti:2016lq}
{Rosotti} G.~P., {Juhasz} A., {Booth} R.~A., {Clarke} C.~J.,
  2016{\natexlab{b}}, \mnras, 459, 2790

\bibitem[{{Sahlmann} {et~al}\mbox{.}(2011){Sahlmann}, {S{\'e}gransan},
  {Queloz}, {Udry}, {Santos}, {Marmier}, {Mayor}, {Naef}, {Pepe}, \&
  {Zucker}}]{Sahlmann:2011qy}
{Sahlmann} J. {et~al.}, 2011, \aap, 525, A95

\bibitem[{{Sahlmann}, {Triaud} \& {Martin}(2015){Sahlmann}, {Triaud}, \&
  {Martin}}]{Sahlmann:2015xy}
{Sahlmann} J., {Triaud} A.~H.~M.~J., {Martin} D.~V., 2015, \mnras, 447, 287

\bibitem[{{Santerne} {et~al}\mbox{.}(2014){Santerne}, {H{\'e}brard}, {Deleuil},
  {Havel}, {Correia}, {Almenara}, {Alonso}, {Arnold}, {Barros}, {Behrend},
  {Bernasconi}, {Boisse}, {Bonomo}, {Bouchy}, {Bruno}, {Damiani},
  {D{\'{\i}}az}, {Gravallon}, {Guillot}, {Labrevoir}, {Montagnier}, {Moutou},
  {Rinner}, {Santos}, {Abe}, {Audejean}, {Bendjoya}, {Gillier}, {Gregorio},
  {Martinez}, {Michelet}, {Montaigut}, {Poncy}, {Rivet}, {Rousseau}, {Roy},
  {Suarez}, {Vanhuysse}, \& {Verilhac}}]{Santerne:2014oq}
{Santerne} A. {et~al.}, 2014, \aap, 571, A37

\bibitem[{{Santerne} {et~al}\mbox{.}(2016){Santerne}, {Moutou}, {Tsantaki},
  {Bouchy}, {H{\'e}brard}, {Adibekyan}, {Almenara}, {Amard}, {Barros},
  {Boisse}, {Bonomo}, {Bruno}, {Courcol}, {Deleuil}, {Demangeon},
  {D{\'{\i}}az}, {Guillot}, {Havel}, {Montagnier}, {Rajpurohit}, {Rey}, \&
  {Santos}}]{Santerne:2016lr}
---, 2016, \aap, 587, A64

\bibitem[{{Schlaufman}(2010)}]{Schlaufman:2010fk}
{Schlaufman} K.~C., 2010, \apj, 719, 602

\bibitem[{{Schneider} {et~al}\mbox{.}(2011){Schneider}, {Dedieu}, {Le Sidaner},
  {Savalle}, \& {Zolotukhin}}]{Schneider:2011lr}
{Schneider} J., {Dedieu} C., {Le Sidaner} P., {Savalle} R., {Zolotukhin} I.,
  2011, \aap, 532, A79

\bibitem[{{Schwarz}(1978)}]{Schwarz:1978zz}
{Schwarz} G., 1978, Annals of Statistics, 6, 461

\bibitem[{{Seager} \& {Deming}(2010)}]{Seager:2010kx}
{Seager} S., {Deming} D., 2010, \araa, 48, 631

\bibitem[{{Sestito} \& {Randich}(2005)}]{Sestito:2005ys}
{Sestito} P., {Randich} S., 2005, \aap, 442, 615

\bibitem[{{Skrutskie} {et~al}\mbox{.}(2006){Skrutskie}, {Cutri}, {Stiening},
  {Weinberg}, {Schneider}, {Carpenter}, {Beichman}, {Capps}, {Chester},
  {Elias}, {Huchra}, {Liebert}, {Lonsdale}, {Monet}, {Price}, {Seitzer},
  {Jarrett}, {Kirkpatrick}, {Gizis}, {Howard}, {Evans}, {Fowler}, {Fullmer},
  {Hurt}, {Light}, {Kopan}, {Marsh}, {McCallon}, {Tam}, {Van Dyk}, \&
  {Wheelock}}]{Skrutskie:2006kx}
{Skrutskie} M.~F. {et~al.}, 2006, \aj, 131, 1163

\bibitem[{{Southworth} {et~al}\mbox{.}(2009){Southworth}, {Hinse},
  {J{\o}rgensen}, {Dominik}, {Ricci}, {Burgdorf}, {Hornstrup}, {Wheatley},
  {Anguita}, {Bozza}, {Novati}, {Harps{\o}e}, {Kj{\ae}rgaard}, {Liebig},
  {Mancini}, {Masi}, {Mathiasen}, {Rahvar}, {Scarpetta}, {Snodgrass}, {Surdej},
  {Th{\"o}ne}, \& {Zub}}]{Southworth:2009vn}
{Southworth} J. {et~al.}, 2009, \mnras, 396, 1023

\bibitem[{{Sozzetti} {et~al}\mbox{.}(2014){Sozzetti}, {Giacobbe}, {Lattanzi},
  {Micela}, {Morbidelli}, \& {Tinetti}}]{Sozzetti:2014qy}
{Sozzetti} A., {Giacobbe} P., {Lattanzi} M.~G., {Micela} G., {Morbidelli} R.,
  {Tinetti} G., 2014, \mnras, 437, 497

\bibitem[{{Spalding} \& {Batygin}(2015)}]{Spalding:2015ve}
{Spalding} C., {Batygin} K., 2015, ArXiv e-prints

\bibitem[{{Sterne}(1939)}]{Sterne:1939fj}
{Sterne} T.~E., 1939, \mnras, 99, 451

\bibitem[{{Stetson}(1987)}]{Stetson:1987kl}
{Stetson} P.~B., 1987, \pasp, 99, 191

\bibitem[{{Sutherland} \& {Fabrycky}(2016)}]{Sutherland:2016vn}
{Sutherland} A.~P., {Fabrycky} D.~C., 2016, \apj, 818, 6

\bibitem[{{Teyssandier} \& {Ogilvie}(2016)}]{Teyssandier:2016kx}
{Teyssandier} J., {Ogilvie} G.~I., 2016, \mnras, 458, 3221

\bibitem[{{Thies} {et~al}\mbox{.}(2011){Thies}, {Kroupa}, {Goodwin},
  {Stamatellos}, \& {Whitworth}}]{Thies:2011yq}
{Thies} I., {Kroupa} P., {Goodwin} S.~P., {Stamatellos} D., {Whitworth} A.~P.,
  2011, \mnras, 417, 1817

\bibitem[{{Torres}, {Andersen} \& {Gim{\'e}nez}(2010){Torres}, {Andersen}, \&
  {Gim{\'e}nez}}]{Torres:2010uq}
{Torres} G., {Andersen} J., {Gim{\'e}nez} A., 2010, \aapr, 18, 67

\bibitem[{{Tregloan-Reed} \& {Southworth}(2013)}]{Tregloan-Reed:2013yq}
{Tregloan-Reed} J., {Southworth} J., 2013, \mnras, 431, 966

\bibitem[{{Triaud}(2011)}]{Triaud:2011fk}
{Triaud} A.~H.~M.~J., 2011, \aap, 534, L6

\bibitem[{{Triaud} {et~al}\mbox{.}(2010){Triaud}, {Collier Cameron}, {Queloz},
  {Anderson}, {Gillon}, {Hebb}, {Hellier}, {Loeillet}, {Maxted}, {Mayor},
  {Pepe}, {Pollacco}, {S{\'e}gransan}, {Smalley}, {Udry}, {West}, \&
  {Wheatley}}]{Triaud:2010fr}
{Triaud} A.~H.~M.~J. {et~al.}, 2010, \aap, 524, A25

\bibitem[{{Triaud} {et~al}\mbox{.}(2015){Triaud}, {Gillon}, {Ehrenreich},
  {Herrero}, {Lendl}, {Anderson}, {Collier Cameron}, {Delrez}, {Demory},
  {Hellier}, {Heng}, {Jehin}, {Maxted}, {Pollacco}, {Queloz}, {Ribas},
  {Smalley}, {Smith}, \& {Udry}}]{Triaud:2015fk}
---, 2015, \mnras, 450, 2279

\bibitem[{{Triaud} {et~al}\mbox{.}(2009){Triaud}, {Queloz}, {Bouchy}, {Moutou},
  {Collier Cameron}, {Claret}, {Barge}, {Benz}, {Deleuil}, {Guillot},
  {H{\'e}brard}, {Lecavelier Des {\'E}tangs}, {Lovis}, {Mayor}, {Pepe}, \&
  {Udry}}]{Triaud:2009qy}
---, 2009, \aap, 506, 377

\bibitem[{{Triaud} {et~al}\mbox{.}(2011){Triaud}, {Queloz}, {Hellier},
  {Gillon}, {Smalley}, {Hebb}, {Collier Cameron}, {Anderson}, {Boisse},
  {H{\'e}brard}, {Jehin}, {Lister}, {Lovis}, {Maxted}, {Pepe}, {Pollacco},
  {S{\'e}gransan}, {Simpson}, {Udry}, \& {West}}]{Triaud:2011vn}
---, 2011, \aap, 531, A24

\bibitem[{{Ward}(1997)}]{Ward:1997kx}
{Ward} W.~R., 1997, \icarus, 126, 261

\bibitem[{{Winn} \& {Fabrycky}(2015)}]{Winn:2015lr}
{Winn} J.~N., {Fabrycky} D.~C., 2015, \araa, 53, 409

\bibitem[{{Winn} {et~al}\mbox{.}(2009){Winn}, {Johnson}, {Albrecht}, {Howard},
  {Marcy}, {Crossfield}, \& {Holman}}]{Winn:2009lr}
{Winn} J.~N., {Johnson} J.~A., {Albrecht} S., {Howard} A.~W., {Marcy} G.~W.,
  {Crossfield} I.~J., {Holman} M.~J., 2009, \apjl, 703, L99

\bibitem[{{Wright} {et~al}\mbox{.}(2011){Wright}, {Fakhouri}, {Marcy}, {Han},
  {Feng}, {Johnson}, {Howard}, {Fischer}, {Valenti}, {Anderson}, \&
  {Piskunov}}]{Wright:2011fj}
{Wright} J.~T. {et~al.}, 2011, \pasp, 123, 412

\bibitem[{{Wu}, {Murray} \& {Ramsahai}(2007){Wu}, {Murray}, \&
  {Ramsahai}}]{Wu:2007ve}
{Wu} Y., {Murray} N.~W., {Ramsahai} J.~M., 2007, \apj, 670, 820

\bibitem[{{Zacharias} {et~al}\mbox{.}(2004){Zacharias}, {Monet}, {Levine},
  {Urban}, {Gaume}, \& {Wycoff}}]{Zacharias:2004fk}
{Zacharias} N., {Monet} D.~G., {Levine} S.~E., {Urban} S.~E., {Gaume} R.,
  {Wycoff} G.~L., 2004, in Bulletin of the American Astronomical Society,
  Vol.~36, American Astronomical Society Meeting Abstracts, p. 1418

\end{thebibliography}

\appendix
\section{Journal of Observations}\label{app:obs}

\onecolumn

\begin{longtable}{ccccccc}
\caption{Radial velocities of WASP-53 obtained with CORALIE before its recent upgrade. BJD is the barycentric Julian date -- 2\,450\,000 days. $V_{\rm rad}$ is the radial velocity obtained by fitting a cross-correlation function with a Gaussian, $\sigma_{\rm RV}$ is the error on $V_{\rm rad}$. FWHM is the full width at half maximum of the cross-correlation function, and contrast is the amplitude.}\label{}\\
\hline
\hline 
BJD	& $V_{\rm rad}$	& $\sigma_{\rm RV}$	& FWHM	& contrast	& slope bisector span	& exposure	\\
(days)		&	(km s$^{-1}$)	&	(km s$^{-1}$)		&	(km s$^{-1}$)	&		(\%)	&	(km s$^{-1}$)	& (sec)   \\     
\hline
\endfirsthead
\hline
\hline 
JDB	& $V_{\rm rad}$	& $\sigma_{\rm RV}$	& FWHM	& contrast	& slope bisector span	& exposure	\\
(days)		&	(km s$^{-1}$)	&	(km s$^{-1}$)		&	(km s$^{-1}$)	&		(\%)	&	(km s$^{-1}$)	& (sec)   \\     
\hline
\endhead
\hline
\multicolumn{6}{l}{Table continues next page...}\\
\hline
\endfoot
\hline
\endlastfoot 
55535.622498&-4.38964&0.01414&7.76355&37.811&0.00572&1800.679\\
55563.633734&-4.75981&0.01675&7.83813&38.764&-0.01915&1800.736\\
55565.620787&-4.10113&0.02222&7.85708&29.332&0.06569&1800.743\\
55583.571236&-4.53580&0.01833&7.76400&37.735&-0.01908&1800.754\\
55586.577359&-4.45769&0.01936&7.88629&36.602&0.03945&1800.755\\
55588.564415&-3.81514&0.02958&7.88939&36.377&0.10109&1800.756\\
55591.537016&-3.85825&0.01594&7.85784&37.219&0.01104&1800.759\\
55593.534932&-4.43614&0.02059&7.88560&36.521&0.01531&1800.758\\
55595.533798&-3.86707&0.02407&7.86921&36.778&0.01591&1800.859\\
55596.534633&-4.37790&0.01454&7.89625&37.250&0.02008&1800.739\\
55599.531272&-4.19644&0.02246&7.82083&35.628&0.04899&1800.775\\
55600.531357&-4.32590&0.02605&7.81524&34.950&0.02073&1800.736\\
55613.530286&-4.35247&0.03126&7.75947&35.915&0.02255&1800.830\\
55614.520174&-3.80772&0.03751&7.98353&12.241&0.08784&1800.768\\
55619.515487&-4.11077&0.03829&7.86610&12.606&0.07253&1800.726\\
55624.511532&-3.81662&0.04668&7.99027&31.484&0.16010&1800.767\\
55769.931969&-3.69384&0.03211&7.83406&40.671&-0.01480&1800.816\\
55770.811670&-3.50787&0.01869&7.84814&40.222&0.05379&1800.797\\
55772.921899&-3.88919&0.02320&7.76734&20.208&0.03801&1800.748\\
55777.849432&-3.68608&0.01545&7.81802&39.166&-0.01838&1800.821\\
55782.883847&-3.89564&0.01539&7.75827&40.337&0.05484&1800.740\\
55802.854702&-3.81609&0.01659&7.79265&40.844&0.00480&1800.739\\
55806.876427&-3.49432&0.01352&7.77050&40.390&-0.00408&1800.777\\
55809.812708&-3.62453&0.01453&7.85935&40.282&-0.03700&1800.739\\
55810.864649&-3.61455&0.01691&7.82177&40.031&0.00585&1800.798\\
55811.834443&-4.11173&0.01961&7.82148&39.929&-0.00983&1800.777\\
55825.779642&-3.93644&0.01342&7.83048&39.001&-0.03320&1800.718\\
55826.823506&-3.49648&0.02008&7.81726&39.820&-0.10400&1800.818\\
55834.845196&-4.07174&0.01455&7.82559&40.420&0.02738&1800.758\\
55852.655157&-3.71694&0.01847&7.89498&39.010&-0.02472&1800.759\\
55869.776795&-3.46744&0.01511&7.84842&39.370&-0.03676&1800.736\\
55887.725010&-4.01495&0.01502&7.76766&39.004&0.02644&1800.754\\
55889.585568&-3.46497&0.01504&7.78976&38.829&-0.01253&1800.755\\
55910.593618&-3.84547&0.01446&7.80091&39.026&-0.05197&1800.778\\
55952.575538&-3.44630&0.01294&7.90378&38.024&0.02529&1800.688\\
55974.534361&-4.02707&0.02632&7.83726&38.734&-0.02393&1800.748\\
56101.929818&-3.49841&0.02543&7.87117&40.333&-0.02903&1800.799\\
56108.902853&-3.68569&0.03173&7.87649&39.586&0.02577&1800.940\\
56130.929240&-3.54201&0.02272&7.82397&39.617&-0.10018&1800.817\\
56158.920126&-3.87751&0.02491&7.83489&39.943&-0.04714&1800.781\\
56165.794126&-3.99424&0.02180&7.84562&35.852&-0.00917&1800.759\\
56166.840915&-3.81564&0.01755&7.86863&39.564&-0.03721&1800.777\\
56182.742965&-4.11729&0.01762&7.83280&38.786&0.04594&1800.798\\
56184.735109&-3.50543&0.01419&7.81771&39.385&-0.00971&1800.758\\
56186.764241&-3.81243&0.01735&7.89344&39.090&0.05377&1800.778\\
56190.675191&-3.46799&0.02295&7.92760&37.779&0.04893&1800.820\\
56196.864653&-3.71479&0.01309&7.83567&38.860&-0.02100&2700.667\\
56235.613045&-4.09012&0.02545&7.83788&37.690&0.01474&1800.777\\
56245.697375&-4.06430&0.01518&7.90888&37.534&0.00491&1800.756\\
56264.722841&-3.84938&0.01631&7.88340&39.189&0.03949&1800.749\\
56309.616625&-3.61756&0.01619&7.91453&40.096&-0.03900&1800.739\\
56335.551640&-3.93777&0.02599&8.01659&39.469&0.01497&1800.758\\
56460.921265&-4.08678&0.01861&7.87168&43.357&-0.00289&1800.756\\
56490.864375&-4.05183&0.02101&7.86144&40.429&-0.05535&1800.755\\
56505.803479&-3.53226&0.01677&7.88179&41.428&0.04581&1800.799\\
56516.868634&-4.11013&0.02259&7.86987&41.506&-0.04888&1800.857\\
56538.754792&-3.46848&0.02937&7.87498&40.897&0.01746&1800.779\\
56543.888666&-3.98696&0.01538&7.87143&39.824&0.02301&1800.763\\
56544.693786&-3.50813&0.03240&7.85548&41.040&0.06695&1800.802\\
56545.715662&-3.63146&0.01620&7.94485&37.812&-0.00368&1800.819\\
56562.781196&-3.94366&0.01417&7.82567&40.715&-0.00693&1800.786\\
56563.802018&-3.97705&0.01494&7.91152&40.932&0.00453&1800.836\\
56565.713332&-3.68586&0.01541&7.84101&40.339&0.00271&1800.777\\
56567.702293&-3.59801&0.02820&7.86619&41.893&0.03071&1800.775\\
56591.740189&-3.50574&0.01941&7.79994&39.462&0.00814&1800.719\\
56603.739663&-3.82508&0.01907&7.84847&40.983&0.01256&1800.779\\
56644.683453&-3.51463&0.01493&7.76010&38.600&-0.01970&1800.778\\
56682.557464&-4.10162&0.01957&7.85501&40.062&-0.00216&1800.770\\
56817.930262&-4.05370&0.02181&7.86074&41.914&0.01356&1602.786\\
56877.826276&-4.07700&0.01910&7.80224&38.824&0.01261&1800.058\\
56878.777852&-3.63053&0.03452&7.88373&34.977&0.04216&1800.685\\
56920.870295&-4.12036&0.01832&7.83603&39.391&-0.03675&1800.677\\
56961.797598&-3.54257&0.01781&7.78155&38.574&0.05712&1800.601\\
\end{longtable}

\begin{longtable}{ccccccc}
\caption{Radial velocities of WASP-53 obtained with CORALIE after its recent upgrade. BJD is the barycentric Julian date -- 2\,450\,000 days. $V_{\rm rad}$ is the radial velocity obtained by fitting a cross-correlation function with a Gaussian, $\sigma_{\rm RV}$ is the error on $V_{\rm rad}$. FWHM is the full width at half maximum of the cross-correlation function, and contrast is the amplitude.}\label{}\\
\hline
\hline 
BJD	& $V_{\rm rad}$	& $\sigma_{\rm RV}$	& FWHM	& contrast	& slope bisector span	& exposure	\\
(days)		&	(km s$^{-1}$)	&	(km s$^{-1}$)		&	(km s$^{-1}$)	&		(\%)	&	(km s$^{-1}$)	& (sec)   \\     
\hline
\endfirsthead
\hline
\hline 
JDB	& $V_{\rm rad}$	& $\sigma_{\rm RV}$	& FWHM	& contrast	& slope bisector span	& exposure	\\
(days)		&	(km s$^{-1}$)	&	(km s$^{-1}$)		&	(km s$^{-1}$)	&		(\%)	&	(km s$^{-1}$)	& (sec)   \\     
\hline
\endhead
\hline
\multicolumn{6}{l}{Table continues next page...}\\
\hline
\endfoot
\hline
\endlastfoot 
56989.740666&-3.92092&0.02509&7.82426&42.096&-0.11250&1800.683\\
57001.640882&-3.53647&0.02300&7.73609&42.361&0.00145&1800.780\\
57003.640065&-4.11709&0.01937&7.71262&41.800&-0.01624&1800.075\\
57004.682396&-3.59763&0.02263&7.74968&41.276&-0.00541&1800.932\\
57012.592977&-3.75798&0.02566&7.80047&42.030&-0.03335&1800.233\\
57027.631091&-3.72146&0.03574&7.68990&38.087&0.04302&1800.770\\
57063.528866&-4.02456&0.05181&7.77355&41.187&-0.02410&1800.102\\
57065.534295&-3.74442&0.03302&7.76903&41.060&-0.00757&1800.943\\
57192.938473&-3.84251&0.03767&7.74347&45.120&0.01185&1800.861\\
57205.925822&-3.96730&0.04649&7.82306&42.941&0.03207&1800.925\\
57261.823294&-4.14072&0.04184&7.70847&43.479&0.00495&1800.924\\
57341.740033&-3.88956&0.05660&7.82722&43.813&-0.08024&1800.895\\
57362.678177&-3.49435&0.01850&7.75263&43.012&-0.00904&2700.715\\
57367.607494&-4.16776&0.02672&7.72596&43.081&0.00961&2700.804\\
57381.547386&-3.92186&0.02486&7.66936&41.940&-0.00042&2700.686\\
57389.610441&-3.61683&0.02988&7.76154&43.413&-0.02558&1800.083\\
57413.581920&-4.09479&0.03288&7.77853&41.771&0.04996&2700.626\\
57587.836041&-3.52915&0.02814&7.79372&41.294&-0.06745&1800.382\\
57616.796911&-3.78390&0.02024&7.67173&42.253&-0.03829&1800.431\\
57681.574518&-4.02549&0.03158&7.69350&43.049&0.03337&1800.382\\
57689.706629&-3.71986&0.02108&7.72289&42.465&0.02337&1800.354\\
57691.675531&-4.13251&0.01765&7.72530&42.658&-0.01465&1800.394\\
57711.743027&-4.18272&0.03598&7.68311&42.736&-0.03452&1800.352\\
57713.676572&-3.56151&0.02072&7.80813&42.301&-0.01114&1800.353\\
57713.698411&-3.54916&0.02373&7.77659&42.830&-0.04693&1800.372\\
\end{longtable}

\begin{longtable}{ccccccc}
\caption{Radial velocities of WASP-53 obtained with HARPS. BJD is the barycentric Julian date -- 2\,450\,000 days. $V_{\rm rad}$ is the radial velocity obtained by fitting a cross-correlation function with a Gaussian, $\sigma_{\rm RV}$ is the error on $V_{\rm rad}$. FWHM is the full width at half maximum of the cross-correlation function, and contrast is the amplitude.}\label{}\\
\hline
\hline 
BJD	& $V_{\rm rad}$	& $\sigma_{\rm RV}$	& FWHM	& contrast	& slope bisector span	& exposure	\\
(days)		&	(km s$^{-1}$)	&	(km s$^{-1}$)		&	(km s$^{-1}$)	&		(\%)	&	(km s$^{-1}$)	& (sec)   \\     
\hline
\endfirsthead
\hline
\hline 
JDB	& $V_{\rm rad}$	& $\sigma_{\rm RV}$	& FWHM	& contrast	& slope bisector span	& exposure	\\
(days)		&	(km s$^{-1}$)	&	(km s$^{-1}$)		&	(km s$^{-1}$)	&		(\%)	&	(km s$^{-1}$)	& (sec)   \\     
\hline
\endhead
\hline
\multicolumn{6}{l}{Table continues next page...}\\
\hline
\endfoot
\hline
\endlastfoot 
55802.804248&-3.87536&0.01428&6.35185&48.819&0.01421&900.000\\
55802.902737&-3.77124&0.00615&6.44231&49.068&0.01648&900.000\\
55803.735877&-3.46797&0.00592&6.43227&49.138&0.02630&900.000\\
55803.876858&-3.49415&0.01413&6.44454&48.445&0.03957&900.000\\
55825.652716&-4.00995&0.00918&6.43519&50.675&-0.00062&900.006\\
55825.842017&-3.90801&0.00695&6.40550&50.365&0.03042&900.006\\
55826.656742&-3.47586&0.00682&6.42903&50.419&0.03065&900.006\\
55826.837326&-3.44345&0.00407&6.45292&49.300&0.02401&900.006\\
55827.646391&-3.74232&0.01085&6.42916&48.557&0.03541&600.000\\
55827.656368&-3.74253&0.00678&6.46451&48.782&0.01240&900.006\\
55827.669238&-3.75174&0.00631&6.49718&48.731&0.01620&900.006\\
55827.680986&-3.73814&0.00634&6.46326&48.738&0.05819&900.001\\
55827.692515&-3.74724&0.00573&6.44915&49.065&0.00441&900.006\\
55827.703846&-3.77343&0.00561&6.42803&49.040&0.01460&900.006\\
55827.715374&-3.78034&0.00546&6.47191&48.934&0.01938&900.006\\
55827.726810&-3.78879&0.00536&6.45148&49.091&0.00973&900.001\\
55827.738546&-3.79623&0.00512&6.43309&49.178&0.00681&900.006\\
55827.749681&-3.81542&0.00512&6.43640&49.235&0.01215&900.006\\
55827.761637&-3.80869&0.00518&6.44554&49.108&0.02622&900.006\\
55827.773177&-3.80569&0.00438&6.43072&49.362&0.02610&900.006\\
55827.784612&-3.81191&0.00451&6.43123&49.319&0.01551&900.006\\
55827.796036&-3.82316&0.00414&6.44297&49.391&0.03097&900.006\\
55827.807576&-3.83545&0.00448&6.44773&49.167&0.01898&900.006\\
55827.819324&-3.84303&0.00454&6.43978&49.196&0.03451&900.006\\
55827.830760&-3.84995&0.00553&6.47049&48.899&0.00391&900.006\\
55827.842091&-3.84996&0.00523&6.47341&48.922&0.00081&900.006\\
55828.639175&-4.09963&0.00613&6.47747&48.749&0.01814&900.006\\
55828.880860&-4.04309&0.00432&6.47502&49.127&0.01328&900.006\\
55831.640517&-4.07244&0.00490&6.49495&48.903&0.00429&900.006\\
55831.862305&-4.09707&0.00367&6.44317&49.351&0.03159&900.006\\
56108.917970&-3.67055&0.01289&6.48511&48.048&0.01989&900.000\\
56109.931976&-4.08058&0.00352&6.42686&49.455&0.01035&1800.000\\
56158.807248&-3.82192&0.00781&6.45609&48.928&0.02814&600.000\\
56159.819284&-4.04984&0.01179&6.43642&49.040&0.01204&600.006\\
56190.816064&-3.45497&0.00746&6.40310&49.253&0.02240&899.999\\
56191.702180&-3.70159&0.00717&6.40858&49.055&0.05467&799.999\\
56191.713245&-3.70009&0.00584&6.40354&49.321&0.02801&900.000\\
56191.724577&-3.69652&0.00596&6.39529&49.155&0.05629&900.000\\
56191.736348&-3.72787&0.00617&6.41546&49.184&0.01366&899.999\\
56191.747992&-3.72226&0.00600&6.41604&49.269&0.01401&900.000\\
56191.759624&-3.72504&0.00565&6.37832&49.470&0.04169&900.000\\
56191.771060&-3.72870&0.00555&6.39147&49.478&0.02959&900.000\\
56191.782391&-3.74305&0.00534&6.39205&49.326&0.01126&900.000\\
56191.794012&-3.74230&0.00596&6.39070&49.423&0.03308&900.000\\
56191.805748&-3.74590&0.00535&6.39200&49.398&0.01416&900.006\\
56191.817172&-3.76612&0.00464&6.40073&49.412&0.00858&899.999\\
56191.828724&-3.77051&0.00455&6.39737&49.586&0.01365&899.999\\
56191.840043&-3.79008&0.00529&6.40980&49.416&0.01943&900.000\\
56191.851583&-3.78008&0.00543&6.41797&49.150&0.03092&900.006\\
56191.863100&-3.78198&0.00526&6.40679&49.445&0.02901&900.006\\
56191.874628&-3.79791&0.00667&6.40276&49.074&0.02995&900.006\\
56191.886168&-3.80101&0.00761&6.47274&48.383&0.02179&900.006\\
56191.897592&-3.80062&0.00746&6.45821&48.542&0.03297&900.006\\
56191.909027&-3.82351&0.00717&6.48428&47.642&-0.00093&900.000\\
56192.821879&-4.07292&0.00787&6.41754&48.966&0.05738&600.006\\
56193.794605&-3.56742&0.01278&6.43435&48.521&-0.07894&599.999\\
56215.823377&-4.10629&0.00762&6.50823&48.183&0.00481&900.000\\
56221.537217&-3.74752&0.01691&6.51008&49.285&0.02560&899.999\\
56239.592460&-3.87591&0.00851&6.45703&48.675&-0.00723&600.004\\
56256.602476&-3.60375&0.01122&6.44653&47.974&0.02296&600.000\\
56257.580498&-3.53847&0.00792&6.44921&48.513&0.02276&600.008\\
56264.522496&-3.70152&0.00630&6.42133&48.725&0.02209&700.000\\
56264.535284&-3.70567&0.00475&6.44658&49.066&0.03476&900.000\\
56264.547494&-3.71370&0.00443&6.45767&49.036&0.00891&900.001\\
56264.559264&-3.72208&0.00418&6.43897&49.154&0.01169&900.001\\
56264.571220&-3.72100&0.00415&6.45931&49.191&0.02381&900.000\\
56264.583638&-3.72408&0.00410&6.43960&49.306&0.02411&900.001\\
56264.595350&-3.73021&0.00391&6.43981&49.248&0.00984&900.001\\
56264.607201&-3.73698&0.00406&6.45517&49.099&0.04027&900.001\\
56264.619145&-3.74812&0.00392&6.43708&49.233&0.01867&900.000\\
56264.630926&-3.76424&0.00387&6.43620&49.161&0.01727&900.001\\
56264.643541&-3.77312&0.00419&6.45208&49.142&0.02687&900.000\\
56264.655415&-3.78187&0.00454&6.44031&48.969&-0.00318&900.000\\
56264.667243&-3.77665&0.00507&6.47638&48.785&0.02202&900.001\\
56264.679106&-3.78417&0.00526&6.45092&48.850&0.02116&900.000\\
56264.700945&-3.80013&0.00563&6.44897&48.770&-0.00597&900.000\\
56307.533598&-3.69537&0.00686&6.45089&48.804&0.01119&600.001\\
56323.548244&-3.45869&0.00931&6.50816&48.367&0.03162&600.001\\
56460.933662&-4.06790&0.01142&6.50197&48.107&0.02895&600.000\\
56567.891937&-3.50754&0.00974&6.50537&48.057&0.03808&900.000\\
56608.727539&-3.69785&0.01233&6.50649&49.948&-0.01752&900.001\\
56927.722485&-4.07286&0.00838&6.44451&50.116&0.04495&900.001\\
56928.789883&-3.51368&0.00559&6.43032&49.834&0.02100&900.001\\
\end{longtable}

\begin{longtable}{ccccccc}
\caption{Radial velocities of WASP-81 obtained with CORALIE before its recent upgrade. BJD is the barycentric Julian date - 2\,450\,000 days. $V_{\rm rad}$ is the radial velocity obtained by fitting a cross-correlation function with a Gaussian, $\sigma_{\rm RV}$ is the error on $V_{\rm rad}$. FWHM is the full width at half maximum of the cross-correlation function, and contrast is the amplitude.}\label{}\\
\hline
\hline 
BJD	& $V_{\rm rad}$	& $\sigma_{\rm RV}$	& FWHM	& contrast	& slope bisector span	& exposure	\\
(days)		&	(km s$^{-1}$)	&	(km s$^{-1}$)		&	(km s$^{-1}$)	&		(\%)	&	(km s$^{-1}$)	& (sec)   \\     
\hline
\endfirsthead
\hline
\hline 
JDB	& $V_{\rm rad}$	& $\sigma_{\rm RV}$	& FWHM	& contrast	& slope bisector span	& exposure	\\
(days)		&	(km s$^{-1}$)	&	(km s$^{-1}$)		&	(km s$^{-1}$)	&		(\%)	&	(km s$^{-1}$)	& (sec)   \\     
\hline
\endhead
\hline
\multicolumn{6}{l}{Table continues next page...}\\
\hline
\endfoot
\hline
\endlastfoot 
55833.506244&-60.15011&0.02609&7.87708&28.619&-0.02494&1800.760\\
55834.511338&-60.31782&0.01861&8.06275&28.959&0.00051&1800.778\\
55835.513064&-60.23827&0.02606&7.95214&29.089&-0.10707&1800.742\\
55851.553175&-60.35013&0.02264&7.95814&26.734&-0.00195&1800.720\\
55852.509304&-60.17890&0.02674&7.99858&27.682&0.01749&1800.738\\
55856.541346&-60.33501&0.02613&7.94386&25.704&0.08807&1800.760\\
55859.509870&-60.34664&0.02357&7.88504&28.508&-0.00745&1800.780\\
55862.507592&-60.34739&0.02726&8.04740&28.210&0.02566&1800.736\\
56050.916807&-60.48137&0.02097&8.07575&29.178&-0.02474&1800.735\\
56067.826597&-60.58673&0.02138&8.03943&30.584&0.00355&1800.755\\
56068.860596&-60.68491&0.02516&8.05512&30.442&0.03245&1800.776\\
56069.893252&-60.54821&0.03319&8.04313&30.912&-0.07498&1800.735\\
56075.788615&-60.51078&0.02761&8.02011&30.192&-0.01121&1800.778\\
56076.818326&-60.73022&0.03529&8.07559&30.163&-0.04915&1800.798\\
56101.786777&-60.65253&0.02870&7.97893&30.248&-0.05570&1800.841\\
56103.782366&-60.71255&0.01969&8.04937&30.321&0.05327&1800.796\\
56108.832891&-60.72716&0.03353&8.05096&30.385&-0.06728&1800.796\\
56133.765892&-60.75905&0.02643&8.07510&30.056&-0.04689&1800.777\\
56135.751516&-60.64218&0.03190&8.09815&28.651&-0.14222&1800.753\\
56147.744152&-60.73914&0.04144&7.95476&26.863&-0.04738&2700.746\\
56151.718366&-60.56578&0.02640&7.99750&29.524&0.08671&2700.607\\
56154.664701&-60.64301&0.05536&8.11592&29.581&-0.05746&1800.799\\
56158.595576&-60.75601&0.02555&7.92280&30.002&-0.02869&1800.841\\
56181.617460&-60.64235&0.01367&7.97978&29.506&0.01627&2700.606\\
56182.616959&-60.80774&0.01615&7.95422&29.259&0.01257&2700.707\\
56204.525708&-60.82325&0.02655&8.06633&27.984&-0.01402&2700.603\\
56230.518367&-60.64910&0.02851&7.95769&22.694&-0.00547&2700.696\\
56431.902050&-60.84674&0.02090&7.98365&30.384&-0.00998&2700.679\\
56455.806063&-60.72687&0.02870&7.95465&29.802&-0.02273&1800.797\\
56487.786778&-60.74249&0.02033&7.98515&30.749&0.00328&2700.713\\
56509.710623&-60.66513&0.01397&7.95360&30.834&-0.02276&2700.732\\
56511.692477&-60.85952&0.02225&7.94794&30.979&0.01657&1800.741\\
56518.621811&-60.71971&0.01770&7.97615&30.608&0.00907&1620.506\\
56531.555446&-60.63273&0.03202&7.97035&29.748&0.00804&2700.637\\
56547.652096&-60.58724&0.01827&8.03302&30.090&0.02345&2700.599\\
56548.506904&-60.61018&0.01534&7.99134&28.749&-0.03196&2700.680\\
56556.531108&-60.54362&0.05433&8.01616&29.137&-0.01045&2700.631\\
56560.494763&-60.71382&0.03762&7.96320&29.254&0.01284&1800.774\\
56565.533770&-60.70908&0.01949&7.86719&29.094&-0.02711&2700.712\\
56573.536534&-60.65052&0.01500&7.97570&29.996&-0.00324&2700.713\\
56585.548076&-60.46305&0.01694&7.98314&29.017&0.05210&2700.649\\
56595.517658&-60.51555&0.01739&7.94433&29.102&0.00994&2700.571\\
56602.513782&-60.25658&0.02280&8.11640&27.697&0.02679&2700.797\\
56610.522915&-60.19533&0.02343&7.99369&28.294&0.07335&2700.613\\
56764.896812&-58.50304&0.01351&8.03570&30.083&0.00910&2700.766\\
56776.887472&-58.68466&0.02343&7.97764&30.796&0.02733&2700.763\\
56804.815200&-58.96099&0.02049&7.95361&30.061&-0.00476&2700.910\\
56811.785740&-58.86942&0.01887&8.04819&30.119&0.03930&2700.909\\
56835.751223&-59.01187&0.03321&8.10950&28.842&-0.03052&2700.800\\
56856.750436&-59.28657&0.01680&8.01407&29.501&-0.03712&2700.187\\
56886.668064&-59.46175&0.01926&7.92853&30.839&-0.03399&2700.767\\
56920.621957&-59.54393&0.02171&8.03739&27.815&-0.01603&2700.428\\
56954.527818&-59.78127&0.02459&7.99078&24.160&0.00379&2699.955\\
\end{longtable}

\begin{longtable}{ccccccc}
\caption{Radial velocities of WASP-81 obtained with CORALIE before its recent upgrade. BJD is the barycentric Julian date - 2\,450\,000 days. $V_{\rm rad}$ is the radial velocity obtained by fitting a cross-correlation function with a Gaussian, $\sigma_{\rm RV}$ is the error on $V_{\rm rad}$. FWHM is the full width at half maximum of the cross-correlation function, and contrast is the amplitude.}\label{}\\
\hline
\hline 
BJD	& $V_{\rm rad}$	& $\sigma_{\rm RV}$	& FWHM	& contrast	& slope bisector span	& exposure	\\
(days)		&	(km s$^{-1}$)	&	(km s$^{-1}$)		&	(km s$^{-1}$)	&		(\%)	&	(km s$^{-1}$)	& (sec)   \\     
\hline
\endfirsthead
\hline
\hline 
JDB	& $V_{\rm rad}$	& $\sigma_{\rm RV}$	& FWHM	& contrast	& slope bisector span	& exposure	\\
(days)		&	(km s$^{-1}$)	&	(km s$^{-1}$)		&	(km s$^{-1}$)	&		(\%)	&	(km s$^{-1}$)	& (sec)   \\     
\hline
\endhead
\hline
\multicolumn{6}{l}{Table continues next page...}\\
\hline
\endfoot
\hline
\endlastfoot 
57186.791639&-60.24486&0.03042&7.95261&34.569&0.01499&2700.846\\
57194.736148&-60.23109&0.04093&7.89406&34.000&0.06709&2700.943\\
57211.722549&-60.44937&0.06100&7.91874&33.627&-0.06934&2700.764\\
57256.542816&-60.39378&0.04425&7.97087&32.551&-0.07702&2700.014\\
57271.550777&-60.56686&0.04207&7.93436&32.440&-0.11635&2452.822\\
57294.561176&-60.47338&0.03198&7.90803&32.325&0.03218&2700.157\\
57324.515825&-60.41128&0.05112&7.96576&32.327&0.01866&2700.768\\
57584.816334&-60.83346&0.09427&8.04004&33.592&0.00022&600.528\\
57595.700710&-60.85481&0.05770&7.87866&32.934&-0.16166&1800.381\\
57650.537435&-60.75600&0.02171&7.81752&32.403&-0.05037&2700.072\\
57652.595356&-60.87248&0.01754&7.90775&32.227&-0.03345&2700.069\\
57661.566328&-60.75291&0.01948&7.98951&32.002&-0.09351&2700.037\\
57680.545006&-60.81445&0.04449&7.85147&31.815&0.02024&1800.422\\
57682.504365&-60.89638&0.02700&7.86095&32.269&-0.03440&1800.000\\
\end{longtable}

\begin{longtable}{ccccccc}
\caption{Radial velocities of WASP-81 obtained with HARPS. BJD is the barycentric Julian date -- 2\,450\,000 days. $V_{\rm rad}$ is the radial velocity obtained by fitting a cross-correlation function with a Gaussian, $\sigma_{\rm RV}$ is the error on $V_{\rm rad}$. FWHM is the full width at half maximum of the cross-correlation function, and contrast is the amplitude. One datum, which was not used in the analysis, is highlighted with an asterisk.}\label{}\\
\hline
\hline 
BJD	& $V_{\rm rad}$	& $\sigma_{\rm RV}$	& FWHM	& contrast	& slope bisector span	& exposure	\\
(days)		&	(km s$^{-1}$)	&	(km s$^{-1}$)		&	(km s$^{-1}$)	&		(\%)	&	(km s$^{-1}$)	& (sec)   \\     
\hline
\endfirsthead
\hline
\hline 
JDB	& $V_{\rm rad}$	& $\sigma_{\rm RV}$	& FWHM	& contrast	& slope bisector span	& exposure	\\
(days)		&	(km s$^{-1}$)	&	(km s$^{-1}$)		&	(km s$^{-1}$)	&		(\%)	&	(km s$^{-1}$)	& (sec)   \\     
\hline
\endhead
\hline
\multicolumn{6}{l}{Table continues next page...}\\
\hline
\endfoot
\hline
\endlastfoot 
6403.902377&-60.75558&0.01422&6.77096&36.778&0.00282&600.000\\
6407.886489&-60.90138&0.01261&6.79512&36.264&0.02080&600.001\\
6411.902064*&-73.15969&0.09454&1.71111&2.286&789.3231&600.000\\
6438.878243&-60.76828&0.01674&6.81762&37.631&-0.00578&900.000\\
6454.847018&-60.87117&0.01534&6.72971&36.706&-0.03163&600.001\\
6457.848776&-60.77664&0.02008&6.77591&36.474&0.00183&600.001\\
6459.922794&-60.93969&0.01386&6.78047&36.319&-0.02629&600.001\\
6510.582235&-60.74526&0.01857&6.81107&36.977&0.06192&600.001\\
6510.591263&-60.70229&0.01341&6.82417&36.599&0.00540&600.001\\
6510.601795&-60.72294&0.00980&6.75473&36.746&-0.00742&900.002\\
6510.613242&-60.73820&0.00957&6.85763&36.762&-0.01984&900.001\\
6510.624769&-60.72645&0.00900&6.77831&36.900&-0.03973&900.000\\
6510.636309&-60.73310&0.00930&6.79062&36.931&-0.04372&900.001\\
6510.647628&-60.74584&0.00988&6.77485&36.935&0.00680&900.001\\
6510.659260&-60.74602&0.00985&6.78777&37.029&0.00791&900.001\\
6510.670810&-60.75529&0.01062&6.81317&37.005&0.01163&900.001\\
6510.682546&-60.75335&0.01157&6.77487&37.079&-0.04139&900.000\\
6510.693877&-60.74670&0.01072&6.78601&36.671&-0.02780&900.002\\
6510.705324&-60.74468&0.01284&6.76216&36.814&0.02584&900.001\\
6510.717072&-60.75111&0.01365&6.77729&36.867&0.01253&900.001\\
6510.728495&-60.75617&0.01286&6.80405&36.945&-0.03819&900.001\\
6510.740046&-60.78206&0.01203&6.75481&37.010&-0.00396&900.000\\
6510.751689&-60.77077&0.01090&6.79623&36.943&-0.00076&900.001\\
6510.763124&-60.75516&0.01073&6.76794&37.020&-0.01169&900.001\\
6510.774455&-60.75574&0.01039&6.79067&36.959&0.00137&900.001\\
6510.786527&-60.78525&0.01380&6.74017&36.801&-0.01100&900.000\\
6511.701655&-60.81984&0.01049&6.77291&36.841&-0.03984&900.001\\
6564.490867&-60.51083&0.01246&6.78860&36.623&-0.02331&600.001\\
6565.479782&-60.69072&0.01538&6.75616&35.156&-0.01447&600.000\\
6736.904747&-58.61384&0.01996&6.74955&37.023&-0.06442&900.001\\
6761.912737&-58.54571&0.01394&6.80035&37.818&-0.03933&900.001\\
6801.852737&-58.92484&0.03752&6.73650&39.960&0.02316&900.001\\
6927.561062&-59.65603&0.01215&6.81727&37.578&-0.01936&900.002\\
\end{longtable}

\twocolumn
\section{the visual companion to WASP-81}\label{app:obs}

\begin{figure}  
\center
\includegraphics[width= 0.45\textwidth]{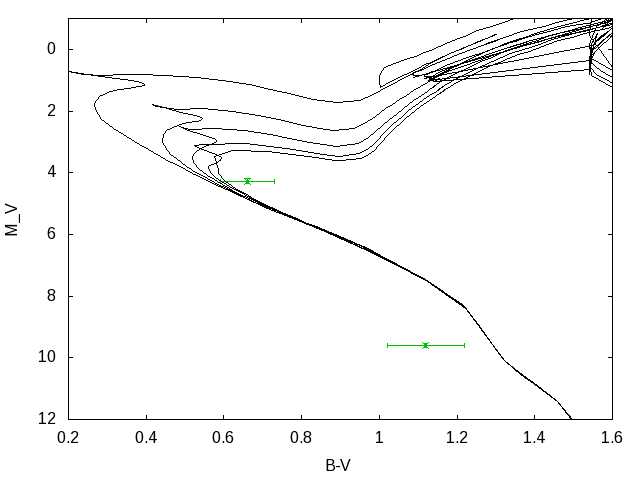}
\caption{Hertzsprung--Russell diagram showing the relative positions of WASP-81A and WASP-81B assuming a similar distance. The pair is unlikely to be related. Models are from \citet{Marigo:2008lr}.
}\label{fig:w81b_hr}  
\end{figure}

\section{Models applied to the photometric data}\label{app:models}

\begin{table*}
\begin{center}
\scriptsize{
\begin{tabular}{cccccclc}
\hline
\hline 
System	& Date	& Instrument	& Filter	& $T_{exp}$	&$N_p$	& Baseline	& CF	\\
		&		&			&		&			&		& function 	&      \\     
\hline
WASP-53	& 2011-07-22  & EulerCAM	& Gunn $r$'  		& 90s	& 110	& $p(t^2+xy)$						& 1.3  \\  
WASP-53	& 2011-09-03  & EulerCAM 	& Gunn $r$'  		& 120s	& 84		& $p(t^2+xy)$						& 1.8  \\ 
WASP-53	& 2011-09-13  & TRAPPIST	& blue blocking		& 12s	& 744	& $p(t^2)+o$						& 1.5  \\  
WASP-53	& 2011-09-13  & EulerCAM 	& Gunn $r$'  		& 80s	& 127	& $p(t^2+xy+ \sin (P_{3}t+T_{0,3}))$		& 1.7  \\ 
WASP-53	& 2011-09-23  & EulerCAM 	& Gunn $r$'  		& 180s	& 77		& $p(t^2+xy +\sin (P_{4}t+T_{0,4}))$		& 2.3  \\  
WASP-53	& 2011-10-26  & EFOSC2		& Gunn $r$' 		& 150s	& 54		& $p(t^2)$							& 1.0  \\  
WASP-53	& 2012-07-30  & TRAPPIST 	& blue blocking		& 12s	& 603	& $p(t^2)$							& 1.5 \\  
WASP-53	& 2012-11-03  & TRAPPIST 	& $I+z'$			& 15s	& 464	& $p(t^2)$							& 1.0  \\ \noalign {\smallskip}

WASP-81	& 2011-09-26  & TRAPPIST 	& $I+z'$			& 12s	& 287	& $p(t^2)+o$						& 0.9  \\ 
WASP-81	& 2012-05-20  & TRAPPIST 	& $I+z'$			& 25s	& 372	& $p(t^2)+o$						& 1.1  \\ 
WASP-81	& 2012-05-31  & TRAPPIST 	& $I+z'$			& 12s	& 812	& $p(t^2)+o$						& 1.4  \\ 
WASP-81	& 2012-06-19  & TRAPPIST 	& $blue blocking$	& 8s		& 644	& $p(t^2)+o$						& 1.5  \\ 
WASP-81	& 2012-07-08  & EulerCAM	& Gunn $r$'  		& 120s	& 116	& $p(t^2)$							& 1.9  \\  
WASP-81	& 2012-07-19  & TRAPPIST 	& $I+z'$			& 20s	& 501	& $p(t^2)+o$						& 1.6  \\ 
WASP-81	& 2012-09-24  & EulerCAM	& Gunn $r$'  		& 120s	& 132	& $p(t^2)$							& 1.4  \\  
WASP-81	& 2013-07-07  & TRAPPIST 	& blue blocking		& 10s	& 917	& $p(t^2)+o$						& 1.9  \\ 
WASP-81	& 2013-08-06  & EulerCAM	& Gunn $r$'  		& 80s	& 248	& $p(t^2)$							& 1.4  \\  
WASP-81	& 2013-08-06  & TRAPPIST 	& blue blocking		& 10s	& 1082	& $p(t^2)+o$						& 1.3  \\ 

\hline
\end{tabular}}
\caption{Photometric time-series used in this work. For each light curve this table shows the date of 
acquisition, the instrument and filter used, the exposure time $T_{\rm exp}$, the number of data points, the baseline function selected for 
our global analysis (see Sec.~\ref{sec:analysis}), and the error correction factor $CF$ used in our global analysis. For the baseline function, $p(\epsilon^N)$ denotes, respectively, a $N$-order polynomial function of
 time ($\epsilon=t$), $x$ and $y$ positions ($\epsilon=xy$); $o$ denotes an offset at the time of a meridian flip of TRAPPIST (see \citealt{Gillon:2012fj}). On two instances we also fit
 a sinusoidal baseline of the form $\sin(Pt + T_0)$ where $P$ is the period, and $T_0$ is the phase. }\label{tab:phot}
\end{center}
\end{table*}

\begin{figure*}  
\begin{center}
\begin{subfigure}[b]{0.33\textwidth}
	\caption{EulerCam $r'$ 2011-07-22}
	\includegraphics[width=\textwidth]{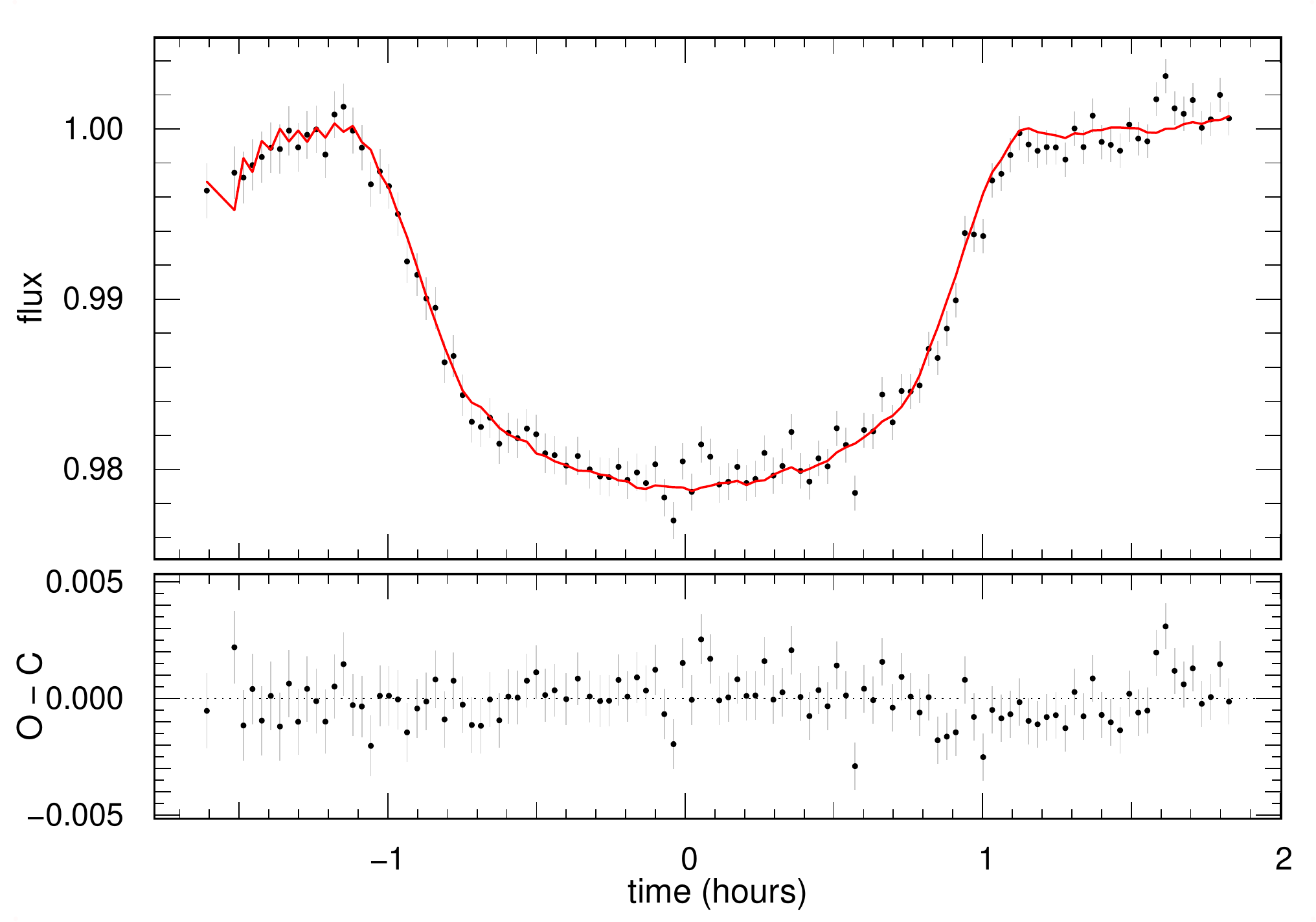}
\end{subfigure}
\begin{subfigure}[b]{0.33\textwidth}
	\caption{EulerCam $r'$ 2011-09-03}
	\includegraphics[width=\textwidth]{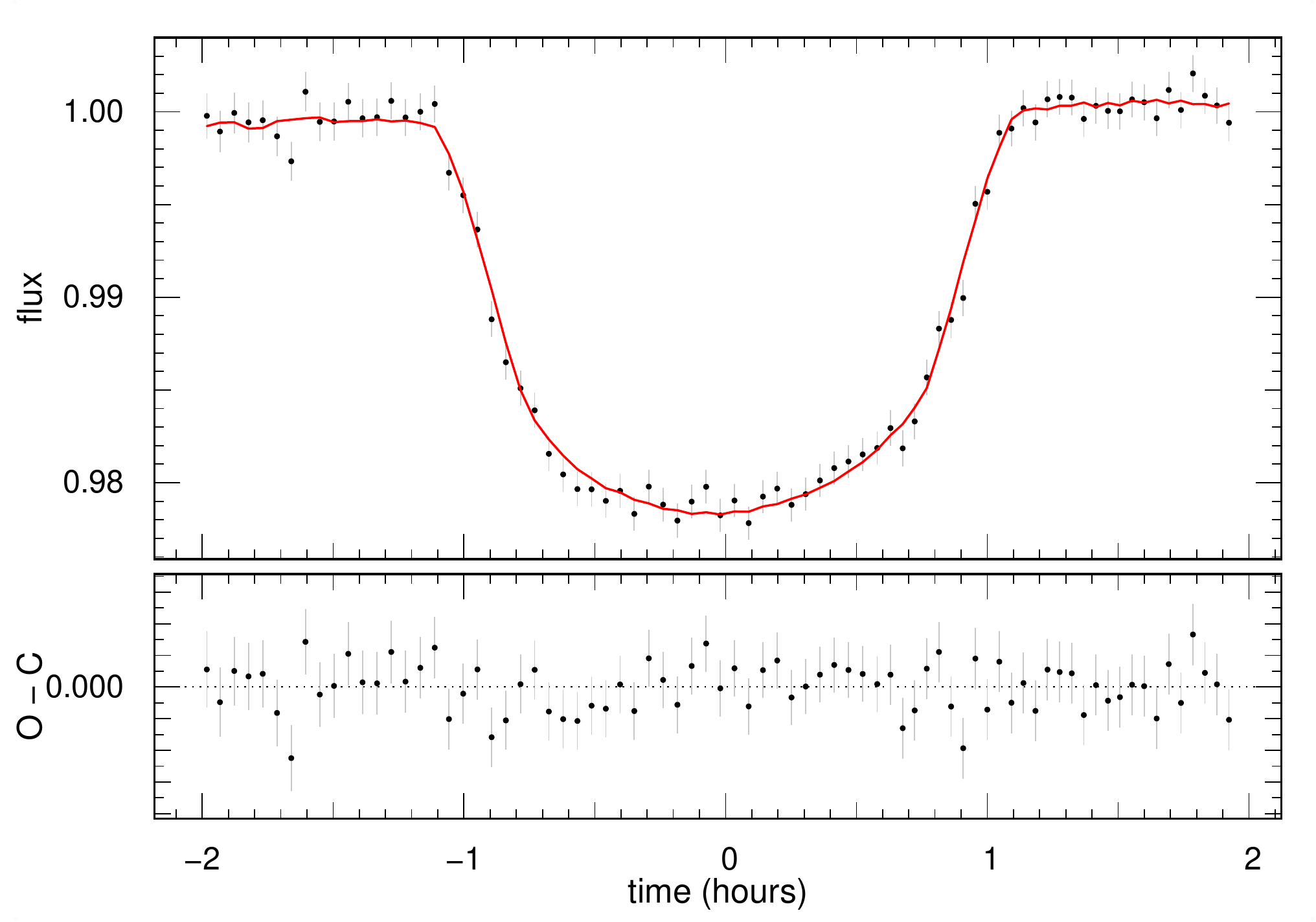}
\end{subfigure}
\begin{subfigure}[b]{0.33\textwidth}
	\caption{TRAPPIST $BB$ 2011-09-13}
	\includegraphics[width=\textwidth]{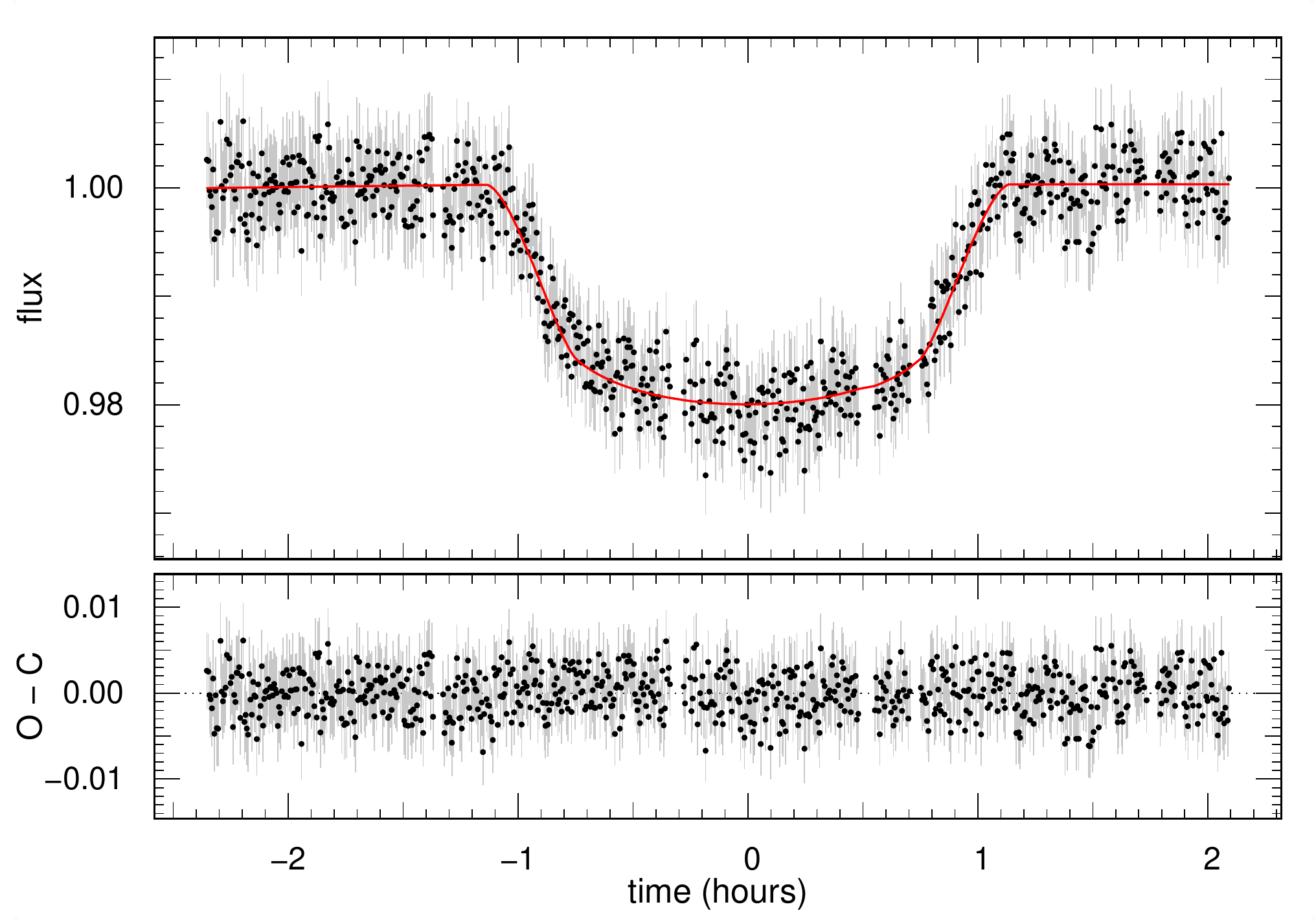}
\end{subfigure}

\begin{subfigure}[b]{0.33\textwidth}
	\caption{EulerCam $r'$ 2011-09-13}
	\includegraphics[width=\textwidth]{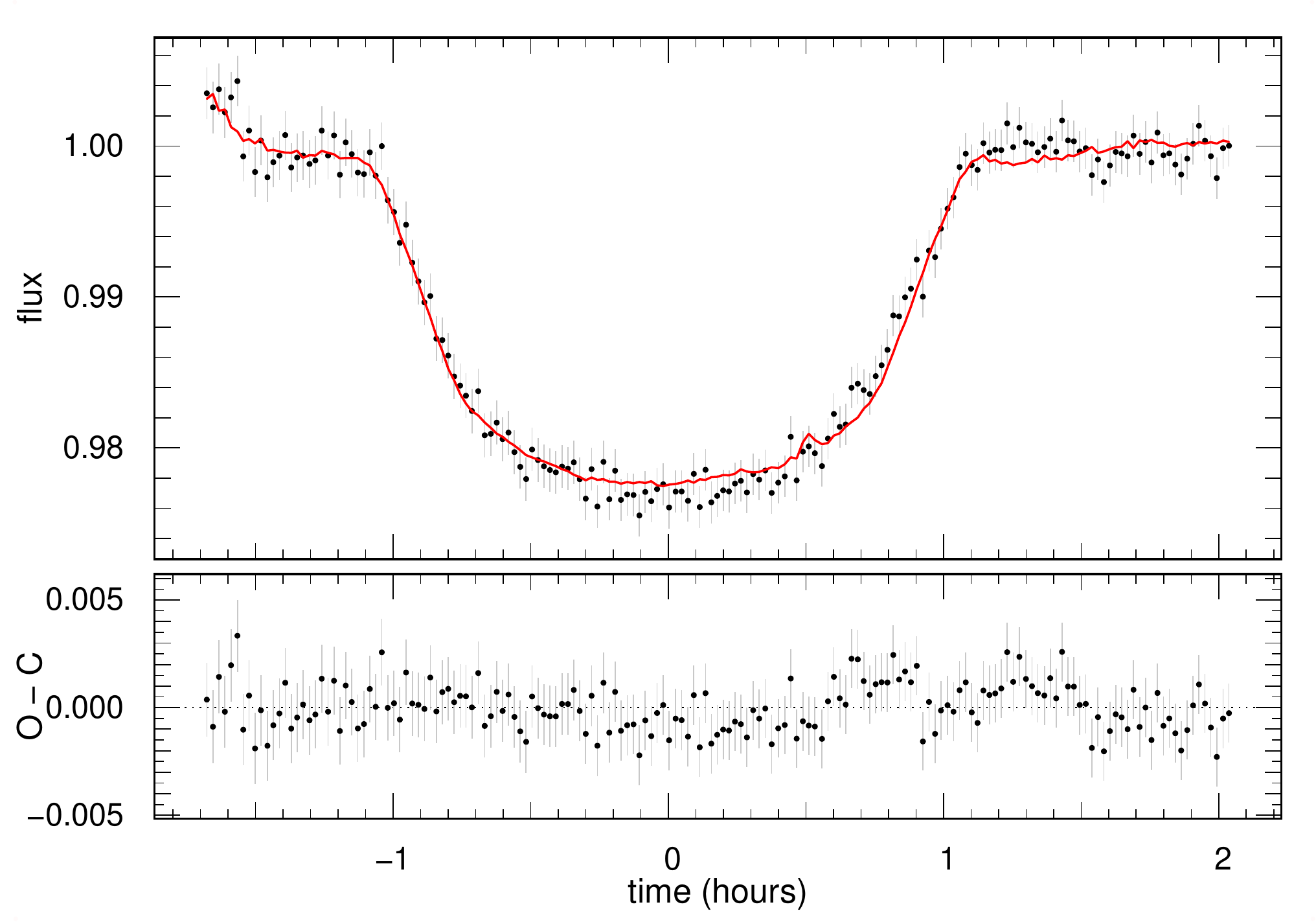}
\end{subfigure}
\begin{subfigure}[b]{0.33\textwidth}
	\caption{EulerCam $r'$ 2011-09-23}
	\includegraphics[width=\textwidth]{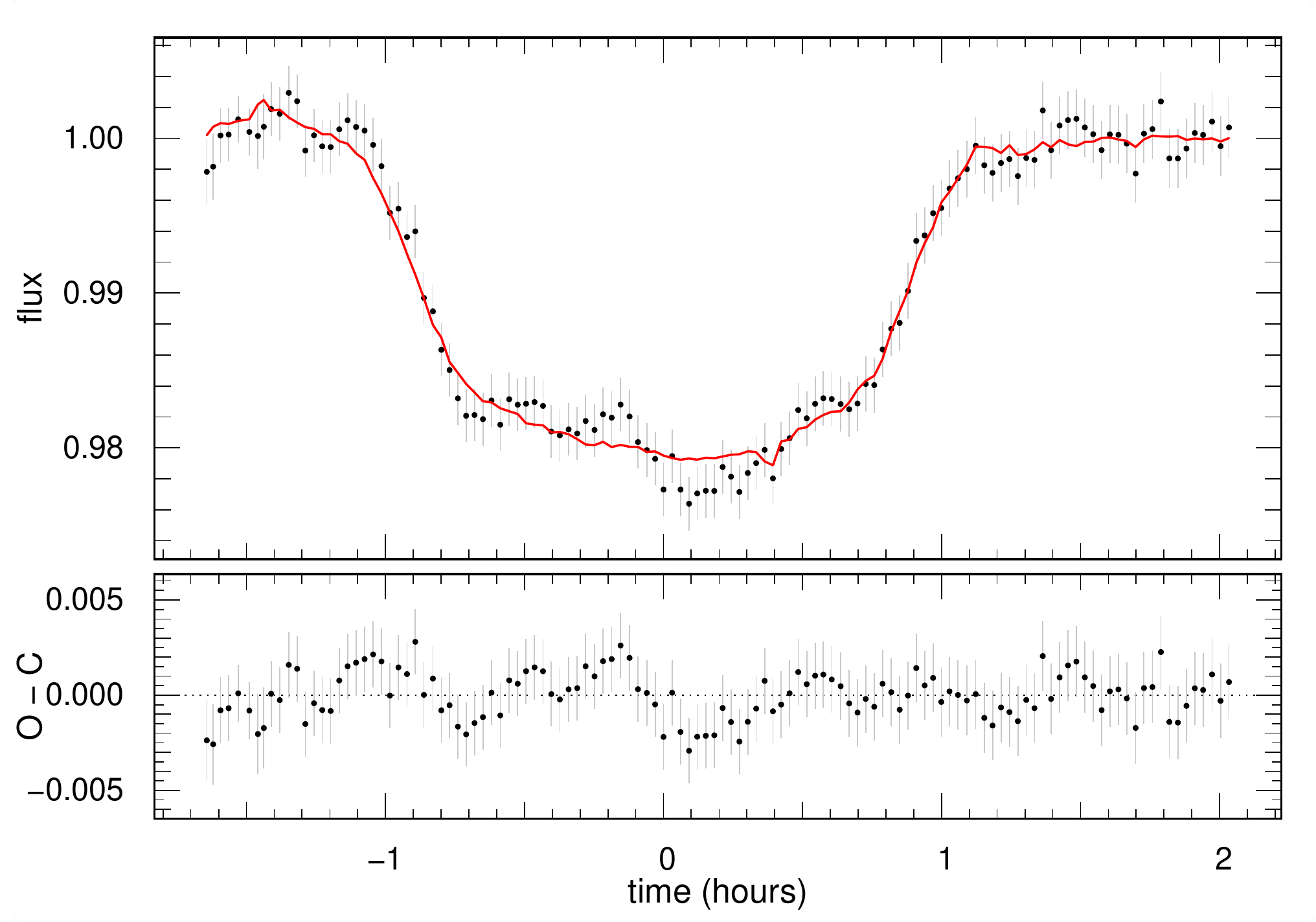}
\end{subfigure}
\begin{subfigure}[b]{0.33\textwidth}
	\caption{TRAPPIST $BB$ 2011-09-23}
	\includegraphics[width=\textwidth]{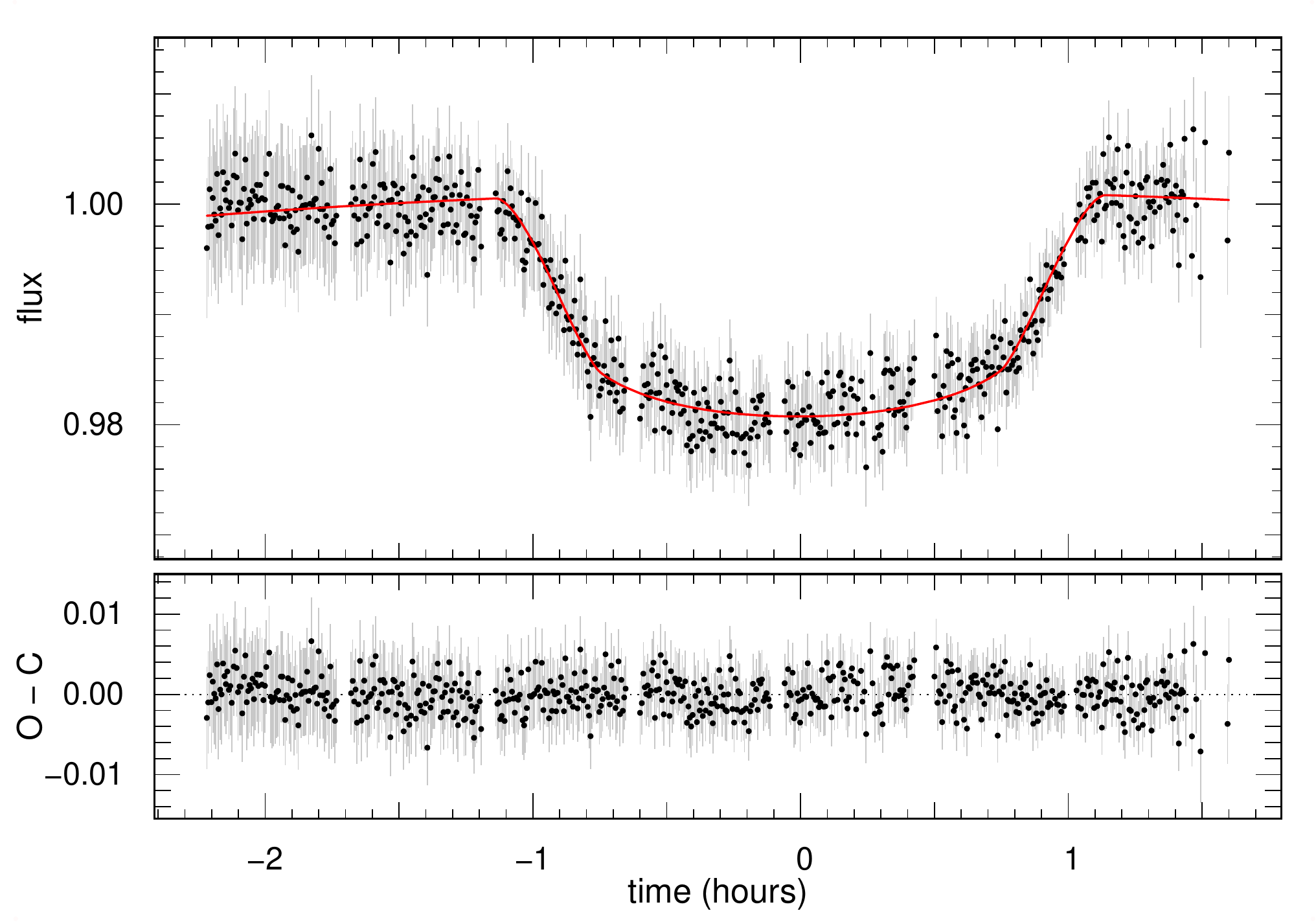}
\end{subfigure}

\begin{subfigure}[b]{0.33\textwidth}
	\caption{EFOSC2 $r'$ 2011-10-26}
	\includegraphics[width=\textwidth]{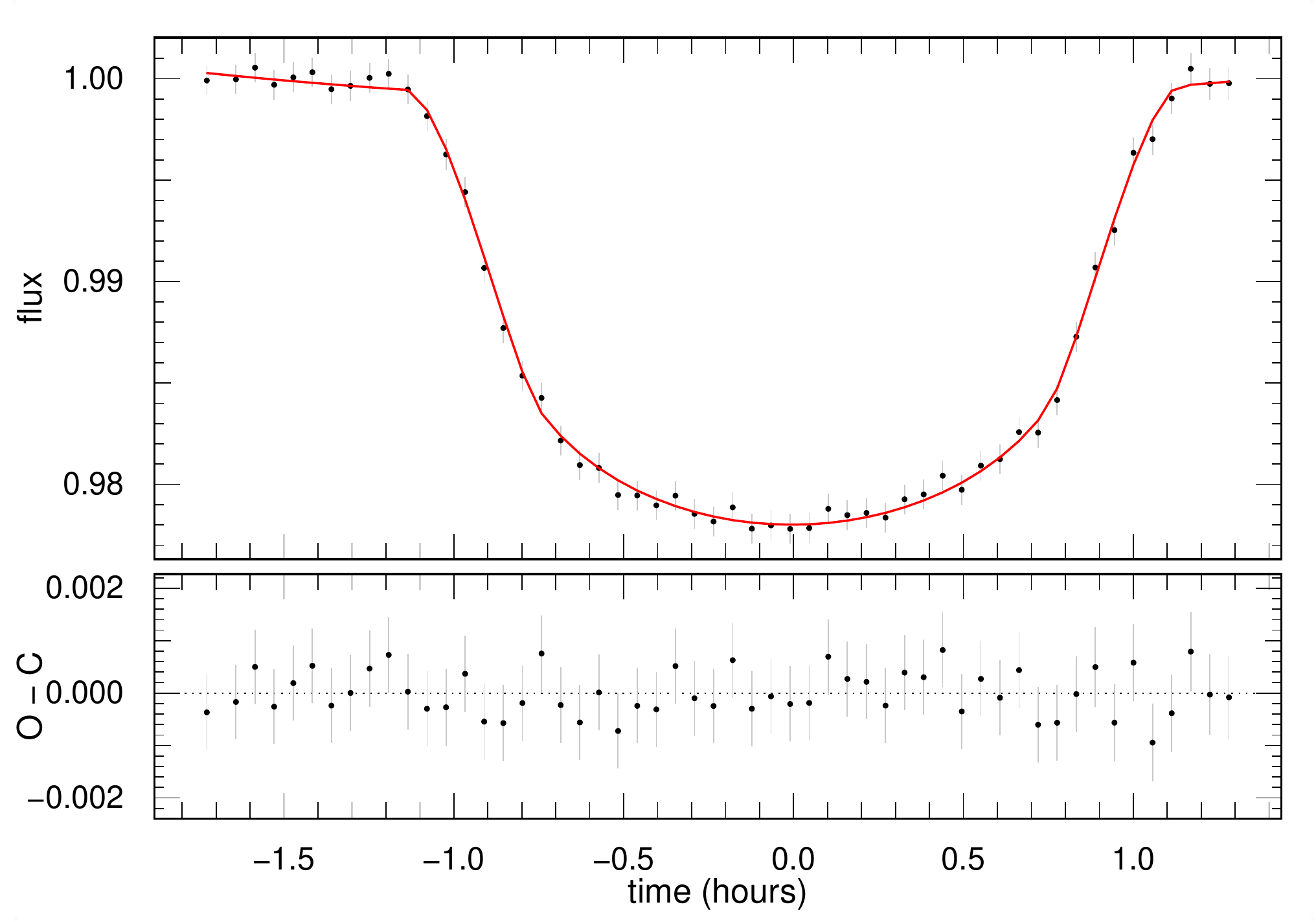}
\end{subfigure}
\begin{subfigure}[b]{0.33\textwidth}
	\caption{TRAPPIST $I+z'$ 2012-07-30}
	\includegraphics[width=\textwidth]{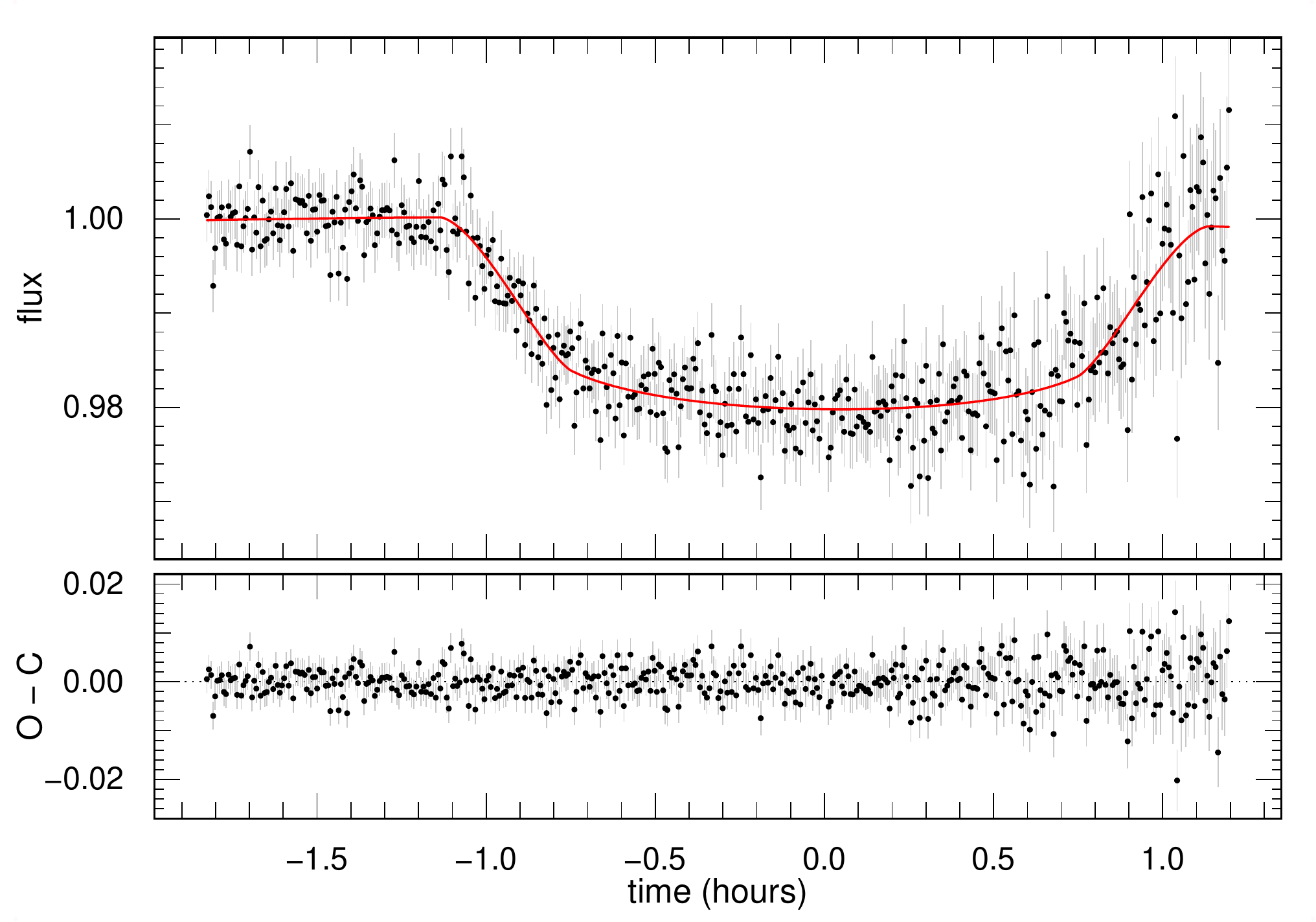}
\end{subfigure}
\begin{subfigure}[b]{0.33\textwidth}
	\includegraphics[width=\textwidth]{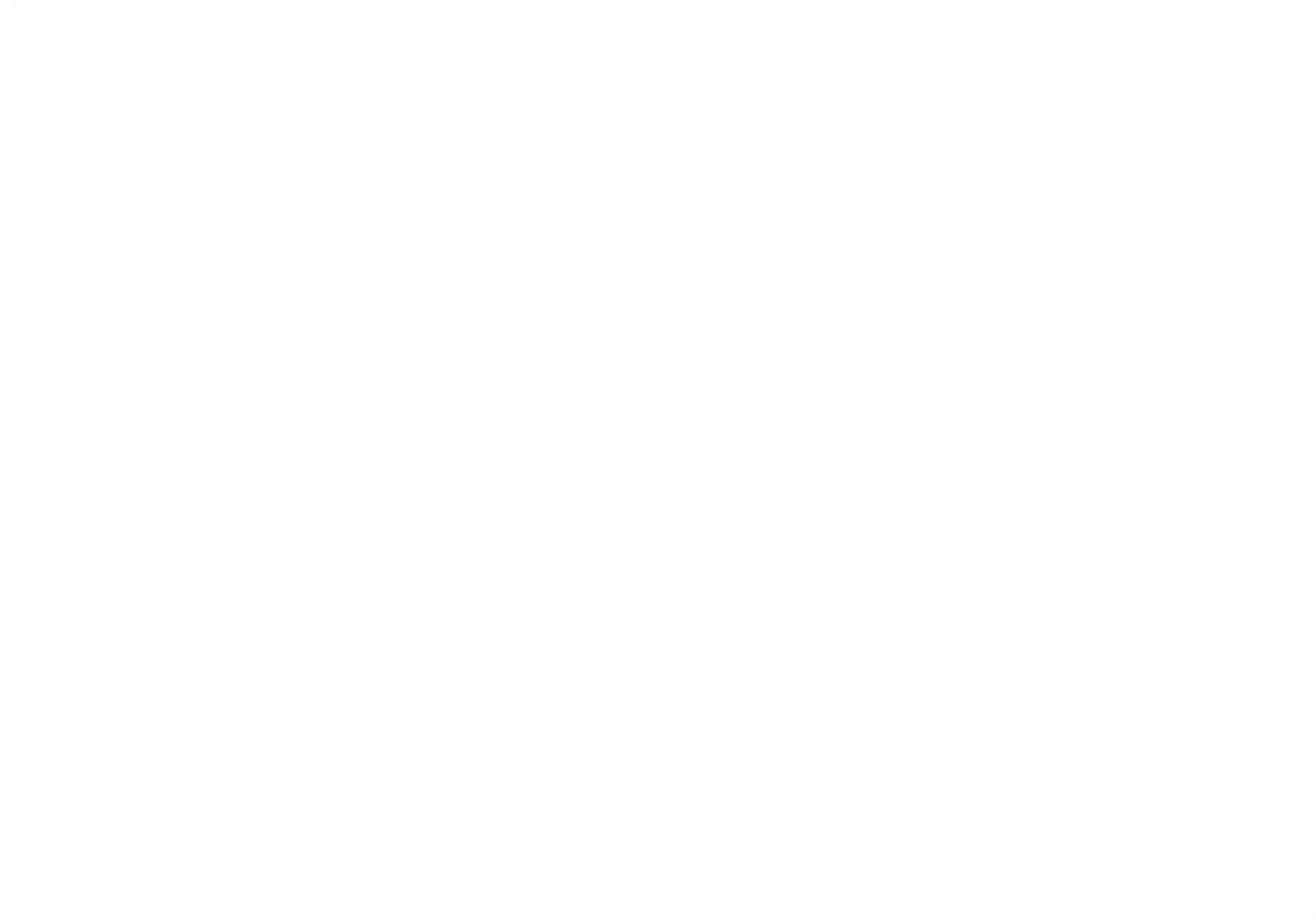}
\end{subfigure}

\caption{Flux as a function of time, centred around mid-transit time of WASP-53b. The red line shows the full model, including the detrending.
}\label{fig:phot53raw}  
\end{center}
\end{figure*}

\begin{figure*}  
\begin{center}
\begin{subfigure}[b]{0.33\textwidth}
	\caption{TRAPPIST $I+z'$ 2011-09-26}
	\includegraphics[width=\textwidth]{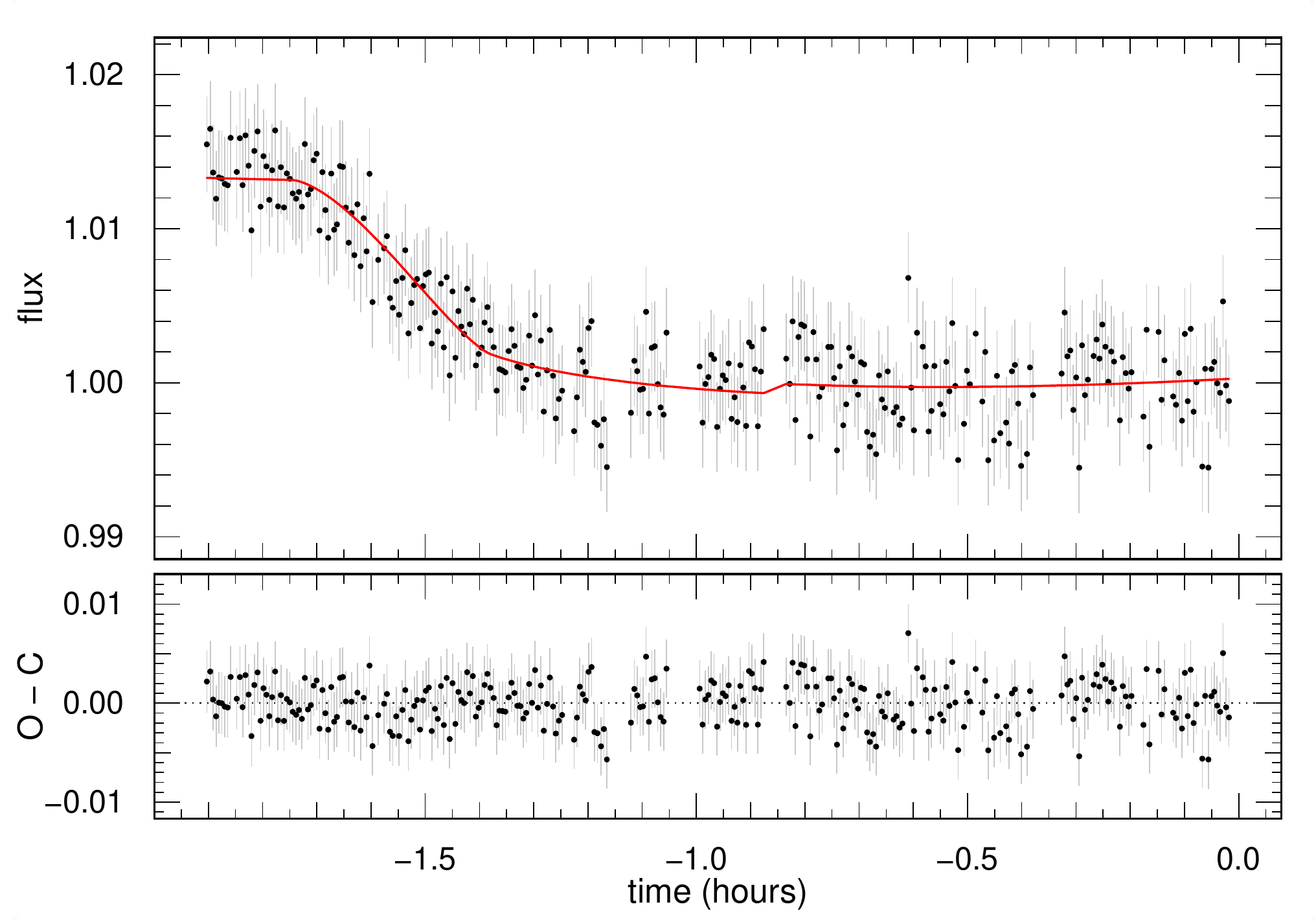}
\end{subfigure}
\begin{subfigure}[b]{0.33\textwidth}
	\caption{TRAPPIST $I+z'$ 2012-05-20}
	\includegraphics[width=\textwidth]{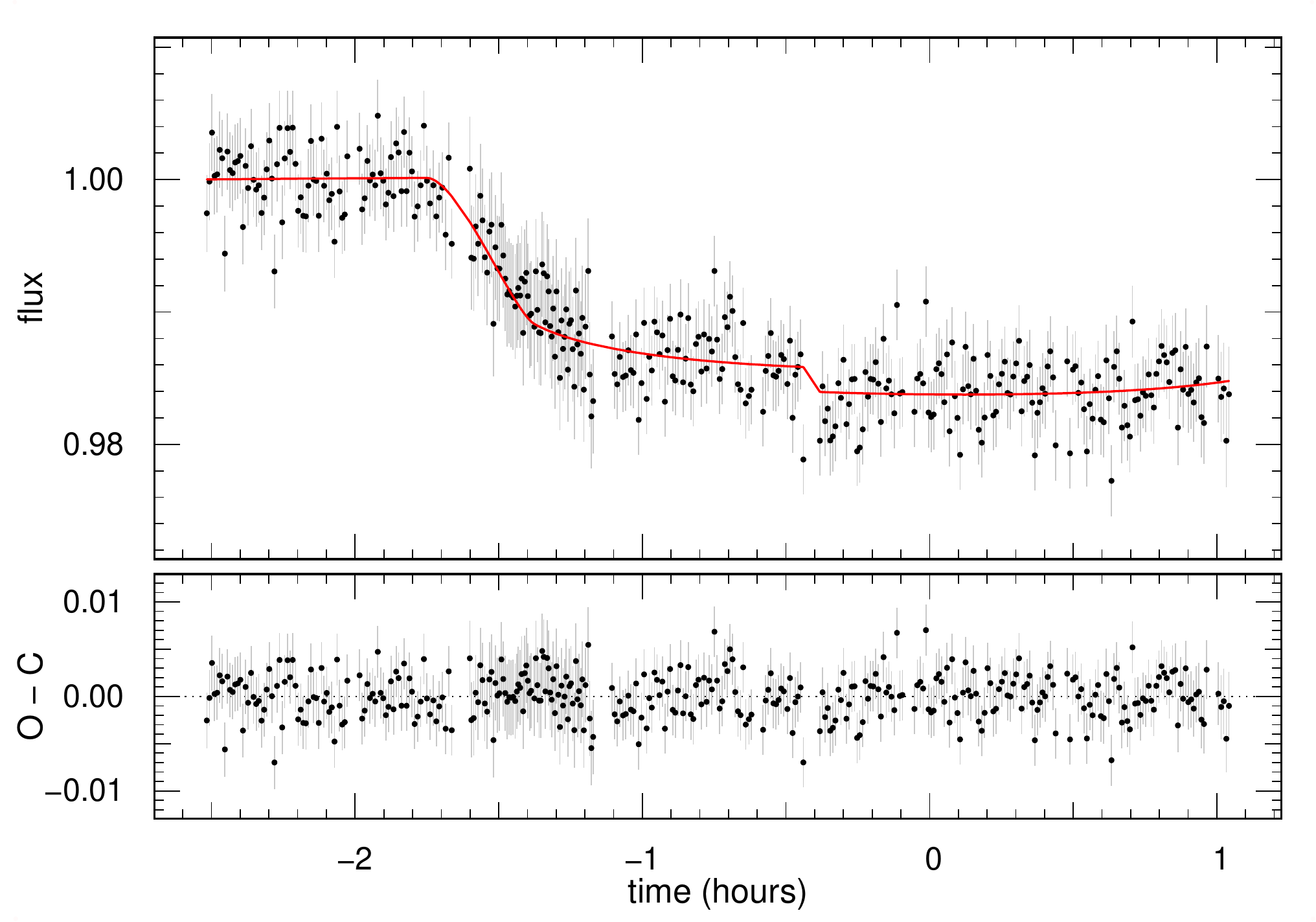}
\end{subfigure}
\begin{subfigure}[b]{0.33\textwidth}
	\caption{TRAPPIST $I+z'$ 2012-05-31}
	\includegraphics[width=\textwidth]{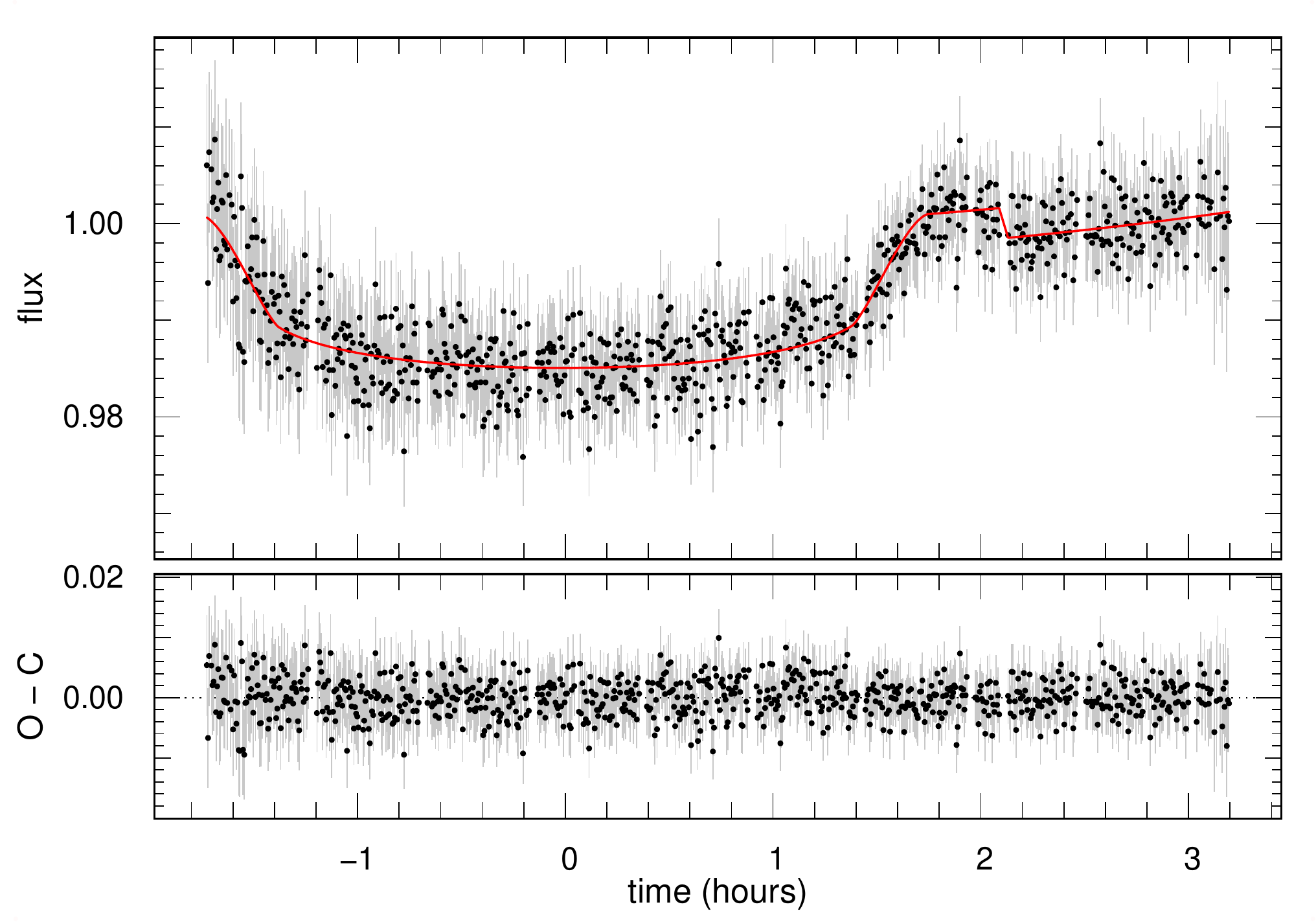}
\end{subfigure}

\begin{subfigure}[b]{0.33\textwidth}
	\caption{TRAPPIST $I+z'$ 2012-06-19}
	\includegraphics[width=\textwidth]{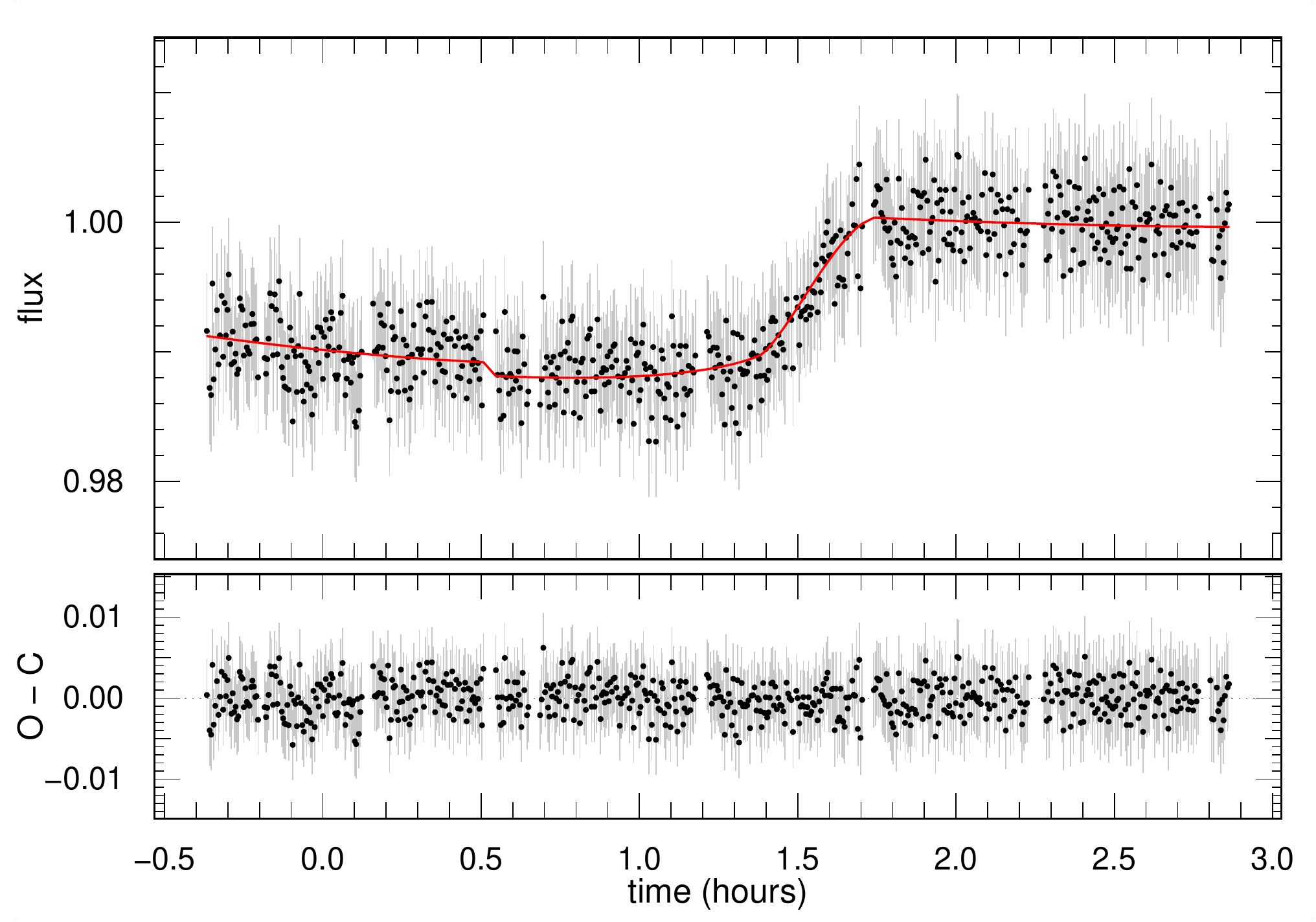}
\end{subfigure}
\begin{subfigure}[b]{0.33\textwidth}
	\caption{EulerCam $r'$  2012-07-08}
	\includegraphics[width=\textwidth]{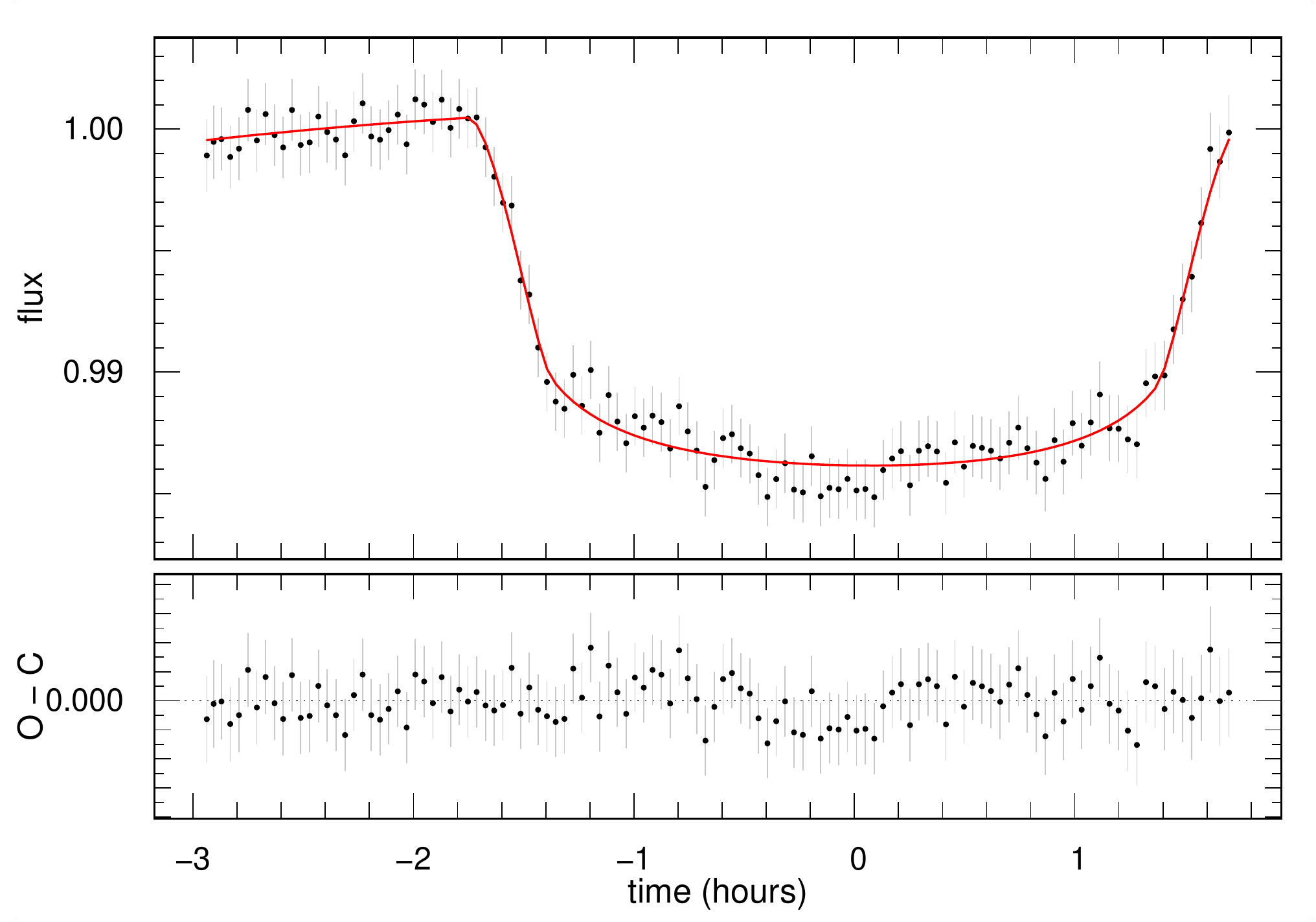}
\end{subfigure}
\begin{subfigure}[b]{0.33\textwidth}
	\caption{TRAPPIST $I+z'$ 2012-07-19}
	\includegraphics[width=\textwidth]{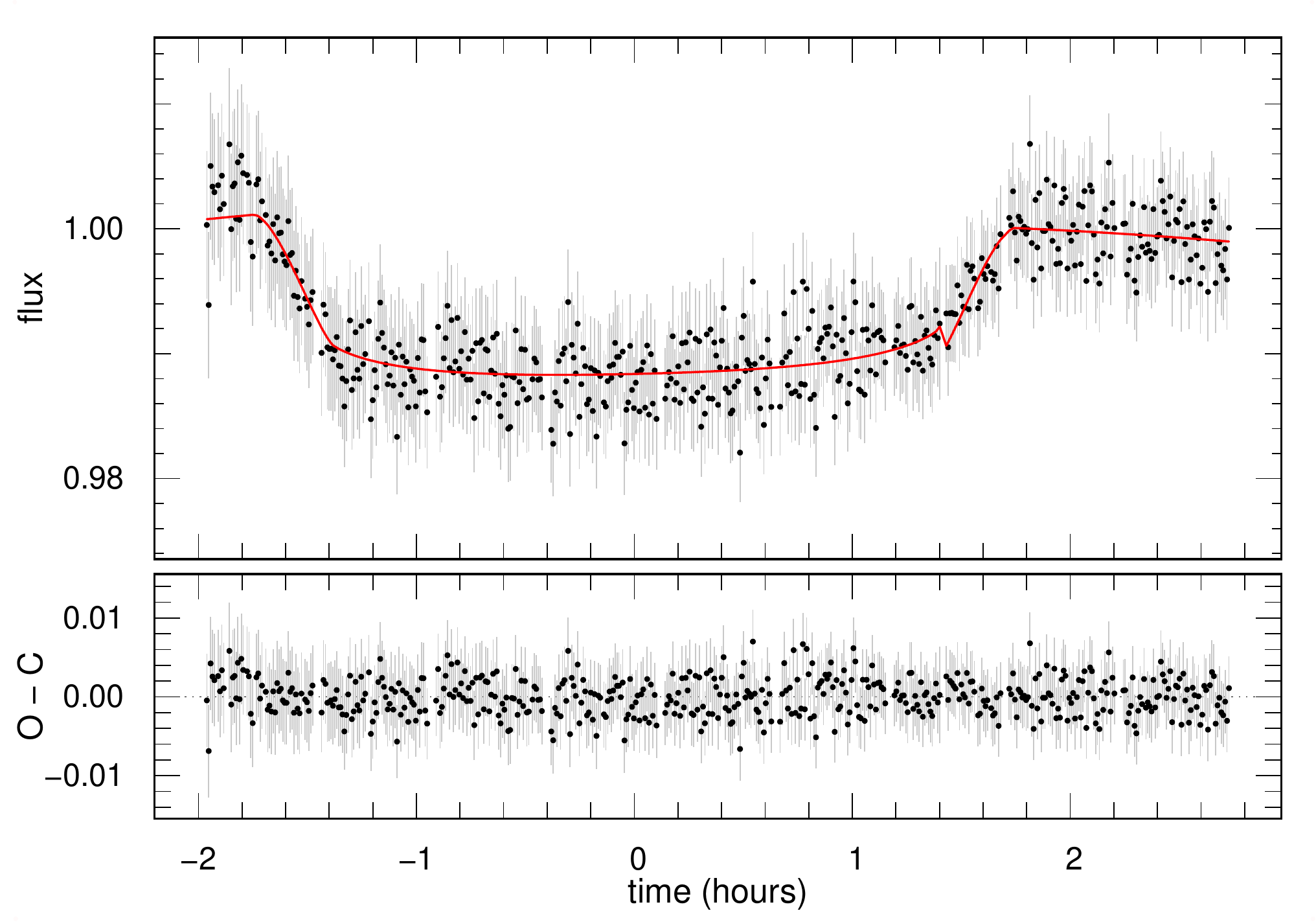}
\end{subfigure}

\begin{subfigure}[b]{0.33\textwidth}
	\caption{EulerCam $r'$   2012-09-24}
	\includegraphics[width=\textwidth]{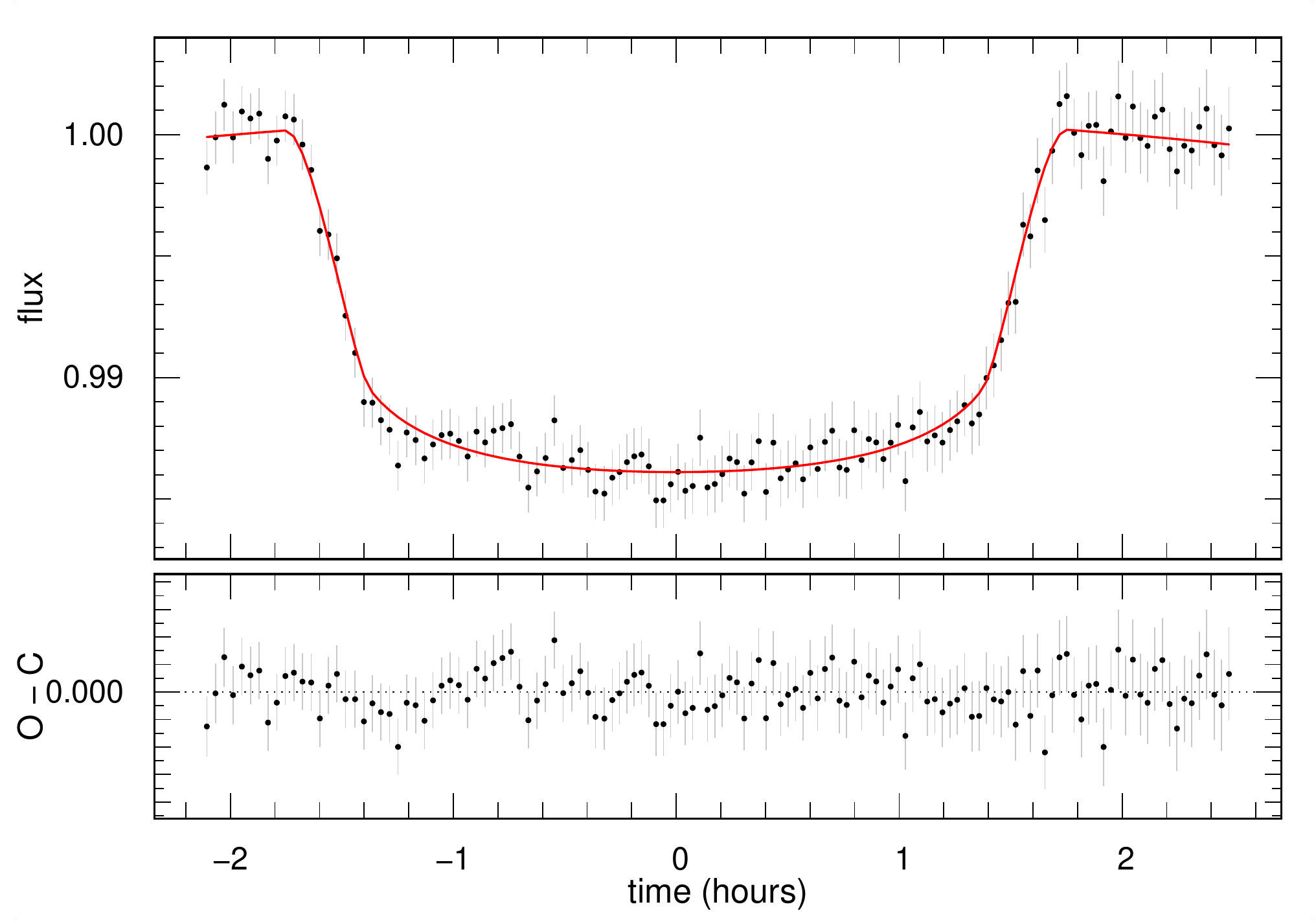}
\end{subfigure}
\begin{subfigure}[b]{0.33\textwidth}
	\caption{TRAPPIST $BB$ 2013-07-07}
	\includegraphics[width=\textwidth]{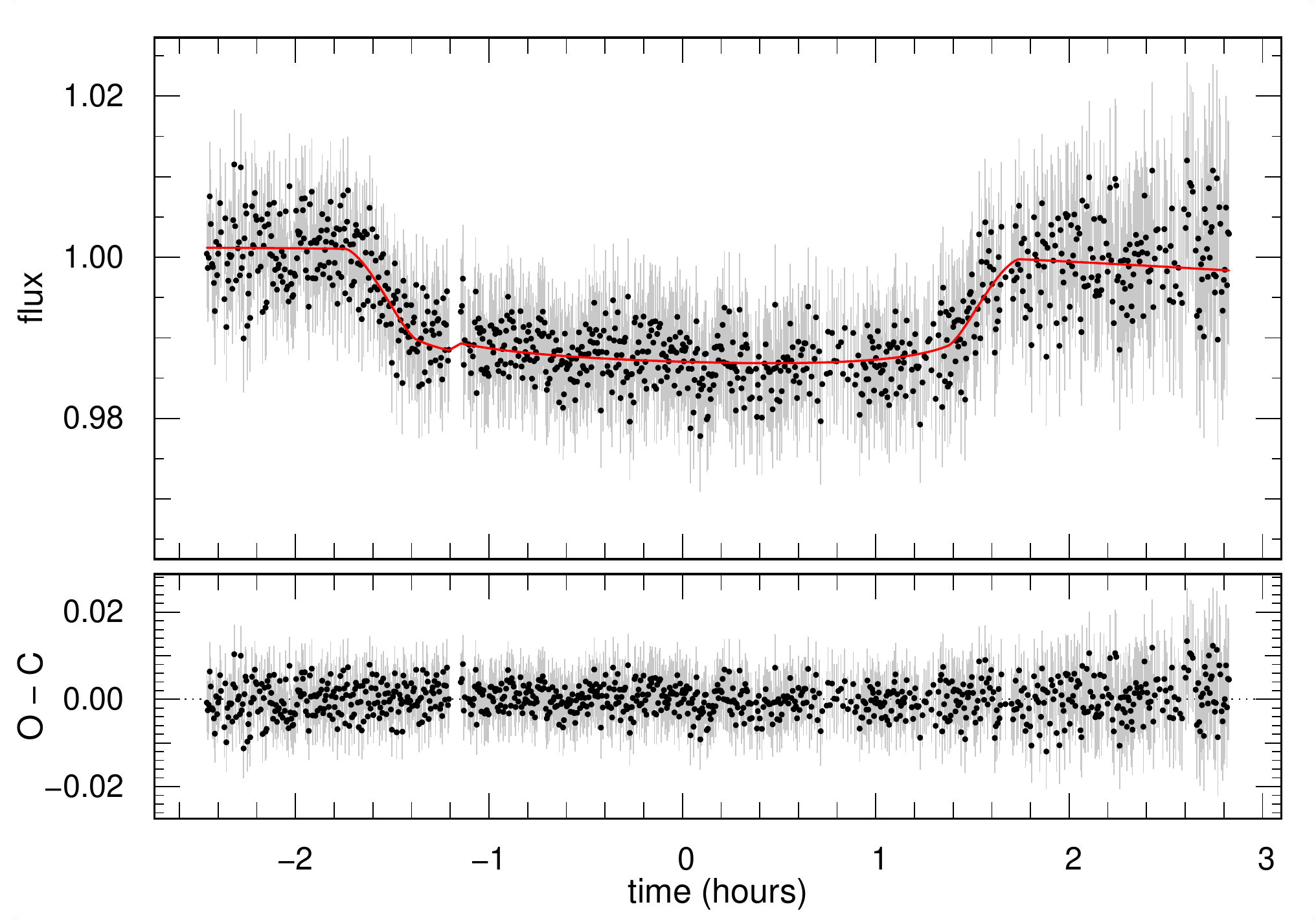}
\end{subfigure}
\begin{subfigure}[b]{0.33\textwidth}
	\caption{EulerCam $r'$  2013-08-06}
	\includegraphics[width=\textwidth]{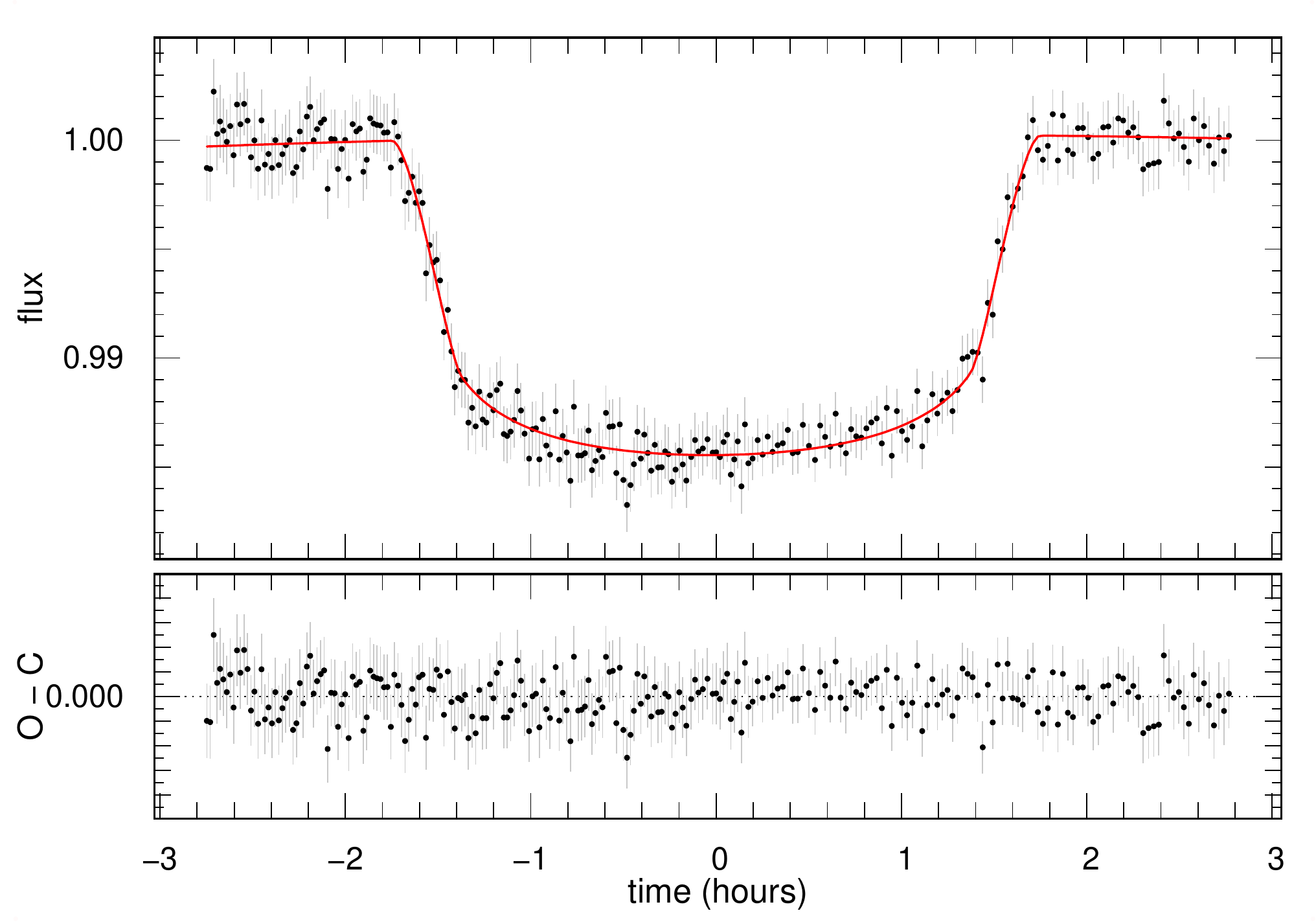}
\end{subfigure}

\begin{subfigure}[b]{0.33\textwidth}
	\caption{TRAPPIST $BB$ 2013-08-06}
	\includegraphics[width=\textwidth]{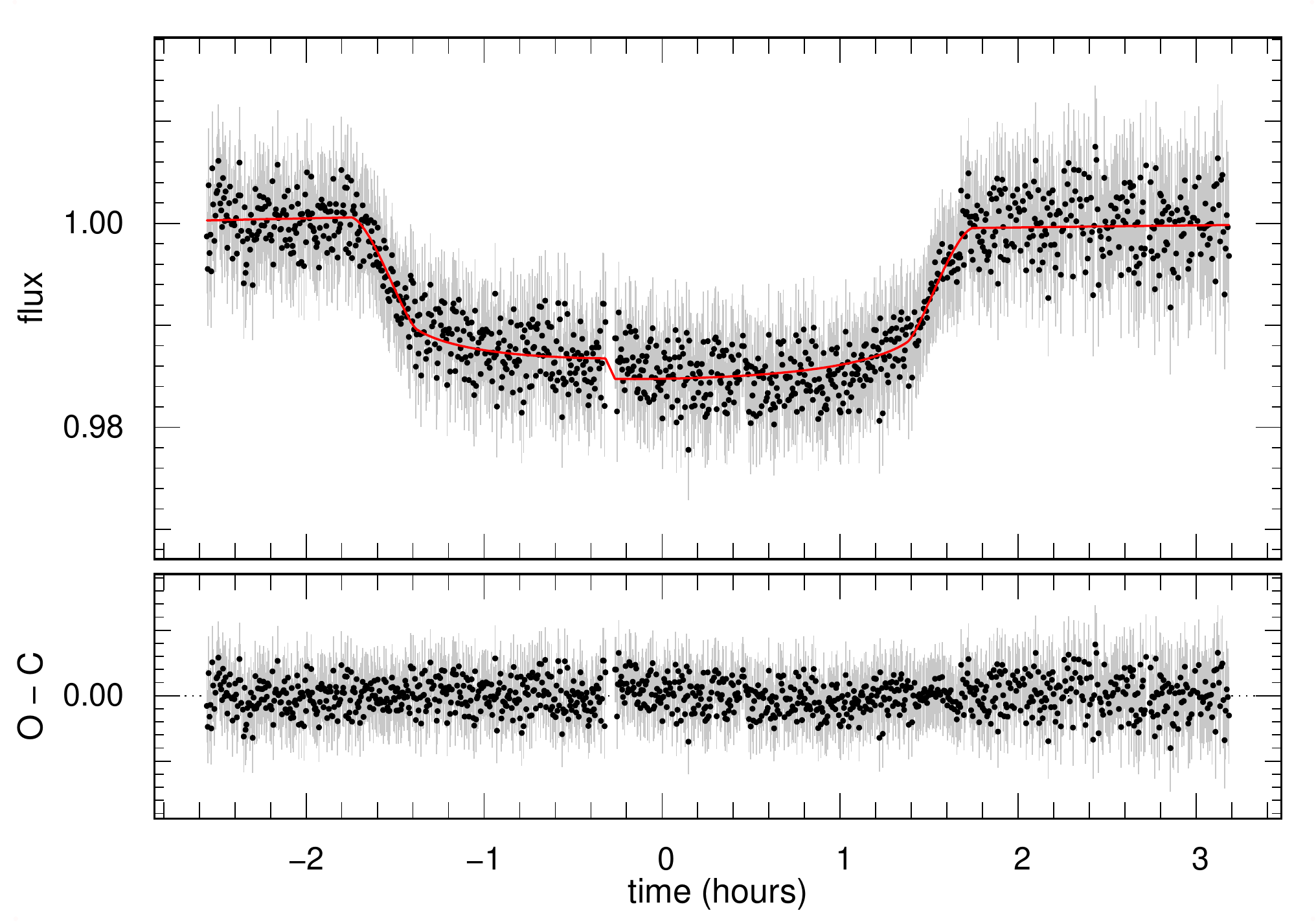}
\end{subfigure}
\begin{subfigure}[b]{0.33\textwidth}
	\includegraphics[width=\textwidth]{white.pdf}
\end{subfigure}
\begin{subfigure}[b]{0.33\textwidth}
	\includegraphics[width=\textwidth]{white.pdf}
\end{subfigure}

\caption{Flux as a function of time, centred around mid-transit time of WASP-81b. The red line shows the full model, including the detrending.
}\label{fig:phot81raw}  
\end{center}
\end{figure*}

\newpage

\bsp

\label{lastpage}

\end{document}